\documentclass[aps,prb,twocolumn,nofootinbib, superscriptaddress,10pt]{revtex4-2}
\usepackage{amsmath,amsthm,amsfonts,amssymb,amscd,bbold,mathrsfs,bm}
\usepackage[T1]{fontenc}
\usepackage{dcolumn}
\usepackage{physics}
\usepackage{graphicx}
\usepackage{epstopdf}
\usepackage{pgf}
\usepackage{tabularx}
\usepackage{subfigure}
\usepackage{hyperref}
\usepackage{simpler-wick}
\usepackage{tikz}
\usetikzlibrary{arrows.meta,positioning}
\usetikzlibrary{shadows}
\usetikzlibrary{calc}
\usetikzlibrary{positioning}

\newcommand*{\me}[3]{[#1]^{#2}_{#3}}
\begin{document}
\title{Inter-band coherence effects in disordered crystals: beyond the non-crossing approximation}
\author{Zhanning Wang}
\affiliation{School of Physics, The University of New South Wales, Sydney 2052, Australia}
\author{James H. Cullen}
\affiliation{School of Physics, The University of New South Wales, Sydney 2052, Australia}
\author{Roberto Raimondi}
\affiliation{Dipartimento di Matematica e Fisica, Universit\`a Roma Tre, Via della Vasca Navale 84, 00146 Roma, Italy}
\author{Dimitrie Culcer}
\affiliation{School of Physics, The University of New South Wales, Sydney 2052, Australia}
\date{\today}
\begin{abstract}
We develop a quantum kinetic theory for Bloch electrons driven by a uniform dc electric field, extending the non-equilibrium density matrix formalism beyond the non-crossing approximation. Such a procedure is necessary in order to capture correctly steady-state contributions nominally of zeroth order in the disorder strength, which compete with the intrinsic contributions to non-equilibrium processes. These terms have been shown diagrammatically to be important in the anomalous Hall effect, and are expected to play an important role in the spin- and orbital Hall effects, as well as related magnetic phenomena. Here we demonstrate the correct procedure for incorporating the relevant crossing terms into the density matrix kinetic equation (DMKE) approach, which is the natural language for describing inter-band coherence effects involving band geometry and extrinsic scattering, as well as their decomposition into Fermi surface and Fermi sea contributions. We work in the length gauge, with an impurity potential $V$ given by Gaussian white noise, and account for impurity scattering up to the fourth order in the disorder potential. We show that the higher-order terms in the potential give rise to a collision integral whose connected $V^4$ part separates into self energy corrections, ladder type vertex renormalization, and crossed quantum interference contributions. The construction follows from an iterative solution for the impurity induced density matrix fluctuations and explicitly removes disconnected impurity pairings to avoid double counting. We discuss in detail the correspondence between the resulting density matrix kinetic equation and the Keldysh formalism. As an example, we apply the formalism to the two-dimensional massive Dirac fermion model, obtaining analytical expressions for the single particle lifetime, the transport relaxation time, and the longitudinal conductivity at the level of the Born approximation. Building on that, we evaluate the anomalous Hall conductivity, showing that crossed impurity processes generate an extrinsic contribution of order $\tau^0$ that coexists with the intrinsic Berry curvature term; for Gaussian white noise impurity in this model the $\Psi$-type term cancels, while the $X$-type term remains finite. Our approach provides a blueprint for incorporating band geometry and crossed disorder corrections into multi-band transport, opening the door to a consistent DMKE description of spin-, pseudospin-, orbital- and valley-related phenomena in the steady state, and potential extensions into the non-linear regime.
\end{abstract}
\maketitle
\section{Introduction}
\label{Section_1_Introduction}

Recent decades have brought to the forefront of condensed matter physics research a host of non-equilibrium phenomena in which inter-band dynamics either dominate the response to an electric field or play a significant role that  must be accounted for thoroughly. In linear response such phenomena include the anomalous, spin, valley and orbital Hall effects, together with closely related magnetic phenomena such as the spin-orbit and orbital torque, while in non-linear response the rectification, shift and injection currents, as well as higher harmonic generation, routinely require a detailed analysis of inter-band contributions. The natural theoretical framework for describing this physics relies on the density matrix, given that the density matrix kinetic equation (DMKE) systematically accounts for quantum geometric terms involving e.g. the Berry curvature and the quantum metric tensor, as well as Zitterbewegung, and inter-band coherence effects induced by disorder. A thorough treatment of the interplay between diagonal and off-diagonal elements in the density matrix is vital in capturing the underlying physics and determining the correct response to an electric field. The density matrix approach, where Berry connection effects and impurity scattering can be treated together, enables a transparent separation of intrinsic, side-jump, and skew contributions with their impurity scaling and semiclassical correspondence \cite{Raimondi2006, Culcer20102, Gao2014, Du2021, Atencia2022, Valet2023}. In the weakly disordered quasiparticle regime, it is known that quantum kinetic (both density matrix and Keldysh) formulations reproduce the standard non-crossing ladder structure of Kubo-Keldysh theory, including the correspondence between ladder vertex corrections and the kinetic collision terms and the separation into band diagonal populations and inter-band coherence \cite{Raimondi2025, Valet2025}.

It is now well established that intrinsic terms reflecting inter-band coherence are accompanied by disorder corrections that appear to be of zeroth order in the disorder strength, which may cancel or overwhelm the intrinsic contributions. Such disorder corrections appear already in the Born approximation and are captured by the generic density matrix formalism of Refs~\cite{Culcer2017, Rhonald2022, Rhonald2023}
via an \textit{anomalous driving term} and associated corrections accounting for skew scattering, side jump, and electric field effects on the collision integral. Nevertheless the seminal work of Ado \textit{et al} identified a series of terms beyond the non-crossing approximation that significantly alter the disorder contribution and the parameter dependence of response functions \cite{Ado2015, Ado2016}.
Since then additional works have highlighted the role of crossed impurity processes \cite{Ado2017, Zhang2023}.
Physically, the terms beyond the non-crossing (ladder) approximation come from coherent impurity pair multiple scattering, the electron can scatter from two impurities in different orders, and the interference of these amplitudes generates an antisymmetric scattering component relevant for Hall responses.
Indicating with $V$ the impurity potential, at order $V^{4}$ the above crossed impurity processes appear in the form of the $X$ and $\Psi$ diagrams \cite{Ado2015,Ado2016,Ado2017}, which produce an extrinsic Hall term of order $\tau^{0}$ that coexists with the intrinsic Berry-curvature contribution.
Whereas the structure of these corrections is clear in the diagrammatic formalism, the equivalent procedure has not been formulated in the DMKE framework. This is the task of the present effort.

In this work we develop a quantum kinetic treatment of impurity scattering for Bloch electrons driven by a uniform DC electric field within the DMKE approach. We provide a generic non-equilibrium formalism that treats longitudinal and transverse responses to an external perturbation on the same microscopic level, while consistently incorporating band geometry, inter-band coherence, and impurity scattering beyond the non-crossing approximation. We decompose the single particle density matrix into its impurity averaged component and impurity induced fluctuations, and construct the collision integral by iteratively solving for these fluctuations in powers of the disorder potential for Gaussian white noise impurities. At second order we recover the Born collision operator and obtain the single particle and transport relaxation times. We then extend the construction to the connected $V^{4}$ order, where the collision operator separates into self-energy, ladder type, and crossed contributions, with disconnected pairings removed to avoid double counting. Importantly, we provide an explicit term-by-term dictionary to the Keldysh contour expansion: the commutator structure of the DMKE maps to the placement of impurity vertices on the forward/backward branches, and the connected $V^4$ collision operator reproduces the standard self-energy, ladder-vertex, and crossed $X/\Psi$ topologies.
The Keldysh non-equilibrium Green function formalism offers a diagrammatic expansion in impurity, clarifying the roles of self energy, ladder, and crossed diagrams and enabling a derivation of the associated quantum kinetic equation \cite{Gorini2010, Mirco2016, Raimondi2018}. As an illustration, we apply the formalism to the two-dimensional massive Dirac model, deriving analytic results for $\tau$, $\tau_{\mathrm{tr}}$, and $\sigma_{xx}$ at the Born level and evaluating the intrinsic and impurity contributions to the anomalous Hall conductivity.
Our goal is to extend the length gauge DMKE into a multiband transport theory that incorporates connected impurity-pair scattering beyond the non-crossing approximation, thereby capturing the $\tau^{0}$ disorder corrections that compete with intrinsic band geometry terms.
The central new element is the connected $V^{4}$ collision operator containing the crossed $X/\Psi$ processes.  For Gaussian white noise in the massive Dirac model the $\Psi$-diagram cancels by symmetry while the $X$-diagram survives.
These crossed contributions as an explicit operator can be upgraded into nonlinear field expansions of the density matrix and to Wannier first principles implementations, facilitating both spin and orbital current responses.

The development of the DMKE formalism provides a powerful tool for treating non-equilibrium phenomena in non-equilibrium multi-band disordered systems, addressing a vast and growing experimental and computational effort. Over the past several decades, precision DC transport measurements have demonstrated the important role of higher-order impurity scattering and the associated quantum interference corrections across a broad range of solid state systems, including conventional ferromagnetic metals \cite{Lee2004,Miyasato2007,Tian2009,Nagaosa2010}, heavy 5d transition metals with strong spin-orbit coupling \cite{Morota2011,Niimi2011,Sinova2015,Musha2019}, topological insulators \cite{Chen2010, Culcer2010, Kim2011, Steinberg2011, Qiong2019, Dmitry2025}, and Kagome lattice metals \cite{Linda2018, Chen2021, Liu2025}. When a uniform electric field is applied to these systems, sizable transverse Hall conductivity is measured, inspiring the development of various theoretical descriptions including anomalous Hall effects \cite{Jungwirth2002, Culcer2003, Cullen2021, Atencia2023}, spin Hall effects \cite{Sinova2004, Culcer2004, Bhalla2021, Hongyang2024, Toshihiro2024}, and more recently, the orbital Hall effects \cite{Dongwook2018, Choi2023, Lyalin2023, Tang2024, Atencia2024, Hong2024, James2025}. This rich Hall physics demonstrates the superposition of the crystal symmetries, band structure geometry, spin-orbit couplings, impurity scattering mechanisms, temperature effects, and electron-electron correlations, while the dependence of the quasiparticle relaxation time $\tau$ has been shown to be less significant apart from skew scattering dominated regimes \cite{Wunderlich2005, Onoda2006, Miyasato2007, Onoda2008, Xiao2010, Niimi2015}. 

The paper is organized as follows. In Sec.~\ref{Section_2_Model_and_theory} we introduce the model assumptions and establish the working formalism, emphasizing the structure of the kinetic equation and its collision operator. The salient new features of the kinetic equation introduced by the crossing terms are discussed in Sec.~\ref{sec:disc}. In Sec.~\ref{Section_3_Results_and_discussions} the kinetic equation is applied to the anomalous Hall effect of massive Dirac fermions. We conclude in Sec.~\ref{Section_4_Conclusion} with a summary and outlook.

\section{Theoretical Framework}
\label{Section_2_Model_and_theory}

We consider Bloch electrons in a periodic crystal field driven by a uniform DC electric field. The impurities are modeled as randomly distributed short range potentials. Electrons are described by the single particle density matrix $\rho$, which may include mean-field theory corrections. The crystal field and many-body effects are included in the mean-field band Hamiltonian $H_0$, which follows the band structure of the material:
\begin{equation}
H_0 \ket{m,\bm{k}} = \varepsilon^m_{\bm{k}} \ket{m,\bm{k}} \,,
\end{equation}
where $\ket{m,\bm{k}}$ is the Bloch state with band index $m$ and crystal momentum $\bm{k}$, and $\varepsilon^{m}_{\bm{k}}$ is the corresponding band energy.
The electron Bloch state, in the continuum normalization, can be expressed as:
\begin{equation}
\ket{m,\bm{k}} = \frac{\mathrm{e}^{\mathrm{i}\,\bm{k} \cdot \bm{x}}}{(2\pi)^{d/2}} \ket{u^{m}_{\bm{k}}} \,,
\end{equation}
where $d$ denotes the dimension of the system.
In this convention, the unit-cell volume is $\Omega_0$, and the normalization of the Bloch state will be $\braket{m,\bm{k}}{m',\bm{k}'} = \delta_{mm'} \delta(\bm{k}-\bm{k}')$.
Therefore, the unit cell wavefunction follows $\braket{u^m_{\bm{k}}}{u^m_{\bm{k}}} = \Omega_0$, the integration region is over a unit cell.
In all our calculations, we assume that the band structure $\varepsilon^m_{\bm{k}}$ and the unit cell wavefunction $u^m_{\bm{k}}(\bm{x})$ is already known from the first-principles calculations, tight-binding calculations, or effective $\bm{k} \cdot \bm{p}$ Hamiltonian.

The electric field is treated in the length gauge. The length-gauge formulation is naturally compatible with implementations in a Wannier basis \cite{Culcer2017, Sekine2017}. In the crystal momentum representation the electrostatic potential is conveniently implemented by using the expression for the position operator $\bm{x}$ in the Bloch representation $\mel**{m,\bm{k}}{\hat{\bm{x}}}{m',\bm{k}'}$:
\begin{equation}
\mathrm{i}\, \delta_{mm'} \nabla_{\bm{k}}\delta(\bm{k} - \bm{k}') + \frac{\mathrm{i}}{\Omega_0} \mel*{u^m_{\bm{k}}}{\nabla_{\bm{k}}}{u^{m'}_{\bm{k}'}} \delta(\bm{k} - \bm{k}') \,.
\end{equation}
The first term is the intra-band contribution, and the second term includes the off-diagonal Berry connection defined as $\me{\mathcal{A}}{mm'}{\bm{k}\bm{k}'} = \mathrm{i}\, \mel*{u^m_{\bm{k}}}{\nabla_{\bm{k}}}{u^{m'}_{\bm{k}'}}/\Omega_0$, where we use the square bracket to indicate the matrix element of an operator between the state $\ket{m,\bm{k}}$ and $\ket{m',\bm{k}'}$, and this notation will be used through our manuscript.
The electric field term in the Hamiltonian is:
\begin{equation}
H_E = e E_x x \,,
\end{equation}
where $-e$ is the electron charge, $E_x$ is the electric field along the $x$ direction.
We assume that the electric field is weak enough so that the linear response theory applies, and the system is close to equilibrium.

The impurity potential is modeled as short range potentials:
\begin{equation}
V(\bm{x}) = V_0 \sum_{i} \delta(\bm{x} - \bm{X}_i) - n_{\text{imp}} V_0 \,,
\end{equation}
where $V_0$ is the strength of a single impurity, $\bm{X}_i$ is the position of the $i$-th impurity, and $n_{\text{imp}}$ is the impurity density.
The second term $- n_{\text{imp}} V_0$ is a constant background potential which is absorbed into the mean-field Hamiltonian $H_0$ in later calculations, so that $\expval*{V(\bm{x})} = 0$.
We assume a Gaussian white noise impurity ensemble characterized by:
\begin{equation}
\expval*{V(\bm{x}) V(\bm{x}')} = n_{\text{imp}} V_0^2 \delta(\bm{x} - \bm{x}') \,.
\end{equation}
All the higher order correlation functions can be decomposed into the product of the second order correlation functions according to Wick's theorem.
The matrix element of the impurity potential in the Bloch representation is denoted as $\mel{m,\bm{k}}{V}{m',\bm{k}'} = \me{V}{mm'}{\bm{k}\bm{k}'}$ in our manuscript.
The second order correlation function of the impurity potential matrix element $\expval*{\me{V}{m_1m_2}{\bm{k}_1\bm{k}_2} \me{V}{m_3m_4}{\bm{k}_3\bm{k}_4}}$ can be expressed as:
\begin{equation}
\frac{n_{\text{imp}}}{(2\pi)^d \Omega_0^2} V_0^2 \delta(\bm{k}_1 + \bm{k}_3 - \bm{k}_2 - \bm{k}_4) \braket{u^{m_1}_{\bm{k}_1}}{u^{m_2}_{\bm{k}_2}} \braket{u^{m_3}_{\bm{k}_3}}{u^{m_4}_{\bm{k}_4}} \,,
\end{equation}

\subsection{DMKE approach}

The kinetic equation of the electron density matrix $\rho$ can be written as:
\begin{equation}
\pdv{\rho}{t} + \frac{i}{\hbar} [H_0 + H_E + V, \rho] = 0 \,.
\end{equation}
To solve the kinetic equation, we make the following ansatz:
\begin{equation}
\rho = \expval*{\rho} + g \,,
\end{equation}
where $\expval*{\rho}$ is the impurity-averaged density matrix, and $g$ is the fluctuation part due to the impurity configurations satisfying $\expval*{g} = 0$.
Taking the impurity average of the kinetic equation, we have:
\begin{equation}\label{Eq_average_rho_equation}
\pdv{\expval*{\rho}}{t} + \frac{\mathrm{i}}{\hbar} [H_0 + H_E, \expval*{\rho}] + \frac{\mathrm{i}}{\hbar} \expval*{[V, g]} = 0 \,.
\end{equation}
Here, we notice that fluctuation part $g$ will only affect the dynamics of the average part $\expval*{\rho}$ through the term $\expval*{[V, g]}$, which is denoted as the collision integral:
\begin{equation}
\mathcal{I} = -\frac{\mathrm{i}}{\hbar} \expval*{[V, g]} \,.
\end{equation}
To solve the collision integral, we need to find $g$ first, which can be solved from the following equation:
\begin{equation}\label{Eq_equation_for_g}
\pdv{g}{t} + \frac{\mathrm{i}}{\hbar} [H_0 + H_E + V, g] + \frac{\mathrm{i}}{\hbar} [V, \expval*{\rho}] - \frac{\mathrm{i}}{\hbar} \expval*{[V, g]} = 0 \,.
\end{equation}
Since the evolution of the system due to the electric field is much slower than the microscopic impurity scatterings, we can ignore the electric field term in Eq.~\eqref{Eq_equation_for_g} when solving for $g$.
Therefore, we will solve the fluctuation part $g$ iteratively in terms of the impurity potential $V$ to obtain the collision integral $\mathcal{I}$ up to the fourth order of $V$, where the important quantum interference effects are captured even in the Gaussian white noise impurity model.

In order to get familiar with the structure of the perturbation theory, it is convenient to express Eq.~\eqref{Eq_equation_for_g} in the following form (neglecting the electric field term for the time being):
\begin{equation}
\mathcal{L} g = \eta_V +\mathcal{M}_V(g) \,,
\end{equation}
where the Liouville operator is defined as:
\begin{equation}
\mathcal{L} \equiv  \pdv{g}{t} + \frac{\mathrm{i}}{\hbar} [H_0 , g] \,.
\end{equation}
The inhomogeneous term is:
\begin{equation}
\eta_V \equiv -\frac{\mathrm{i}}{\hbar} [V, \expval*{\rho}]  \,,
\end{equation}
and finally the linear operator $\mathcal{M}_V$ is given by:
\begin{equation}
\mathcal{M}_V \left( g \right) \equiv -\left( \frac{\mathrm{i}}{\hbar} [V, g] -\frac{\mathrm{i}}{\hbar} \expval*{[V, g]}\right) \,.
\end{equation}
It is not hard to see then that the $n$-th correction to $g$ is obtained as
\begin{equation}
g^{(n)}=\mathcal{L}^{-1}\mathcal{M}_V \left(\dots \mathcal{L}^{-1} \mathcal{M}_V \left( \mathcal{L}^{-1}\eta_V\right)  \right),
\end{equation}
where there are exactly $n$ occurrences of the inverse Liouville operator $\mathcal{L}^{-1}$ and exactly $n-1$ occurrences of the operator $\mathcal{M}_V$.
It is also useful to keep in mind that each occurrence of $\mathcal{M}_V$ yields two distinct terms.

At first order, the explicit solution $g^{(1)}(t)$ is:
\begin{equation}
-\frac{\mathrm{i}}{\hbar} \lim_{\eta \to 0} \int_{-\infty}^t U_{t,t_1} [V, \expval**{\rho(t_1)}] U^\dagger_{t,t_1} \mathrm{e}^{\eta t_1} \dd{t_1} \,,
\end{equation}
where $U_{t,t_1} = \exp[-\mathrm{i}\, H_0 (t-t_1)/\hbar]$ is the free time evolution operator without impurity potential.
The collision integral to the second order of the impurity potential $\mathcal{I}^{(2)}(t)$ is:
\begin{equation}
-\frac{1}{\hbar^2} \int_{-\infty}^t \expval**{[V, U_{t,t_1} [V, \expval**{\rho(t_1)}] U^\dagger_{t,t_1}]} \mathrm{e}^{\eta t_1} \dd{t_1} \,.
\end{equation}
We have included the adiabatic switching factor $\mathrm{e}^{\eta t_1}$, which is taken to $0^+$, to ensure the convergence of the time integral.
The $\expval*{\rho(t_1)}$ is assumed to vary slowly with time compared with the scattering process, so we adopt the Markov approximation by replacing $\expval*{\rho(t_1)}$ with $\expval*{\rho(t)}$ in the collision integral.
While the DC transport properties like the steady state longitudinal and Hall conductivities are not affected by this approximation, some temporal response properties like the weak (anti-)localizations and Altshuler-Aronov corrections may not be captured correctly \cite{Zala2001,Adroguer2015,Liu2017}.

Up to the second order of the impurity potentials, the quantum kinetic equation is solved in the Born approximation.
We first project the collision integral $\mathcal{I}^{(2)}(t)$ onto the basis of the eigenstates of the unperturbed Hamiltonian $H_0$, and then use the Sokhotski-Plemelj formula to evaluate the time integral.
The principal value part induced energy shift is ignored in our calculations.
Focusing on the on-shell scattering processes, the diagonal part collision integral $\mel*{m,\bm{k}}{I^{(2)}}{m,\bm{k}}$ can be expressed as:
\begin{equation}
\begin{aligned}\label{Eq: Second order collision integral}
\frac{n_{\text{imp}} V_0^2}{\hbar (2\pi)^{d-1}\Omega_0^2} &\sum_{m_1} \int \norm{\braket{u^{m}_{\bm{k}}}{u^{m_1}_{\bm{k}_1}}}^2 (\me{\expval*{\rho(t)}}{m_1m_1}{\bm{k}_1\bm{k}_1} \\
&\quad - \me{\expval*{\rho(t)}}{mm}{\bm{k}\bm{k}}) \delta(\varepsilon^{m_1}_{\bm{k}_1} - \varepsilon^{m}_{\bm{k}}) \dd[d]{k_1} \,.
\end{aligned}
\end{equation}
From the collision integral, we can identify the scattering time $1/\tau^{m}_{\bm{k}}$ as:
\begin{equation}\label{Eq: Born relaxation time}
\frac{n_{\text{imp}} V_0^2}{\hbar (2\pi)^{d-1}\Omega_0^2} \sum_{m_1} \int \norm{\braket{u^{m}_{\bm{k}}}{u^{m_1}_{\bm{k}_1}}}^2 \delta(\varepsilon^{m_1}_{\bm{k}_1} - \varepsilon^{m}_{\bm{k}}) \dd[d]{k_1} \,,
\end{equation}
and the collision kernel $W_{\ket{n,\bm{k}} \to \ket{m,\bm{k}_1}}$ as:
\begin{equation}\label{Eq: Born collision kernel}
\frac{n_{\text{imp}} V_0^2}{\hbar (2\pi)^{d-1}\Omega_0^2} \norm{\braket{u^{m}_{\bm{k}}}{u^{m_1}_{\bm{k}_1}}}^2 \delta(\varepsilon^{m_1}_{\bm{k}_1} - \varepsilon^{m}_{\bm{k}}) \,.
\end{equation}
While the scattering time captures the isotropic scatterings and the particle coherence time, the transport current, whose momentum relaxation concerns about the direction of the velocity, needs to be treated using the transport relaxation time $1/\tau^{m}_{\text{tr},\bm{k}}$:
\begin{equation}\label{Eq_Born_transport_relaxation_time}
\hspace{-0.8cm}
\frac{n_{\text{imp}} V_0^2}{\hbar (2\pi)^{d-1}\Omega_0^2} \sum_{m_1} \int \norm{\braket{u^{m}_{\bm{k}}}{u^{m_1}_{\bm{k}_1}}}^2 h(\theta) \delta(\varepsilon^{m_1}_{\bm{k}_1} - \varepsilon^{m}_{\bm{k}}) \dd[d]{k_1} \,.
\end{equation}
Here, the angle factor is $h(\theta) = 1 - \cos(\phi-\phi_1)$, where $\phi$ and $\phi_1$ are the angles of the velocity $\bm{v}^{m}_{\bm{k}}$ and $\bm{v}^{m_1}_{\bm{k}_1}$ respectively.
The inclusion of this angle factor is equivalent to the velocity vertex corrections of the $l=1$ spherical harmonics models in the Bethe-Salpeter equation in the Kubo-Green formalism by resummation of the ladder diagrams.
Therefore, we will use the transport relaxation time $\tau^{m}_{\text{tr},\bm{k}}$ in the following calculations of the longitudinal and Hall conductivities.
We notice that several important models can already be captured by this framework, with appropriate inclusion of the off-diagonal contributions and Berry phases, including the quantum corrections to the orbital Hall effects, and spin Hall effects \cite{Culcer2017,Hong2025}.

However, the second-order Born approximation, proportional to $\tau^{-1}$, is not sufficient to capture the quantum interference effects introduced by the crossing impurity scattering processes, this is particularly important in the evaluation of the Hall conductivities and system with strong anisotropy.
In our impurity model, the lowest-order quantum interference effect appears in the fourth-order in $V^4$, which is proportional to $\tau^{0}$.
Therefore, we need to evaluate the fourth-order collision integral $\mathcal{I}^{(4)}(t)$ to capture these effects.
Following the similar procedures as in the second-order case, we can solve the $g(t)$ iteratively from Eq.~\eqref{Eq_equation_for_g}.
In the second order, we have:
\begin{widetext}
\begin{equation}\label{Eq_g_solution_order_2}
\begin{aligned}
g^{(2)}(t) =
& \left(\frac{\mathrm{i}}{\hbar}\right)^2 \lim_{\eta \to 0} \int_{-\infty}^t \int_{-\infty}^{t_1} U_{t,t_1} [V, U_{t_1,t_2} [V, \expval*{\rho(t_2)}] U^\dagger_{t_1,t_2}] U^\dagger_{t,t_1} \mathrm{e}^{\eta (t_1 + t_2)} \dd{t_2} \dd{t_1} \\
& - \left(\frac{\mathrm{i}}{\hbar}\right)^2 \lim_{\eta \to 0} \int_{-\infty}^t \int_{-\infty}^{t_1} U_{t,t_1} \expval**{[V, U_{t_1,t_2} [V, \expval*{\rho(t_2)}] U^\dagger_{t_1,t_2}]} U^\dagger_{t,t_1} \mathrm{e}^{\eta (t_1 + t_2)} \dd{t_2} \dd{t_1}
\end{aligned}
\end{equation}
\end{widetext}
By recalling that at the $n$-th order we have $n-1$ occurrences of the operator $\mathcal{M}_V$, the two terms above originate from the two distinct terms of $\mathcal{M}_V$.

Then we can evaluate the third-order collision integral $\mathcal{I}^{(3)}(t)$ by substituting Eq.~\eqref{Eq_g_solution_order_2} into the definition of the collision integral.
However, due to the Gaussian white noise nature of our impurity model, the third order collision integral vanishes after taking the impurity average, i.e., $\mathcal{I}^{(3)}(t) = 0$.
At third order, we have two occurrences of the operator $\mathcal{M}_V$ and we expect four terms.
The third order explicit solution is:
\begin{widetext}
\begin{equation}\label{Eq_g_solution_order_3}
\begin{aligned}
g^{(3)}(t)&=-\left( \frac{\mathrm{i}}{\hbar} \right)^3 \lim_{\eta \to 0} \int_{-\infty}^t \int_{-\infty}^{t_1} \int_{-\infty}^{t_2} U_{t,t_1} [V, U_{t_1,t_2} [V, U_{t_2,t_3} [V, \expval*{\rho(t)}] U^\dagger_{t_2,t_3}] U^\dagger_{t_1,t_2}] U^\dagger_{t,t_1} \mathrm{e}^{\eta (t_1 + t_2 + t_3)} \dd{t_3} \dd{t_2} \dd{t_1} \\
& + \left( \frac{\mathrm{i}}{\hbar} \right)^3 \lim_{\eta \to 0} \int_{-\infty}^t \int_{-\infty}^{t_1} \int_{-\infty}^{t_2} U_{t,t_1} [V, U_{t_1,t_2} \expval*{[V, U_{t_2,t_3} [V, \expval*{\rho(t)}] U^\dagger_{t_2,t_3}]} U^\dagger_{t_1,t_2}] U^\dagger_{t,t_1} \mathrm{e}^{\eta (t_1 + t_2 + t_3)} \dd{t_3} \dd{t_2} \dd{t_1} \\
& + \left( \frac{\mathrm{i}}{\hbar} \right)^3 \lim_{\eta \to 0} \int_{-\infty}^t \int_{-\infty}^{t_1} \int_{-\infty}^{t_2} U_{t,t_1} \expval*{[V, U_{t_1,t_2} [V, U_{t_2,t_3} [V, \expval*{\rho(t)}] U^\dagger_{t_2,t_3}] U_0^\dagger(t_1, t_2)]} U^\dagger_{t,t_1} \mathrm{e}^{\eta (t_1 + t_2 + t_3)} \dd{t_3} \dd{t_2} \dd{t_1} \\
& - \left( \frac{\mathrm{i}}{\hbar} \right)^3 \lim_{\eta \to 0} \int_{-\infty}^t \int_{-\infty}^{t_1} \int_{-\infty}^{t_2} U_{t,t_1} \expval*{[V, U_{t_1,t_2} \expval*{[V, U_{t_2,t_3} [V, \expval*{\rho(t)}] U^\dagger_{t_2,t_3}]} U^\dagger_{t_1,t_2}]} U^\dagger_{t,t_1} \mathrm{e}^{\eta (t_1 + t_2 + t_3)} \dd{t_3} \dd{t_2} \dd{t_1}
\end{aligned}
\end{equation}
\end{widetext}
Again, the impurity average of an odd number of impurity potentials is zero, therefore, the third and fourth terms vanish. 
Finally, we can write down the fourth-order collision integral:
\begin{widetext}
\begin{align}\label{Eq_I4_Term1}
\mathcal{I}^{(4)}(t) = & \left( \frac{\mathrm{i}}{\hbar} \right)^4 \lim_{\eta \to 0} \int_{-\infty}^t \int_{-\infty}^{t_1} \int_{-\infty}^{t_2} \langle [V, U_{t,t_1} [V, U_{t_1,t_2} [V, U_{t_2,t_3} [V, \expval*{\rho(t)}] U^\dagger_{t_2,t_3}] U^\dagger_{t_1,t_2}] U^\dagger_{t,t_1}] \rangle \mathrm{e}^{\eta (t_1 + t_2 + t_3)} \dd{t_3} \dd{t_2} \dd{t_1} + \\ \label{Eq_I4_Term2}
& -\left( \frac{\mathrm{i}}{\hbar} \right)^4 \lim_{\eta \to 0} \int_{-\infty}^t \int_{-\infty}^{t_1} \int_{-\infty}^{t_2} \langle [V, U_{t,t_1} [V, U_{t_1,t_2} \expval*{[V, U_{t_2,t_3} [V, \expval*{\rho(t)}] U^\dagger_{t_2,t_3}]} U^\dagger_{t_1,t_2}] U^\dagger_{t,t_1}] \rangle \mathrm{e}^{\eta (t_1 + t_2 + t_3)} \dd{t_3} \dd{t_2} \dd{t_1}
\end{align}
\end{widetext}
All the important quantum interference effects are encoded in the four-fold commutator and the four point impurity average taking the form of $\expval*{V V V V}$.
Eq.~\eqref{Eq_I4_Term1} includes all the possible pairings of the impurities, capturing all the connected and disconnected contributions to the scattering processes, while the second term Eq.~\eqref{Eq_I4_Term2} can be simplified to $\mathcal{I}^{(2)}[\mathcal{I}^{(2)}]$, which spans all the disconnected contributions in the fourth-order scattering processes.
Our formalism already considers this double counting of the disconnected contribution by subtracting the $-\mathrm{i} \expval*{[V,g]}/\hbar$ term in Eq.~\eqref{Eq_equation_for_g}, therefore, the collision integral $\mathcal{I}^{(4)}$ only includes connected coherent fourth-order scatterings.

Now, we have all the components to write down the quantum kinetic equation of the average density matrix $\expval*{\rho}$ up to the fourth order of the impurity potential:
\begin{equation}
\pdv{\expval*{\rho}}{t} + \frac{\mathrm{i}}{\hbar} [H_0 + H_E, \expval*{\rho}] + \mathcal{I}^{(2)}[\expval*{\rho}] + \mathcal{I}^{(4)}[\expval*{\rho}] = 0 \,.
\end{equation}
Following the strategy described in Ref.~\cite{Culcer2017}, we can consider the diagonal terms $n_E$ and the off-diagonal terms $S_E$ of the density matrix separately under the driving of the electric field in the steady state limit.
The diagonal part $n_E$ describes the band populations, which can be further linearized into the Boltzmann equation.
At the Born approximation level, according to Eq.~\eqref{Eq_Born_transport_relaxation_time}, performing the linear response to the electric field, $n_E^{(-1)}$ can be expressed as
\begin{equation}
\mel{n,\bm{k}}{n_E^{(-1)}}{n,\bm{k}} = \frac{e E_x \tau^{n}_{\text{tr},\bm{k}}}{\hbar} \pdv{f_n^{(0)}}{\bm{k}_x} \,,
\end{equation}
where $f_n^{(0)}$ is the equilibrium Fermi-Dirac distribution function, and we used the superscript $(-1)$ to indicate the order of the impurity density $(n_{\text{imp}} V_0^2)^{-1} \propto \tau_{\text{tr}}$.
If we proceed to the fourth order, the quantum kinetic equation for the diagonal part will read:
\begin{equation}
\mathcal{I}^{(4)}[n_E^{(-1)}] + \mathcal{I}^{(2)}[n_E^{(0)}] = 0 \,,
\end{equation}
where all the functional forms of the collision integrals are already known from the previous discussions.
Therefore, the diagonal part of the density matrix at the zeroth order of the impurity potential $n_E^{(0)}$ can be solved as:
\begin{equation}
n_E^{(0)} = - \tau^{n}_{\text{tr},\bm{k}} \mathcal{I}^{(4)}[n_E^{(-1)}] \,.
\end{equation}
That $n_E^{(0)}$ must be zeroth order in the impurity density is now clear: the formal fourth order expression for the $\mathcal{I}^{(4)}$ is compensated by the presence of $n_E^{(-1)}$ and the transport time $\tau^n_{\text{tr},\bm{k}}$.

Guided by the previous discussions, we need to evaluate the collision integrals and their matrix elements respectively.
The first step is to expand the four nested commutators in Eq.~\eqref{Eq_I4_Term1}-\eqref{Eq_I4_Term2}.
Each layer of the commutator yields  a left action and right action of the impurity potential on the density matrix, i.e., $[V,\expval*{\rho(t)}] = V \expval*{\rho (t)} - \expval*{\rho(t)}V$, and, together with  the time evolution operators, forms a complete time evolution loop, which is reminiscent of the structure of the \textit{lesser} Green function in the Keldysh formalism.
There is a sequence of times $t>t_1>t_2 >t_3$ with \textit{retarded} evolution operators $U$ and then the reverse sequence of the \textit{advanced} evolution operators $U^{\dagger}$.
When considering the expansion of the commutators:
\begin{equation}
[V, U_{t,t_1} [V, U_{t_1,t_2} [V, U_{t_2,t_3} [V, \expval*{\rho(t)}] U^\dagger_{t_2,t_3}] U^\dagger_{t_1,t_2}] U^\dagger_{t,t_1}] \,,
\end{equation}
one obtains 16 terms, which can be grouped into three categories according to the relative position of the density matrix $\expval*{\rho(t)}$ and the impurity potential $V$ in the time evolution loop.

The first category includes the terms where all the impurity potentials appear in the time evolution loop before (or after) the density matrix $\expval*{\rho(t)}$:
\begin{align}
& + V U_{t,t_1} V U_{t_1,t_2} V U_{t_2,t_3} V \expval*{\rho(t)} U^\dagger_{t_2,t_3} U^\dagger_{t_1,t_2} U^\dagger_{t,t_1} \\
& + U_{t,t_1} U_{t_1,t_2} U_{t_2,t_3} \expval*{\rho(t)} V U^\dagger_{t_2,t_3} V U^\dagger_{t_1,t_2} V U^\dagger_{t,t_1} V \,.
\end{align}
They yield fourth-order corrections to the \textit{retarded} branch of the time evolution loop and correspond to the self-energy corrections of the diagrammatic theory.
We will refer to them as \textit{retarded-branch corrections} $\mathcal{I}^{(4)}_{\mathrm{I}}$.
Projecting these two terms onto the Bloch eigenbasis, we can perform the time integrals and impurity averages.

We define the following notations to simplify the expressions.
The Wick contractions of the four-point impurity average can be expressed into three terms:
\begin{align}
\mathcal{M}_1 = \expval*{\me{V}{nm_1}{\bm{k}\bm{k}_1} \me{V}{m_1m_2}{\bm{k}_1\bm{k}_2}} \expval*{\me{V}{m_2m_3}{\bm{k}_2\bm{k}_3} \me{V}{m_3n'}{\bm{k}_3\bm{k}'}} \,, \\
\mathcal{M}_2 = \expval*{\me{V}{nm_1}{\bm{k}\bm{k}_1} \me{V}{m_2m_3}{\bm{k}_2\bm{k}_3}} \expval*{\me{V}{m_1m_2}{\bm{k}_1\bm{k}_2} \me{V}{m_3n'}{\bm{k}_3\bm{k}'}} \,, \\
\mathcal{M}_3 = \expval*{\me{V}{nm_1}{\bm{k}\bm{k}_1} \me{V}{m_3n'}{\bm{k}_3\bm{k}'}} \expval*{\me{V}{m_1m_2}{\bm{k}_1\bm{k}_2} \me{V}{m_2m_3}{\bm{k}_2\bm{k}_3}} \,.
\end{align}
In these definitions, $n,\bm{k}$ and $n',\bm{k}'$ are the external band and momentum labels of the matrix element $\mel{n,\bm{k}}{\mathcal{I}^{(4)}_{\mathrm{I}}}{n',\bm{k}'}$, while $m_1,\bm{k}_1$, $m_2,\bm{k}_2$, and $m_3,\bm{k}_3$ are the intermediate labels summed over in the matrix element below.

Next, we define the following factors to simplify the result of the time integrals, where we have replaced the band energy $\varepsilon^{m}_{\bm{k}}$ with its frequency $\omega^{m}_{\bm{k}} = \varepsilon^{m}_{\bm{k}}/\hbar$, and defined the frequency difference $\Delta^{ab}_{\bm{p}\bm{q}} = \omega^{a}_{\bm{p}} - \omega^{b}_{\bm{q}}$:
\begin{align}
\Phi_1 =& \frac{1}{(3\eta + \mathrm{i} \Delta^{m_1n'}_{\bm{k}_1\bm{k}'})(2\eta + \mathrm{i} \Delta^{m_2n'}_{\bm{k}_2\bm{k}'})(\eta + \mathrm{i} \Delta^{m_3n'}_{\bm{k}_3\bm{k}'})}\,, \\
\Phi_8 =& \frac{1}{(3\eta + \mathrm{i} \Delta^{nm_1}_{\bm{k}\bm{k}_1})(2\eta + \mathrm{i} \Delta^{nm_2}_{\bm{k}\bm{k}_2})(\eta + \mathrm{i} \Delta^{nm_3}_{\bm{k}\bm{k}_3})}\,.
\end{align}
The factor $\Phi_1$ multiplies the term in which the density matrix appears as $\me{\expval*{\rho(t)}}{n'n'}{\bm{k}'\bm{k}'}$.
The factor $\Phi_8$ multiplies the term in which the density matrix appears as $\me{\expval*{\rho(t)}}{nn}{\bm{k}\bm{k}}$.
In the latter term, the impurity vertices are on the advanced branch and therefore the corresponding impurity matrix elements appear in the reversed order, as written explicitly below.

Using these notations, and retaining the contribution generated by the band- and momentum-diagonal part of the density matrix, we write the corresponding matrix elements of the fourth-order \textit{retarded-branch corrections} as:
\begin{widetext}
\begin{equation}
\begin{aligned}
\left(\frac{\mathrm{i}}{\hbar}\right)^4 \sum& \left(\expval{\me{V}{nm_1}{\bm{k}\bm{k}_1} \me{V}{m_2m_3}{\bm{k}_2\bm{k}_3}} \expval*{\me{V}{m_1m_2}{\bm{k}_1\bm{k}_2} \me{V}{m_3n'}{\bm{k}_3\bm{k}'}} + \expval*{\me{V}{nm_1}{\bm{k}\bm{k}_1} \me{V}{m_3n'}{\bm{k}_3\bm{k}'}} \expval{\me{V}{m_1m_2}{\bm{k}_1\bm{k}_2} \me{V}{m_2m_3}{\bm{k}_2\bm{k}_3}}\right) \me{\expval{\rho(t)}}{n'n'}{\bm{k}'\bm{k}'}\Phi_1 +\\
&\left(\expval{\me{V}{nm_3}{\bm{k}\bm{k}_3} \me{V}{m_2m_1}{\bm{k}_2\bm{k}_1}} \expval{\me{V}{m_3m_2}{\bm{k}_3\bm{k}_2} \me{V}{m_1n'}{\bm{k}_1\bm{k}'}} + \expval{\me{V}{nm_3}{\bm{k}\bm{k}_3} \me{V}{m_1n'}{\bm{k}_1\bm{k}'}} \expval{\me{V}{m_3m_2}{\bm{k}_3\bm{k}_2} \me{V}{m_2m_1}{\bm{k}_2\bm{k}_1}}\right) \me{\expval{\rho(t)}}{nn}{\bm{k}\bm{k}}\Phi_8
\end{aligned}
\end{equation}
\end{widetext}
Here the sums over $m_1, m_2, m_3$ and $\bm{k}_1, \bm{k}_2, \bm{k}_3$ run over the intermediate band indices and momenta, respectively.
The first square bracket is the $\mathcal{M}_2 + \mathcal{M}_3$ contribution associated with $\Phi_1$, while the second square bracket is the corresponding advanced-branch contribution associated with $\Phi_8$, with the order of the impurity matrix elements fixed by the order of the advanced time evolution operators.

We further note that the disconnected impurity pairings in $\mathcal{M}_1$ given by Eq.~\eqref{Eq_I4_Term2} are already subtracted in our formalism and, therefore, they do not appear in the final expression.
Focusing on the diagonal part of the collision integral, $n'=n$ and $\bm{k}'=\bm{k}$.
In this case, $\Phi_1$ and $\Phi_8$ are conjugate with each other, $\Phi_8=\Phi_1^*$, and therefore $\Phi_1+\Phi_8 = 2 \mathrm{Re}(\Phi_1)$.
The advanced-branch impurity contraction in the second square bracket above is the counterpart of the retarded-branch contraction in the first square bracket, with the internal labels appearing in the order fixed by the advanced branch.
Using the Sokhotski-Plemelj formula, we can perform the limit $\eta \to 0^+$ and get the following expression:
\begin{equation}
\hspace{-0.2cm}
- 2\pi \left(
\frac{\delta(\Delta^{m_1n}_{\bm{k}_1\bm{k}})}{\Delta^{m_2n}_{\bm{k}_2\bm{k}} \Delta^{m_3n}_{\bm{k}_3\bm{k}}} + 
\frac{\delta(\Delta^{m_2n}_{\bm{k}_2\bm{k}})}{\Delta^{m_1n}_{\bm{k}_1\bm{k}} \Delta^{m_3n}_{\bm{k}_3\bm{k}}} +
\frac{\delta(\Delta^{m_3n}_{\bm{k}_3\bm{k}})}{\Delta^{m_2n}_{\bm{k}_2\bm{k}} \Delta^{m_1n}_{\bm{k}_1\bm{k}}}
\right)\,.
\end{equation}
In each of the three terms above, one of the intermediate states (i.e., either $\bm{k}_1$ or $\bm{k}_2$ or $\bm{k}_3$) is on the mass-shell, whereas the other two are off-shell.
This implies that this type of corrections are iterations of the second-order self-energy type corrections and, hence, will not contribute to the transport features.

The second category includes the terms where two impurity potentials dress the \textit{retarded branch} of the time evolution before the density matrix $\expval*{\rho(t)}$, and the other two dress \textit{the advanced branch} of the time evolution after $\expval*{\rho(t)}$:
\begin{align}\label{Eq_I4_TypeII}
&+ V U_{t,t_1} V U_{t_1,t_2} U_{t_2,t_3} \expval*{\rho(t)} V U^\dagger_{t_2,t_3} V U^\dagger_{t_1,t_2} U^\dagger_{t,t_1} \\
&+ V U_{t,t_1} U_{t_1,t_2} V U_{t_2,t_3} \expval*{\rho(t)} V U^\dagger_{t_2,t_3} U^\dagger_{t_1,t_2} V U^\dagger_{t,t_1} \\
&+ V U_{t,t_1} U_{t_1,t_2} U_{t_2,t_3} V \expval*{\rho(t)} U^\dagger_{t_2,t_3} V U^\dagger_{t_1,t_2} V U^\dagger_{t,t_1} \\
&+ U_{t,t_1} V U_{t_1,t_2} V U_{t_2,t_3} \expval*{\rho(t)} V U^\dagger_{t_2,t_3} U^\dagger_{t_1,t_2} U^\dagger_{t,t_1} V \\
&+ U_{t,t_1} V U_{t_1,t_2} U_{t_2,t_3} V \expval*{\rho(t)} U^\dagger_{t_2,t_3} V U^\dagger_{t_1,t_2} U^\dagger_{t,t_1} V \\
&+ U_{t,t_1} U_{t_1,t_2} V U_{t_2,t_3} V \expval*{\rho(t)} U^\dagger_{t_2,t_3} U^\dagger_{t_1,t_2} V U^\dagger_{t,t_1} V \,.
\end{align}
In such situation, the way the contractions of the impurity potentials are carried out leads to different types of contributions.
The first type of contractions is still the self-energy terms, where the two impurity potentials on the same side are contracted, e.g.,:
\begin{equation}\label{Eq_Contraction_1}
\wick{\c V U_{t,t_1} \c V U_{t_1,t_2} U_{t_2,t_3} \expval*{\rho(t)} \c V U^\dagger_{t_2,t_3} \c V U^\dagger_{t_1,t_2} U^\dagger_{t,t_1}} \,.
\end{equation}
These terms carry the momentum conservation delta functions in the form $\delta(\bm{k}-\bm{k}_2)$ and $\delta(\bm{k}'-\bm{k}_2)$, and only affect the conductivities indirectly through the scattering time and energy renormalizations, which are not the focus in this manuscript.
The second type of contractions takes the following form:
\begin{equation}\label{Eq_Contraction_2}
\wick{\c1 V U_{t,t_1} \c2 V U_{t_1,t_2} U_{t_2,t_3} \expval*{\rho(t)} \c2 V U^\dagger_{t_2,t_3} \c1 V U^\dagger_{t_1,t_2} U^\dagger_{t,t_1}} \,,
\end{equation}
whose momentum conservation delta functions are in the form $\delta(\bm{k}_1-\bm{k}_3)$.
The corresponding terms in Eq.~\eqref{Eq_I4_TypeII} in the Bloch eigen-basis can be expressed as:
\begin{align}\label{Eq_I4_TypeII-LadderTerm}
& -\left( \frac{\mathrm{i}}{\hbar} \right)^4 \sum_{\bm{k}_1,\bm{k}_2,\bm{k}_3} \mathcal{M}_3 \me{\expval*{\rho(t)}}{m_2m_2}{\bm{k}_2\bm{k}_2} (\Phi_2 + \Phi_7) \,,
\end{align}
where:
\begin{align}
\Phi_2 =& \frac{1}{(3\eta + \mathrm{i} \Delta^{m_1n'}_{\bm{k}_1\bm{k}'})(2\eta + \mathrm{i} \Delta^{m_2n'}_{\bm{k}_2\bm{k}'})(\eta + \mathrm{i} \Delta^{m_2m_3}_{\bm{k}_2\bm{k}_3})}\,, \\
\Phi_7 =& \frac{1}{(3\eta + \mathrm{i} \Delta^{nm_3}_{\bm{k}\bm{k}_3})(2\eta + \mathrm{i} \Delta^{nm_2}_{\bm{k}\bm{k}_2})(\eta + \mathrm{i} \Delta^{m_2m_3}_{\bm{k}_2\bm{k}_3})}\,.
\end{align}
Performing the limit $\eta \to 0^+$, only following term will lead to the on-shell scattering processes:
\begin{equation}
-\pi \delta(\Delta^{nm_2}_{\bm{k}\bm{k}_2}) \left( \frac{1}{\Delta^{m_1n}_{\bm{k}_1\bm{k}} \Delta^{m_2m_3}_{\bm{k}_2\bm{k}_3}} + \frac{1}{\Delta^{nm_3}_{\bm{k}\bm{k}_3} \Delta^{m_1m_2}_{\bm{k}_1\bm{k}_2}} \right) \,,
\end{equation}
while the others terms are again self-energy renormalization.
The frequency difference denominators should be understood as the Cauchy principal value part.
Since these terms do not involve any anti-symmetric factors, it will not contribute to the Hall conductivity directly as well, but it can affect the longitudinal conductivity through the transport relaxation time renormalization, known as the ladder-type vertex corrections in the Kubo-Green formalism.
The last possible impurity contractions in this category is:
\begin{equation}\label{Eq_Contraction_3}
\wick{\c1 V U_{t,t_1} \c2 V U_{t_1,t_2} U_{t_2,t_3} \expval*{\rho(t)} \c1 V U^\dagger_{t_2,t_3} \c2 V U^\dagger_{t_1,t_2} U^\dagger_{t,t_1}} \,,
\end{equation}
which gives rise to the most important quantum interference effects in our fourth-order collision integral, as the momentum conservation delta functions exhibit a crossing structure $\delta(\bm{k} - \bm{k}_1 + \bm{k}_2 - \bm{k}_3)$.
The corresponding terms in Eq.~\eqref{Eq_I4_TypeII} have the matrix elements:
\begin{align}\label{Eq_I4_TypeII_X}
& - \left( \frac{\mathrm{i}}{\hbar} \right)^4 \sum_{\bm{k}_1,\bm{k}_2,\bm{k}_3} \mathcal{M}_2 \me{\expval*{\rho(t)}}{m_2m_2}{\bm{k}_2\bm{k}_2} \left(\sum_{i=2}^{7} \Phi_i \right) \,,
\end{align}
where the $\Phi_i$ terms are obtained from the time integrals.
We note that this term can give rise to extrinsic skew-scattering contributions to the anomalous Hall effects even when we  confine to the Gaussian white noise impurity model, at the order of $\tau^{0}$, the same at which  the intrinsic and side-jump contributions manifest.

The third category includes the terms where there are  three impurity potentials along {\it retarded branch} before  and only one impurity potential along the {\it advanced branch} after $\expval*{\rho(t)}$, respectively,
\begin{align}\label{Eq_I4_TypeIII}
& - V U_{t,t_1} V U_{t_1,t_2} V U_{t_2,t_3} \expval*{\rho(t)} V U_{t_2,t_3}^{\dagger} U_{t_1,t_2}^{\dagger} U_{t,t_1}^{\dagger} \\
& - V U_{t,t_1} V U_{t_1,t_2} U_{t_2,t_3} V \expval*{\rho(t)} U_{t_2,t_3}^{\dagger} V U_{t_1,t_2}^{\dagger} U_{t,t_1}^{\dagger} \\
& - V U_{t,t_1} U_{t_1,t_2} V U_{t_2,t_3} V \expval*{\rho(t)} U_{t_2,t_3}^{\dagger} U_{t_1,t_2}^{\dagger} V U_{t,t_1}^{\dagger} \\
& - U_{t,t_1} V U_{t_1,t_2} V U_{t_2,t_3} V \expval*{\rho(t)} U_{t_2,t_3}^{\dagger} U_{t_1,t_2}^{\dagger} U_{t,t_1}^{\dagger} V \\
& - V U_{t,t_1} U_{t_1,t_2} U_{t_2,t_3} \expval*{\rho(t)} V U_{t_2,t_3}^{\dagger} V U_{t_1,t_2}^{\dagger} V U_{t,t_1}^{\dagger} \\
& - U_{t,t_1} V U_{t_1,t_2} U_{t_2,t_3} \expval*{\rho(t)} V U_{t_2,t_3}^{\dagger} V U_{t_1,t_2}^{\dagger} U_{t,t_1}^{\dagger} V \\
& - U_{t,t_1} U_{t_1,t_2} V U_{t_2,t_3} \expval*{\rho(t)} V U_{t_2,t_3}^{\dagger} U_{t_1,t_2}^{\dagger} V U_{t,t_1}^{\dagger} V \\
& - U_{t,t_1} U_{t_1,t_2} U_{t_2,t_3} V \expval*{\rho(t)} U_{t_2,t_3}^{\dagger} V U_{t_1,t_2}^{\dagger} V U_{t,t_1}^{\dagger} V
\end{align}
The first possible impurity contractions in this category is:
\begin{equation}
\wick{\c1 V U_{t,t_1} \c2 V U_{t_1,t_2} \c2 V U_{t_2,t_3} \expval*{\rho(t)} \c1 V U^\dagger_{t_2,t_3} U^\dagger_{t_1,t_2} U^\dagger_{t,t_1}} \,,
\end{equation}
which are characterized by the momentum conservation delta function in the form $\delta(\bm{k}_1-\bm{k}_3)$.
The corresponding matrix elements are:
\begin{align}\label{Eq_I4_TypeIII_SelfEnergy2}
& + \left( \frac{\mathrm{i}}{\hbar} \right)^4 \sum_{\bm{k}_1,\bm{k}_2,\bm{k}_3} \mathcal{M}_3 \me{\expval*{\rho(t)}}{m_3m_3}{\bm{k}_3\bm{k}_3} (\Phi_1 + \Phi_2) \,.
\end{align}
The second possible impurity contraction in this category is:
\begin{equation}
\wick{\c V U_{t,t_1} \c V U_{t_1,t_2} \c V U_{t_2,t_3} \expval*{\rho(t)} \c V U^\dagger_{t_2,t_3} U^\dagger_{t_1,t_2} U^\dagger_{t,t_1}} \,,
\end{equation}
They carry the momentum conservation delta function in the form $\delta(\bm{k}-\bm{k}_2)$ and $\delta(\bm{k}'-\bm{k}_2)$, and the corresponding matrix element is:
\begin{equation}\label{Eq_I4_TypeIII_SelfEnergy1}
+ \left( \frac{\mathrm{i}}{\hbar} \right)^4 \sum_{\bm{k}_1,\bm{k}_2,\bm{k}_3} \mathcal{M}_1 \me{\expval*{\rho(t)}}{m_3m_3}{\bm{k}_3\bm{k}_3} (\Phi_3 + \Phi_5) \,.
\end{equation}
We notice that both Eq.~\eqref{Eq_I4_TypeIII_SelfEnergy1} and Eq.~\eqref{Eq_I4_TypeIII_SelfEnergy2} are combinations of self-energy and ladder correction terms.
Taking Eq.~\eqref{Eq_I4_TypeIII_SelfEnergy1} as an example, the $\Phi_3+\Phi_5$ terms can be simplified as:
\begin{equation}
\frac{\pi \delta(\Delta^{m_2m_3}_{\bm{k}_2\bm{k}_3})}{(\Delta^{nm_1}_{\bm{k}\bm{k}_1}+\mathrm{i}\,0^+)(\Delta^{nm_3}_{\bm{k}\bm{k}_3}-\mathrm{i}\,0^+)} \,,
\end{equation}
where the energy conservation part describes the ladder corrections to the density matrix, and the two frequency difference denominators describe the self-energy renormalization.
Finally, the last possible impurity contraction in this category is:
\begin{equation}
\wick{\c1 V U_{t,t_1} \c2 V U_{t_1,t_2} \c1 V U_{t_2,t_3} \expval*{\rho(t)} \c2 V U^\dagger_{t_2,t_3} U^\dagger_{t_1,t_2} U^\dagger_{t,t_1}} \,,
\end{equation}
which give rise to another set of important crossing quantum interference effects, characterized by the momentum conservation delta function in the form $\delta(\bm{k} - \bm{k}_1 + \bm{k}_2 - \bm{k}_3)$.
The corresponding matrix elements are:
\begin{align}\label{Eq_I4_TypeIII_Psi}
+ \left( \frac{\mathrm{i}}{\hbar} \right)^4 \sum_{\bm{k}_1,\bm{k}_2,\bm{k}_3} \mathcal{M}_2 \me{\expval*{\rho(t)}}{m_3m_3}{\bm{k}_3\bm{k}_3} \left(\Phi_1 + \Phi_2 + \Phi_3 + \Phi_5 \right) \,.
\end{align}
Similar to Eq.~\eqref{Eq_I4_TypeII_X}, these terms can also give rise to extrinsic skew-scattering contributions to the anomalous Hall effects at the order of $\tau^{0}$.
This last type of contributions is referred to as the $\Psi$ diagram in the standard perturbation theory.

\subsection{Keldysh approach}

Our density matrix formalism can be directly connected to the Keldysh nonequilibrium Green's function formalism, which provides an alternative way to evaluate the higher-order scattering effects. In order to illustrate this connection, we recall that, starting from the kinetic equation and using the decomposition $\rho = \expval*{\rho} + g$, the impurity averaged dynamics of $\expval*{\rho}$ is governed by the collision integral $\mathcal{I}[\expval*{\rho}] = - \mathrm{i}\,\expval*{[V,g]}/\hbar$, where the multiple layers of the commutators in Eqs.~\eqref{Eq_I4_Term1} and \eqref{Eq_I4_Term2} can be related to the time evolution structure of the Keldysh Green's functions.
Each commutator layer corresponds to a left action (forward branch) or right action (backward branch) of the impurity potential along the time-ordered Keldysh contour.
Up to the fourth order in the impurity potential, the 16 scattering terms generated by the nested commutators  appear as different locations of the impurity vertices along the Keldysh contour.
After impurity averaging within the Gaussian white-noise impurity model, Wick contractions arrange these contributions into non-crossing (self energy and ladder terms)  and crossing two-impurity quantum interference diagrams ($X$- and $\Psi$-terms), while the subtraction built into Eq.~\eqref{Eq_equation_for_g} removes the disconnected fourth order diagrams that factorize into two successive $V^{2}$ vertices, leaving $\mathcal{I}^{(4)}$ as the connected $V^{4}$ scattering contribution.
In this subsection, we will explore these details.

We denote the creation and annihilation operators for an electron in the Bloch state $\ket{m,\bm{k}}$ as $\hat{c}^{\dagger}_{m,\bm{k}}$ and $\hat{c}_{m,\bm{k}}$ respectively, and they satisfy the standard anticommutation relations:
\begin{equation}
\{\hat{c}_{m,\bm{k}}, \hat{c}^{\dagger}_{m',\bm{k}'}\} = \delta_{mm'} \delta_{\bm{k}\bm{k}'} \,.
\end{equation}
Therefore, our single particle density matrix can be expressed as $\me{\rho}{mm'}{\bm{k}\bm{k}'} = \expval*{\hat{c}^{\dagger}_{m',\bm{k}'} \hat{c}_{m,\bm{k}}}$.
 The time evolution is defined along the Keldysh contour, which goes from $-\infty$ to $+\infty$ (upper forward branch denoted as $C_+$ index) and then from $+\infty$ to $-\infty$ (lower backward branch denoted as $C_-$ index).
 
The contour-ordered Keldysh Green function in the Bloch eigenbasis is defined as:
\begin{equation}
\me{G(t,t')}{mm'}{\bm{k}\bm{k}'} = -\frac{\mathrm{i}}{\hbar} \expval*{ \mathcal{T}_{C} \hat{c}_{m,\bm{k}}(t) \hat{c}^{\dagger}_{m',\bm{k}'}(t') } \,,
\end{equation}
where $\mathcal{T}_C$ is the contour time ordering operator, putting operators with later contour time arguments to the left.

The lesser and greater Green's functions are given by:
\begin{align}
&\me{G^<(t,t')}{mm'}{\bm{k}\bm{k}'} = \frac{\mathrm{i}}{\hbar}\,\expval*{ \hat{c}^{\dagger}_{m',\bm{k}'}(t') \hat{c}_{m,\bm{k}}(t) } \,,\\
&\me{G^>(t,t')}{mm'}{\bm{k}\bm{k}'} = -\frac{\mathrm{i}}{\hbar}\,\expval*{ \hat{c}_{m,\bm{k}}(t) \hat{c}^{\dagger}_{m',\bm{k}'}(t') } \,,
\end{align}
where $t'\in C_-$, $t\in C_+$  and $t\in C_-$, $t'\in C_+$, respectively.
Correspondingly, the retarded and advanced Green's functions are defined as:
\begin{equation}
G^{R/A}(t,t') = \pm \Theta(\pm(t-t')) (G^>(t,t') - G^<(t,t')) \,.
\end{equation}
In particular, the density matrix is related to the lesser Green's function by the equal time limit:
\begin{equation}
\me{\rho(t)}{mm'}{\bm{k}\bm{k}'} = -\mathrm{i}\,\hbar \me{G^<(t,t)}{mm'}{\bm{k}\bm{k}'} .
\end{equation}
In this way, the evaluation of the density matrix can be fully translated into the evaluation of the lesser Green's function, which is governed by the Dyson equation.

In this framework, the collision integral can be expressed as the difference between the inflow and outflow scattering rates known as the Kadanoff-Baym equation and arises naturally from the self-energy due to the impurity scattering potential.
By considering the lesser component of the Dyson equation $(G_0^{-1} - \Sigma)*G = I$, we have:
\begin{equation}
(G_0^{-1} - \Sigma^R)*G^< - \Sigma^< * G^A = 0 \,,
\end{equation}
where $\Sigma^R$ ($\Sigma^A$) is the retarded (advanced) self-energy, and $\Sigma^<$ ($\Sigma^>$) is the lesser (greater) self-energy.
We can obtain a similar equation by taking its Hermitian conjugate, whose difference from the above equation gives, in the equal-time limit, the following equation.
\begin{equation}
\pdv{\rho(t)}{t} + \frac{\mathrm{i}}{\hbar} [H_0, \rho(t)] = (\Sigma^< * G^> - \Sigma^> * G^<) \,.
\end{equation}
Therefore, the collision integral in the Keldysh formalism is given by $\mathcal{I}[\rho(t)] =  \Sigma^> * G^<-\Sigma^< * G^>$.
In the Markovian limit, the memoryless approximation is reflected as a rapidly decay of the Green's functions and self-energies in the relative time $t - t'$, while the density matrix varies slowly.
The time convolution becomes local in time, and the collision integral reduces to:
\begin{equation}\label{Eq_Keldysh_collision_integral}
\hspace{-10pt}
\frac{1}{2\pi\hbar} \int \Sigma^>(\omega,t) G^<(\omega,t)-\Sigma^<(\omega,t) G^>(\omega,t) \dd{\omega} \, ,
\end{equation}
with $\omega$ being the Fourier transformed variable of the relative time $t-t'$ and the equal-time limit $t=t'$ is taken at last.
Eq.~\eqref{Eq_Keldysh_collision_integral} is an equivalent statement of the scattering relaxation and is always determined by the current state of the system as used in the last subsection.

At the second order, the self-energy $\me{\Sigma^{(2)}(\omega)}{mm'}{\bm{k}\bm{k}'}$ in the Born approximation is given by:
\begin{equation}\label{Eq_Second_order_self-energy}
\hspace{-10pt}
\sum_{m_1,m_2}\int \expval*{ \me{V}{mm_1}{\bm{k}\bm{k}_1} \me{V}{m_2m'}{\bm{k}_2\bm{k}'} } \me{G(\omega)}{m_1m_2}{\bm{k}_1\bm{k}_2} \dd[d]{k_1} \dd[d]{k_2} \,.
\end{equation}
By using our impurity model to perform the impurity average $\langle \dots \rangle$, and focusing on the band diagonal part in frequency space, we can write $\me{\Sigma^{(2),<}(\omega)}{mm}{\bm{k}\bm{k}}$ as:
\begin{equation}
\frac{n_{\text{imp}} V_0^2}{(2\pi)^d\Omega_0^2} \sum_{m_1} \int \me{G^<(\omega)}{m_1m_1}{\bm{k}_1\bm{k}_1} \norm{\braket{u^m_{\bm{k}}}{u^{m_1}_{\bm{k}_1}}}^2 \dd[d]{k_1} \,,
\end{equation}
where we have used the property that after impurity averaging, the Green's functions are diagonal in momentum space.
Similarly, we can get the greater self-energy $\me{\Sigma^{(2),>}(t,t')}{mm}{\bm{k}\bm{k}}$.
Under the on-shell approximation, the lesser and greater Green's function can be expressed as:
\begin{align}
\me{G^<(\omega)}{mm}{\bm{k}\bm{k}} =& 2 \pi \mathrm{i}\,\me{\rho(t)}{mm}{\bm{k}\bm{k}} \delta(\hbar\omega - \varepsilon^m_{\bm{k}})/\hbar \,,\\
\me{G^>(\omega)}{mm}{\bm{k}\bm{k}} =& -2 \pi \mathrm{i}\,(1 - \me{\rho(t)}{mm}{\bm{k}\bm{k}}) \delta(\hbar\omega - \varepsilon^m_{\bm{k}})/\hbar \,.
\end{align}
Using these expressions, we can recover Eq.~\eqref{Eq_Born_transport_relaxation_time}.

The above derivation shows that our density matrix collision integral is equivalently encoded in the Keldysh Green's functions through Eq.~\eqref{Eq_Keldysh_collision_integral}.
Now, once we go beyond the second order, and try to isolate the connected $V^4$ contributions, we need to construct $\Sigma^{(4)}$ systematically.
This requires expanding the Dyson's equation along the Keldysh contour, where the Wick contractions can organize everything into nonreducible crossing or non crossing terms.
We will see that the multiple layers of the commutators in Eqs.~\eqref{Eq_I4_Term1} and \eqref{Eq_I4_Term2} are automatically encoded in the Keldysh contours: each commutator in the density matrix iteration corresponds to assigning an impurity vertex to the forward or backward branch of the Keldysh contour.

For an observable $O$, its expectation value at time $t$ can be written in the Keldysh contour as a normalized average:
\begin{equation}\label{Eq_Keldysh_Expectation}
\hspace{-10pt}
\expval*{O(t)} = \frac{\tr[\rho(t_0)  T_C \exp\left(-\frac{\mathrm{i}}{\hbar} \int_C V_I(\tau) \dd{\tau}\right) O_C(t)(t) ]}{\tr[ \rho(t_0) T_C \exp\left(-\frac{\mathrm{i}}{\hbar} \int_C V_I(\tau) \dd{\tau}\right)]} \,.
\end{equation}
Here, all operators are in the interaction picture with respect to $H_0$.
The observable operator $O_C(t)$ is inserted at time $t$, which is at the end of the forward branch of the contour, i.e., all the vertices on the forward branch are time ordered before $O$, while those on the backward branch are anti-time ordered after $O$.
The denominator provides the normalization that removes disconnected vacuum contributions.

A clear way to see how commutators appear is to explicitly separate the contour integrals into the two branches.
We append the branch index $a = \pm$ to the impurity potential temporarily, so that $V_I^+(t)$ ($V_I^-(t)$) denotes the impurity potential on the forward (backward) branch of the contour.
Then a contour integral can be written as:
\begin{equation}
\int_C f \dd{t} = \sum_{\alpha = \pm} \alpha \int_{t_0}^t f^{(\alpha)}(t) \dd{t} \,.
\end{equation}
The factor $\alpha$ tracks the orientation of the contour branches.
This separation of the integral reflects the left action and the right action of the impurity potential on the density matrix respectively.

In the first order in $V$, we will get $2$ terms:
\begin{equation}
\frac{\mathrm{i}}{\hbar} \int_{t_0}^t V_I(t_1) O(t) \dd{t_1} - \frac{\mathrm{i}}{\hbar} \int_{t_0}^t O(t) V_I(t_1) \dd{t_1} \,,
\end{equation}
which recovers the commutator structure $[V_I(t_1), O(t)]$.
At the second order, we have:
\begin{equation}
\left( -\frac{\mathrm{i}}{\hbar} \right)^2 \frac{1}{2!} \int_C \int_C \mathcal{T}_C V_I(\tau_1) V_I(\tau_2) O_C(t) \dd{\tau_1} \dd{\tau_2} \,.
\end{equation}
This is equivalent to:
\begin{equation}
\frac{1}{2!} \left(-\frac{\mathrm{i}}{\hbar}\right)^2 \sum_{\alpha,\beta = \pm} \alpha \beta \int_{t_0}^t \mathcal{T}_C V_I^{(\alpha)}(t_1) V_I^{(\beta)}(t_2) O_C(t) \dd{t_1} \dd{t_2} \,.
\end{equation}
The following two identities can help us further simplify the expression:
\begin{equation}
\begin{aligned}
T_C V_I^+(t_1) V_I^+(t_2) = T[ V_I(t_1) V_I(t_2) ] \,, \\
T_C V_I^-(t_1) V_I^-(t_2) = \tilde{T}[ V_I(t_1) V_I(t_2) ] \,.
\end{aligned}
\end{equation}
For those mixed branch terms, we just place one vertex to the left of $O$ and the other to the right.
The factor $1/2!$ is compensated by the  elimination of the double counting of the case $t_1 > t_2$ and $t_2 > t_1$ on the same branch.
If we focus on the domain over the triangular region $t_0 < t_2 < t_1 < t$ as we used in the last subsection, we will recover the standard double commutator structure:
\begin{equation}
\left( -\frac{\mathrm{i}}{\hbar} \right)^2 \int_{t_0}^t \int_{t_0}^{t_1} [V_I(t_1), [V_I(t_2), O(t)]] \dd{t_2} \dd{t_1} \,.
\end{equation}
The same reasoning extends directly to higher order situations.
At fourth order in $V$, each interaction vertex can be placed on either branch, producing $2^4=16$ distinct contour contributions.
These are in one-to-one correspondence with the 16 terms obtained by expanding the fourfold commutator in the density matrix formulation.
This is the precise sense in which the multilayer commutator structure is automatically encoded by the Keldysh contour expansion, providing the natural bridge to the fourth order diagrammatic after impurity averaging and Wick contractions.

Next, we discuss the connection to the standard Keldysh diagrammatic impurity technique.
For Gaussian white noise impurities, the fourth-order average $\expval*{VVVV}$ is evaluated by Wick's theorem, which pairs the four impurity vertices into two contractions.
Different pairings, together with the placement of the vertices on the forward or backward branches relative to the density matrix (observable) insertion, generate distinct terms like in Eq.~\eqref{Eq_Contraction_1}, \eqref{Eq_Contraction_2}, and \eqref{Eq_Contraction_3}.

First, we notice that not all fourth order contributions yield new scattering physical effects.
Two successive Born scatterings, which are those pieces generated automatically by iterating the second order self energy:
\begin{equation}
G = G_0 + G_0 * \Sigma^{(2)} * G_0 + G_0 * \Sigma^{(2)} * G_0 * \Sigma^{(2)} * G_0 + \cdots \,,
\end{equation}
should be excluded.
This is the direct counterpart of the subtraction we did for the terms like $\mathcal{I}^{(2)}[\mathcal{I}^{(2)}[\expval*{\rho}]]$ in the density matrix formalism.
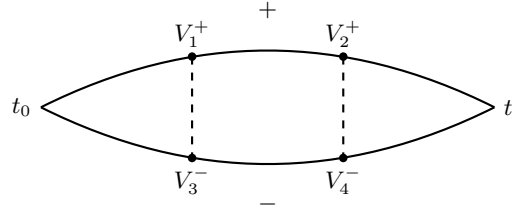
\begin{figure}[t]
\centering
\begin{tikzpicture}[x=1cm,y=1cm,line width=0.8pt]
\draw (0,0) .. controls (2,1) and (4,1) .. (6,0);
\draw (6,0) .. controls (4,-1) and (2,-1) .. (0,0);
\node[left]  at (0,0) {$t_0$};
\node[right] at (6,0) {$t$};
\node[above] at (3,1.05) {$+$};
\node[below] at (3,-1.05) {$-$};
\fill (2,0.67) circle (1.6pt);
\node[above] at (2,0.67) {$V_1^{+}$};
\fill (4,0.67) circle (1.6pt);
\node[above] at (4,0.67) {$V_2^{+}$};
\fill (2,-0.67) circle (1.6pt);
\node[below] at (2,-0.67) {$V_3^{-}$};
\fill (4,-0.67) circle (1.6pt);
\node[below] at (4,-0.67) {$V_4^{-}$};
\draw[dashed] (2,0.67) -- (2,-0.67);
\draw[dashed] (4,0.67) -- (4,-0.67);
\end{tikzpicture}
\caption{Ladder fourth order contribution on the Keldysh contour: two impurity contractions connect vertices on opposite branches, forming a two-rung ladder that generates the leading vertex correction.}
\label{fig:keldysh_contour_reducible}
\end{figure}

Next, we can consider all the ladder corrections as indicated in Fig.~\ref{fig:keldysh_contour_reducible}.
After disorder averaging, a ladder rung is generated by a Wick contraction that connects an impurity vertex on the forward branch with one on the backward branch of the Keldysh contour (relative to the density matrix insertion).
This cross-branch pairing attaches a retarded and an advanced Green's function together, producing a dressing of the velocity vertex.
Keeping a single rung gives the first-order  vertex correction.
Projecting onto the Bloch eigenbasis, we get:
\begin{equation}
\begin{aligned}
\me{\tilde{v}_{x/y}^{(1)}}{nn}{\bm{k}\bm{k}} = \frac{n_{\text{imp}} V_0^2}{(2\pi)^d\Omega_0^2}& \sum_{m} \int \me{G^R}{mm}{\bm{k}_1\bm{k}_1} \me{v_{x/y}^{(0)}}{mm}{\bm{k}_1\bm{k}_1} \\
& \times \me{G^A}{mm}{\bm{k}_1\bm{k}_1} \norm{\braket{u^n_{\bm{k}}}{u^m_{\bm{k}_1}}}^2 \dd[d]{k_1} \,.
\end{aligned}
\end{equation}
If we iterate this process, we can sum up all the ladder diagrams to get the full vertex correction (Bethe-Salpeter equation) to the velocity operator:
\begin{equation}
\begin{aligned}\label{Eq_Vertex_correction_equation}
& \me{\tilde{v}_{x/y}}{nn}{\bm{k}\bm{k}} = \me{v_{x/y}^{(0)}}{nn}{\bm{k}\bm{k}} + \frac{n_{\text{imp}} V_0^2}{(2\pi)^d\Omega_0^2} \sum_{m} \int \me{G^R}{mm}{\bm{k}_1\bm{k}_1} \\
&\quad \times \me{v_{x/y}}{mm}{\bm{k}_1\bm{k}_1} \me{G^A}{mm}{\bm{k}_1\bm{k}_1} \norm{\braket{u^n_{\bm{k}}}{u^m_{\bm{k}_1}}}^2 \dd[d]{k_1} \,.
\end{aligned}
\end{equation}
Here, the Green functions are the impurity averaged ones including the self-energy corrections:
\begin{equation}
\me{G^{R/A}(\omega)}{mm}{\bm{k}\bm{k}} = \frac{1}{\varepsilon - \varepsilon^m_{\bm{k}} \pm \mathrm{i}\,\hbar/(2\tau^m_{\bm{k}})} \,,
\end{equation}
where the principal part is neglected and the self-energy can be evaluated as in Eq.~\eqref{Eq_Second_order_self-energy}.
Using the on-shell relation:
\begin{equation}\label{Eq_Green_function_identity}
\me{G^R}{mm}{\bm{k}\bm{k}} \me{G^A}{mm}{\bm{k}\bm{k}} = \frac{2\pi \tau^m_{\bm{k}}}{\hbar} \delta(\varepsilon - \varepsilon^m_{\bm{k}}) \,,
\end{equation}
into the Bethe-Salpeter equation reduces the momentum integral to the Fermi-surface contribution, and the ladder re-summation reproduces the familiar replacement of the single-particle lifetime by the transport relaxation time in the longitudinal response \cite{Valet2025}.

So far, we have evaluated all the components of the Green function and vertex corrections.
Using Eq.~\eqref{Eq_Keldysh_Expectation}, we can now evaluate the expectation value of the current operator $j_{x/y} = -e v_{x/y}$ up to the fourth order in the impurity potential.
We evaluate the DC conductivity directly from the Keldysh contour expectation value, taking the current operator $j_y$ as the observable insertion at time $t$ (the turning point of the forward branch), which can be expressed as:
\begin{equation}
j_i = \frac{\mathrm{i}e}{(2\pi)^{d+1}} \int \tr[ v_i(\bm{k}) G^<(\epsilon,\bm{k}) ] \dd[d]k \dd{\epsilon} \,.
\end{equation}
The linear response Hall conductivity is obtained by $\sigma_{xy} = \lim_{E_x \to 0} \delta j_y / E_x$, where $\delta j_y$ is the change of the transverse current expectation value linear in the applied electric field $E_x$.

In order to compare with the density matrix collision integral discussed above, we isolate the kinetic Fermi surface part of the Keldysh response, in which the electric field changes the nonequilibrium distribution.
This is the Keldysh counterpart of inserting $n_E^{(-1)}$ into $\mathcal{I}^{(4)}[n_E^{(-1)}]$.
The full electric field variation in the Keldysh formalism also contains spectral insertions on the retarded and advanced Green functions.
Schematically, using $G^< = G^R \Sigma^< G^A$, one has
\begin{equation}
\delta_E G^< = (\delta_E G^R)\Sigma_0^< G_0^A + G_0^R(\delta_E \Sigma^<)G_0^A + G_0^R\Sigma_0^<(\delta_E G^A)
\end{equation}
In the comparison with $\mathcal{I}^{(4)}[n_E^{(-1)}]$, however, we retain only the kinetic lesser component insertion associated with the nonequilibrium distribution:
\begin{equation}
\delta_E G_{\text{kin}}^< = G^R(\varepsilon, \bm{k})\left(\mathrm{i}\,e E_x \pdv{f^{(0)}_\varepsilon}{\varepsilon} v_x(\bm{k}) \tau_{\text{tr}}\right)G^A(\varepsilon, \bm{k})
\end{equation}
This object contains the transport relaxation-time renormalization and the corresponding dressed longitudinal-vertex information.

For products of single particle propagators separated by static impurity vertices, the Langreth rule fixes the analytic structure of the lesser component:
\begin{equation}
\hspace{-1cm}
(G_1 V_1 G_2 V_2 \cdots V_N G_{N + 1})^< = \sum_j G_1^R V_1 \cdots G_j^< \cdots V_N G_{N + 1}^A
\end{equation}
Therefore, whenever the restricted kinetic insertion $\delta_E G_{\text{kin}}^<$ is used, all propagators to its left must be retarded and all propagators to its right must be advanced.
The electric field insertions on $G^R$ and $G^A$ are part of the remaining spectral response and are not included in the restricted comparison below.

For the longitudinal response, the leading order contribution comes from the second order Born approximation with the vertex correction included, which recovers the standard Boltzmann transport result.
This can be supported by a dimensional analysis, $V^2$ calculations scales as $n_{\text{imp}}$, while the $V^4$ contributions scale as $n_{\text{imp}}^2$.
But each retarded-advanced Green function pair produces a factor of $1/n_{\text{imp}}$ due to the lifetime broadening, so the leading order longitudinal response is always at order $n_{\text{imp}}^0$.
Therefore, the longitudinal conductivity can be written as:
\begin{equation}\label{Eq_Keldysh_longitudinal_conductivity}
\begin{aligned}
\sigma_{xx} = \frac{e^2}{(2\pi)^{d+1}} \int \left(-\pdv{f^{(0)}_\varepsilon}{\varepsilon}\right) & \tr_{\text{band}} \left[ v_x G^R v_x \tau_{\text{tr}} G^A \right] \dd[d]{k} \dd{\varepsilon} \,. 
\end{aligned}
\end{equation}
We notice that  the Green function $G^{R/A}$ are impurity averaged.
Therefore, in the on-shell approximation, we will have $G^R G^A = 2\pi \tau \delta(\varepsilon_F - \varepsilon^m_{\bm{k}})/\hbar$ and  this will reproduce the standard Boltzmann transport result for the longitudinal conductivity with the renormalised transport relaxation time $\tau_{\text{tr}}$.

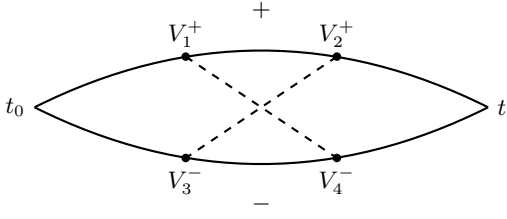
\begin{figure}[t]
\centering
\begin{tikzpicture}[x=1cm,y=1cm,line width=0.8pt]
\draw (0,0) .. controls (2,1) and (4,1) .. (6,0);
\draw (6,0) .. controls (4,-1) and (2,-1) .. (0,0);
\node[left]  at (0,0) {$t_0$};
\node[right] at (6,0) {$t$};
\node[above] at (3,1.05) {$+$};
\node[below] at (3,-1.05) {$-$};
\fill (2,0.67) circle (1.6pt);
\node[above] at (2,0.67) {$V_1^{+}$};
\fill (4,0.67) circle (1.6pt);
\node[above] at (4,0.67) {$V_2^{+}$};
\fill (2,-0.67) circle (1.6pt);
\node[below] at (2,-0.67) {$V_3^{-}$};
\fill (4,-0.67) circle (1.6pt);
\node[below] at (4,-0.67) {$V_4^{-}$};
\draw[dashed] (2,0.67) -- (4,-0.67);
\draw[dashed] (4,0.67) -- (2,-0.67);
\end{tikzpicture}
\caption{X-type crossed impurity contribution on the Keldysh contour: two impurity contractions intersect, yielding a connected fourth-order scattering process.}
\label{fig:keldysh_contour_X}
\end{figure}

Focusing on the crossed impurity contractions, we firstly consider the $X$-diagram shown in Fig.~\ref{fig:keldysh_contour_X}.
This topology is generated by connecting two impurity vertices on the forward branch with two vertices on the backward branch, forming two crossing impurity lines.
The Wick contraction follows the form $\expval*{ \me{V}{m m_1}{\bm{k} \bm{k}_1} \me{V}{m_2m_3}{\bm{k}_2 \bm{k}_3} } \expval*{ \me{V}{m_1m_2}{\bm{k}_1 \bm{k}_2} \me{V}{m_3 m}{\bm{k}_3 \bm{k}}}$.
The $X$-diagram corresponds to the Type-II structure in Eq.~\eqref{Eq_I4_TypeII}, where two impurity vertices lie on the retarded side of the density matrix and two lie on the advanced side.
Therefore, in the restricted kinetic comparison, the lesser insertion is not moved arbitrarily through all propagator segments.
Instead, each $X$-type contribution has the canonical $2R$-lesser-$2A$ analytic structure.
The $X$-diagram contribution to the Hall conductivity can be expressed as:
\begin{equation}\label{Eq_X_Keldysh_conductivity}
\begin{aligned}
\hspace{-0.8cm}
\sigma_{xy}^{X} =& - \frac{e}{E_x}\frac{1}{(2\pi)^{d + 1}} \int \Im \sum_{\alpha \in X} \mathcal{F}^X_\alpha \expval{\me{V}{m m_1}{\bm{k} \bm{k}_1} \me{V}{m_2 m_3}{\bm{k}_2 \bm{k}_3}} \\
& \times \expval{\me{V}{m_1 m_2}{\bm{k}_1 \bm{k}_2} \me{V}{m_3 m}{\bm{k}_3 \bm{k}}} \dd[d]{k} \dd[d]{k_1} \dd[d]{k_2} \dd[d]{k_3} \dd{\varepsilon}
\end{aligned}
\end{equation}
where $\mathcal{F}^X_\alpha$ denotes the band-space trace associated with the $\alpha$-th Type-II ordering and the $X$-type impurity contraction.
A representative term is
\begin{equation}
\mathcal{F}^X_\alpha = \tr_{\text{band}}\left[v_y G^R V_1 G^R V_2 \delta_E G_{\text{kin}}^< V_4 G^A V_3 G^A\right]
\end{equation}
The remaining $X$-type terms are obtained from the other Type-II orderings in Eq.~\eqref{Eq_I4_TypeII} and from the corresponding permutation of impurity labels, while preserving the same retarded-lesser-advanced ordering.

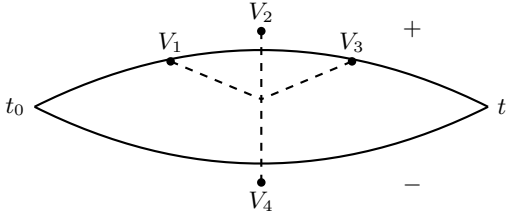
\begin{figure}[t]
\centering
\begin{tikzpicture}[x=1cm,y=1cm,line width=0.8pt]
\draw (0,0) .. controls (2,1) and (4,1) .. (6,0);
\draw (6,0) .. controls (4,-1) and (2,-1) .. (0,0);
\node[left]  at (0,0) {$t_0$};
\node[right] at (6,0) {$t$};
\node[above] at (5,0.8) {$+$};
\node[below] at (5,-0.8) {$-$};
\fill (3,1) circle (1.6pt);
\node[above] at (3,1) {$V_2$};
\fill (3,-1) circle (1.6pt);
\node[below] at (3,-1) {$V_4$};
\fill (1.8,0.6) circle (1.6pt);
\node[above] at (1.8,0.6) {$V_1$};
\fill (4.2,0.6) circle (1.6pt);
\node[above] at (4.2,0.6) {$V_3$};
\draw[dashed] (3,1) -- (3,-1);
\draw[dashed] (1.8,0.6) -- (3,0.1);
\draw[dashed] (4.2,0.6) -- (3,0.1);
\end{tikzpicture}
\caption{$\Psi$-diagram on the Keldysh contour: two impurity contractions intersect, yielding a connected fourth-order scattering process. However, it has a different topology (impurity vertex placement) compared to the $X$-diagram in Fig.~\ref{fig:keldysh_contour_X}.}
\label{fig:keldysh_contour_Psi}
\end{figure}

A similar construction applies to the $\Psi$-diagram shown in Fig.~\ref{fig:keldysh_contour_Psi}.
This topology corresponds to the Type-III structure in Eq.~\eqref{Eq_I4_TypeIII}, where the impurity vertices are distributed as three retarded vertices and one advanced vertex, or conversely as one retarded vertex and three advanced vertices.
In momentum space, it produces a momentum constraint and a band-space ordering different from those of the $X$-diagram.
The $\Psi$-diagram contribution to the Hall conductivity can be expressed as:
\begin{equation}
\begin{aligned}
\sigma_{xy}^{\Psi} =& - \frac{e}{E_x}\frac{1}{(2\pi)^{d + 1}} \int \Im \sum_{\alpha \in \Psi} \mathcal{F}^{\Psi}_{\alpha} \expval{\me{V}{m m_1}{\bm{k} \bm{k}_1} \me{V}{m_2 m_3}{\bm{k}_2 \bm{k}_3}} \\
& \times \expval{\me{V}{m_1 m_2}{\bm{k}_1 \bm{k}_2} \me{V}{m_3 m}{\bm{k}_3 \bm{k}}} \dd[d]{k} \dd[d]{k_1} \dd[d]{k_2} \dd[d]{k_3} \dd{\varepsilon}
\end{aligned}
\end{equation}
In this case, the order of the impurity vertices in the band-space trace is different because of the different topology.
The same Langreth ordering nevertheless applies to the restricted kinetic insertion.
Two representative analytic structures are
\begin{equation}
\mathcal{F}^{\Psi}_{\alpha,1} = \tr_{\text{band}}\left[v_y G^R V_1 G^R V_2 G^R V_3 \delta_E G_{\text{kin}}^< V_4 G^A\right]
\end{equation}
and
\begin{equation}
\mathcal{F}^{\Psi}_{\alpha,2} = \tr_{\text{band}}\left[v_y G^R V_1 \delta_E G_{\text{kin}}^< V_2 G^A V_3 G^A V_4 G^A\right]
\end{equation}
The remaining $\Psi$-type terms are constructed similarly from the Type-III orderings in Eq.~\eqref{Eq_I4_TypeIII}, with every kinetic lesser insertion separating a retarded block on its left from an advanced block on its right.

\section{Discussion}
\label{sec:disc}

The DMKE formalism presented in this work describes the response of Bloch electrons to external perturbations and is equivalent to diagrammatic approaches such as the Kubo, Matsubara, and Keldysh formalisms, as explained in Ref.~\cite{Rhonald2022,Valet2025}. In contrast to semiclassical approaches the DMKE captures disorder effects on the same level as diagrammatic formalisms. However, since the latter have a considerably longer history the strategy for expanding disorder contributions to arbitrary orders has been established to a larger extent than in the DMKE. With this in mind, one motivation behind the present work was to demonstrate that the DMKE approach remains equally powerful when sub-leading disorder effects need to be taken into account. The equivalence with the Keldysh approach is demonstrated explicitly in our work, and the results derived here using Keldysh and DMKE agree with the Kubo diagrammatic derivation of Refs.~\cite{Ado2015,Ado2016}. Here we obtain an additional collision integral which incorporates the sub-leading disorder diagrams explicitly. In fact, as Ref.~\cite{Liu2017} showed for the case of weak localization, once one goes beyond the ladder approximation, the DMKE and diagrammatic formalisms become qualitatively very similar, and one can identify entire series of diagrams that can be converted into scattering terms in the kinetic equation in a similar spirit to that of the Keldysh formalism. In this way one can envisage the DMKE as a complete quantum kinetic theory based on the density matrix.

The DMKE formalism is an important tool in the study of non-equilibrium phenomena. To begin with, it relies on the density matrix, and once the density matrix is found it can be traced with any desired operator to determine its expectation value, making the approach extremely convenient. Likewise, as our earlier work has shown, the DMKE formalism transparently identifies topological terms in non-equilibrium responses such as the Berry curvature and quantum metric tensor, as well as their complex interplay with disorder effects \cite{Cullen2021,Bhalla2023,Pandey2024,Hong2024,Rhonald2025}. Finally, the DMKE naturally distinguishes Fermi surface and Fermi sea contributions. The DMKE formalism can naturally be applied a variety of non-equilibrium effects, including the spin-Hall effect, orbital Hall effect, valley Hall effect, spin torques, and non-linear electromagnetic responses \cite{Bhalla2020, Bhalla2022, Cullen2023, Das2023, Das2024}. We will not consider these in detail here, rather in the following section we illustrate the use of the DMKE by determining the AHE in the massive Dirac fermion model and showing it agrees with the diagrammatic prediction.

In this context some words are in order concerning the lengthy controversy that has surrounded the anomalous Hall effect (AHE) \cite{Culcer2022AHE}.
To begin with, the longitudinal conductivity is dominated by terms $\propto \tau$ in the high-mobility regime. For electric field driven longitudinal transport, the current is typically governed by momentum relaxation processes that are well captured by quasiparticle relaxation time $\tau$ and transport relaxation time $\tau_{\text{tr}}$ in the framework of the semiclassical Boltzmann equation and their corresponding self energy and vertex corrected Born approximations \cite{Lee1985, Rammer1986}. In contrast, the AHE conductivity contains terms that are $\propto \tau$, which come from skew scattering, as well as terms of order $\tau^0$, which are intrinsic as well as disorder-induced. Notably, while conventional skew scattering at third order in the impurity potential leads to a $\tau$ dependence, higher order interference effects can give rise to $\tau^0$ contributions that remain robust in the weak impurity limit. For high-mobility samples currently available the zeroth order terms are expected to dominate (an exhaustive discussion of the various regimes in the AHE is found in Ref.~\cite{Nagaosa2010}). This significant contrast between the longitudinal and transverse conductivities has motivated consistent efforts across theoretical considerations and experimental examinations. Whereas the qualitative picture is considerably clearer at the present stage, the quantitative evaluation of the zeroth-order disorder terms is still developing. The work of Ado \textit{et al} demonstrated that terms of order $1/\tau^2$ must be taken into account in order to obtain the full expression for the zeroth order disorder terms in the conductivity \cite{Ado2015, Ado2016}. The DMKE, as shown in this work, reveals that it is not just inter-band coherence that is important, but quantum interference between disorder lines as well. 

The procedure adopted in the DMKE in linear response is physically insightful in its own right. The leading term in the expansion comes from the band-diagonal part of the density matrix, $n$, and is linear in  $\tau$, denoted by $n^{(-1)}$. The band off-diagonal part, $S$, starts at order zero in $\tau$ and has an intrinsic contribution with topological origins and a disorder contribution which is the result of substituting the leading order $n \propto \tau$ into the inter-band scattering term, itself of order $1/\tau$. This latter term is one of the sub-leading disorder corrections which gives rise to important cancellations in the anomalous and spin-Hall effects in semiconductor systems. Next, an additional sub-leading term is present in the band-diagonal part $n$, which we may refer to as $n^{(0)}$. The equation used to determine this term was initially shown to have the form
\begin{equation}
    J_0 [n^{(0)}] = - J_{\text{inter}} [S^{0}], 
\end{equation}
where $J_0$ refers to the intra-band scattering term, $J_{\text{inter}}$ to the inter-band scattering term, and the RHS acts as the driving term for the LHS in the way the perturbation scheme is formulated. It was subsequently shown that the RHS needs to be supplemented by additional scattering terms in order to account fully for skew scattering and side jump \cite{Rhonald2022}. In this work, we have shown that, in order to recover the results of Ref.~\cite{Ado2017}, an additional source term appears on the RHS, which has the form $ - J_4[n^{(-1)}]$. The scattering operator $J_4$ is formally of order (disorder strength)$^2$, while $n^{(-1)}$ is formally of order (disorder strength)$^{-1}$, making the RHS linear in the disorder strength, and the additional contribution to $n$ formally of zeroth order in the disorder strength, as expected \cite{Kovalev2008, Onoda2008, Kamenev2009, Mirco2016}. This procedure also demonstrates why (i) the sub-leading contribution $n^{(0)}$ requires terms of order $1/\tau^2$ and (ii) no further terms of higher order in the impurity strength are needed, in other words no need for $1/\tau^3$ \cite{Kovalev2009}.

In contrast, even within linear response, the semiclassical Boltzmann picture lacks a strategy for systematically organizing higher order disorder scattering particularly the quantum interference contributions associated with crossed impurity processes\cite{Gorini2012, Xiao2019, Nandy2019, Du2021, Rhonald2024}. Quantum kinetic formulations, derived from the quantum Liouville equation and non-equilibrium Green function methods, provide such a microscopic description by incorporating impurity scattering and interband coherence into a systematic kinetic equation for Bloch electrons \cite{Culcer2017, Bhalla2023, Kamal2023}. In gauge covariant formulations, band geometry enters through Berry connection matrix elements, while impurity appears through a collision integral with the familiar self energy and vertex structures. Upon projecting onto the band diagonal parts in the quasiparticle limit, we recover the semiclassical Boltzmann equation, making the quantum kinetic framework a suitable starting point for our analysis \cite{Aversa1995, Takahiro2016, Xiao2019, Parks2023}.

\section{Application: Anomalous Hall Effect}
\label{Section_3_Results_and_discussions}

In this section, we will apply the quantum kinetic theory developed in the Sec.~\ref{Section_2_Model_and_theory} to evaluate the Hall conductivities in the massive Dirac model, which is an effective model for various two-dimensional materials such as the topological insulators surface states, and monolayer transition metal dichalcogenides.
In the effective massive Dirac model below, the two-component spinors are taken to be unit-normalized pseudospinors.
Their overlaps therefore represent the dimensionless quantities corresponding to $\Omega_0^{-1}\braket{u^m_{\bm{k}}}{u^m_{\bm{k}'}}$ in the general Bloch formalism.
The massive Dirac model Hamiltonian is given by:
\begin{equation}
H_0(\bm{k}) = \hbar v_F (k_x \sigma_x + k_y \sigma_y) + m \sigma_z \,,
\end{equation}
where $v_F$ is the Fermi velocity, $m$ is the mass term, and $\sigma_{x,y,z}$ are the Pauli matrices.
The Hamiltonian can be diagonalized by introducing the following eigenstates:
\begin{align}
& \ket{u^-_{\bm{k}}} = 
\left( -\frac{\varepsilon_0 - m}{\sqrt{2 \varepsilon_0(\varepsilon_0-m)}}, \frac{\hbar v_F (k_x + \mathrm{i}\,k_y)}{\sqrt{2 \varepsilon_0(\varepsilon_0-m)}} \right) \,,\\
& \ket{u^+_{\bm{k}}} = 
\left( \frac{\hbar v_F (k_x-\mathrm{i}\,k_y)}{\sqrt{2 \varepsilon_0(\varepsilon_0-m)}}, \frac{\varepsilon_0 - m}{\sqrt{2 \varepsilon_0(\varepsilon_0-m)}} \right) \,,
\end{align}
corresponding to the eigen-energy $\varepsilon_0^+=+\varepsilon_0$ and $\varepsilon_0^-=-\varepsilon_0$, where $\varepsilon_0 = \sqrt{(\hbar v_F k)^2 + m^2}$.
In all the following discussions, we will consider the electron-doped case with the Fermi energy $\varepsilon_F > m > 0$.

As the first step, we focus on the second-order Born approximation results.
Since the magnitude of the wavevector is conserved in the elastic scattering processes, we only need to evaluate the following unit-cell wavefunction overlap $\norm{\braket{u^+_{\bm{k}}}{u^+_{\bm{k}_1}}}^2$:
\begin{equation}
\frac{1}{2} \left(1+\frac{m^2}{\varepsilon_F^2}\right) + \frac{1}{2} \left(1-\frac{m^2}{\varepsilon_F^2}\right) \cos(\phi - \phi_1) \,.
\end{equation}
Using Eq.~\eqref{Eq: Born relaxation time}, we get:
\begin{equation}
\frac{1}{\tau^+_{\bm{k}}} = \frac{n_{\text{imp}} V_0^2}{2 \hbar^3 v_F^2} \frac{\varepsilon_F^2+m^2}{\varepsilon_F} \,.
\end{equation}
The corresponding transport relaxation time Eq.~\eqref{Eq_Born_transport_relaxation_time}, corrected with the angle factor $(1-\cos(\phi-\phi_1))$, where $\phi-\phi_1 = \bm{k}\cdot\bm{k}_1/(k k_1)$, can be evaluated as:
\begin{equation}
\frac{1}{\tau_{\text{tr}, \bm{k}}^+} = \frac{n_{\text{imp}} V_0^2}{4 \hbar^3 v_F^2} \frac{\varepsilon_F^2 + 3 m^2}{\varepsilon_F}
\end{equation}

In the linear response to the electric field regime, the diagonal part of the density matrix at the leading order of the impurity density at zero temperature $\mel{+,\bm{k}}{n_E^{(-1)}}{+,\bm{k}}$ has the following matrix elements:
\begin{equation}\label{Eq: nE Boltzmann ansatz}
-e E_x \tau_{\text{tr}} \frac{\hbar v_F^2 k_F}{\varepsilon_{F}} \cos\phi_{\bm{k}} \delta(\varepsilon^+_{\bm{k}} - \varepsilon_{F})\,.
\end{equation}
Therefore, we can evaluate the longitudinal conductivity $\sigma_{xx} = -e \mathrm{Tr} ( n_E^{(-1)} v_x)/E_x$.
The velocity operator is given by $v_x = v_F \sigma_x$, and its matrix elements in the Bloch eigen-basis read (we have pinned the Fermi level in the conduction band):
\begin{equation}
\mel{+,\bm{k}}{v_x}{+,\bm{k}} = \frac{\hbar v_F^2 k}{\varepsilon^+_{\bm{k}}} \cos\phi_{\bm{k}} \,.
\end{equation}
Now we are ready to evaluate the longitudinal conductivity:
\begin{equation}
\sigma_{xx} = \frac{e^2}{4\pi \hbar^2} \tau_{\text{tr}} \frac{\varepsilon_F^2 - m^2}{\varepsilon_F} \,.
\end{equation}
We note that this result can be directly mapped to the Drude conductivity formula $\sigma_{xx} = n e^2 \tau_{\text{tr}}/m^*$, where the carrier density $n = (\varepsilon_F^2 - m^2)/(h^2 v_F^2)$, and the effective mass $m^* = \varepsilon_F/v_F^2$.

Now, we can move on to evaluate the Hall conductivity.
The first leading order contribution comes from the intrinsic Berry connection effect, which can be captured by the off-diagonal part of the density matrix at the zeroth order of the impurity density $S_E^{(0)}$:
\begin{equation}
\me{S_E^{(0)}}{mm'}{\bm{k}\bm{k}} = - e E_x \frac{f^{(0)}(\varepsilon^{m'}_{\bm{k}}) - f^{(0)}(\varepsilon^{m}_{\bm{k}})}{\varepsilon^{m}_{\bm{k}} - \varepsilon^{m'}_{\bm{k}}} \me{\mathcal{A}_x}{mm'}{\bm{k}\bm{k}}\,.
\end{equation}
The intrinsic Hall conductivity can be evaluated as $\sigma_{xy}^{\mathrm{int}} = -e \mathrm{Tr} ( S_E^{(0)} v_y )/E_x$, where the off-diagonal matrix elements of the velocity operator read:
\begin{equation}
\me{v_y}{mm'}{\bm{k}\bm{k}} = \frac{\varepsilon^{m}_{\bm{k}} - \varepsilon^{m'}_{\bm{k}}}{\hbar} \me{\mathcal{A}_y}{mm'}{\bm{k}\bm{k}} \,.
\end{equation}
Since we only have two bands, the intrinsic Hall conductivity can be simplified as:
\begin{equation}
\sigma_{xy}^{\mathrm{int}} = \frac{e^2}{(2\pi)^2\hbar} \int \Theta(\varepsilon_F - \varepsilon^+_{\bm{k}}) \Omega^+_z + \Omega^-_z \dd[2]{k} \,,
\end{equation}
where $\Theta(\varepsilon)$ is the Heaviside step function, and the valence band is fully occupied.
The Berry curvature contributions can be evaluated as:
\begin{equation}
\Omega^{n}_z = 2 \Im \sum_{m \neq n} \me{\mathcal{A}_x}{nm}{\bm{k}\bm{k}} \me{\mathcal{A}_y}{mn}{\bm{k}\bm{k}} \,, \quad
\Omega^{\pm}_z = \mp \frac{mv^2}{2\varepsilon_0^3} \,.
\end{equation}
Finally, we get the relaxation time irrelevant intrinsic Hall conductivity contribution:
\begin{equation}
\sigma_{xy}^{\mathrm{int}} = \frac{e^2}{4\pi\hbar} \frac{m}{\varepsilon_F} \,.
\end{equation}

Next, we evaluate the Hall conductivity due to the quantum interference effects in the fourth-order collision integral.
Focusing on the diagonal part of the density matrix, explicitly, we first evaluate the following trace:
\begin{equation}
\sigma^{\mathrm{X}}_{xy} = \frac{e}{E_x} \mathrm{Tr} \left( \tau^{+}_{\text{tr},\bm{k}} \mathcal{I}^{(4)}_{\mathrm{X}}[n_E^{(-1)}] v_y \right) \,,
\end{equation}
where $\mathcal{I}^{(4)}_{\mathrm{X}}$ is given by Eq.~\eqref{Eq_I4_TypeII_X}.
For our two band model, the scattering integral can be further simplified as $\me{\mathcal{I}^{(4)}_{\mathrm{X}}[n_E^{(-1)}]}{++}{\bm{k}\bm{k}}$:
\begin{equation}
\begin{aligned}
& -\sum_{\bm{k}_1,\bm{k}_2,\bm{k}_3} \frac{3eE_x}{\hbar^2} \tau_{\text{tr}} \frac{n_{\text{imp}}^2 V_0^4}{(2\pi)^4} \delta(\bm{k}-\bm{k}_1+\bm{k}_2-\bm{k}_3) \times \\
& \quad\qquad \braket{u^+_{\bm{k}}}{u^+_{\bm{k}_1}} \braket{u^+_{\bm{k}_1}}{u^+_{\bm{k}_2}} \braket{u^+_{\bm{k}_2}}{u^+_{\bm{k}_3}} \braket{u^+_{\bm{k}_3}}{u^+_{\bm{k}}} \times \\
& \quad\qquad \cos(\phi_2) \delta(k_2-k_F) (\Phi_2+\Phi_7)
\end{aligned}
\end{equation}
The velocity operator diagonal matrix elements in the Bloch eigen-basis read:
\begin{equation}
\mel{+,\bm{k}}{v_y}{+,\bm{k}} = \frac{\hbar v_F^2 k}{\varepsilon^+_{\bm{k}}} \sin\phi_{\bm{k}} \,.
\end{equation}
Now, we are ready to evaluate the Hall conductivity contribution from the X diagram:
\begin{equation}
\sigma^{\mathrm{X}}_{xy} = \frac{e\tau_{\text{tr}}}{E_x} \sum_{\bm{k}} \me{\mathcal{I}^{(4)}_{\mathrm{X}}[n_E^{(-1)}]}{++}{\bm{k}\bm{k}} \me{v_y}{++}{\bm{k}\bm{k}} \,,
\end{equation}
which can be further anti-symmetrized to $(\sigma^{\mathrm{X}}_{xy} - \sigma^{\mathrm{X}}_{yx})/2$ to get the Hall conductivity.

Next, we need to evaluate the four layers of wave vector integrations, we first use the momentum conservation delta function to eliminate the $\bm{k}_3$ integration $\bm{k}_3 = \bm{k} - \bm{k}_1 + \bm{k}_2$.
Next, we perform the $\bm{k}_2$ integration by switching to polar coordinates, where we have $k_2 = k_F$ due to the delta function $\delta(k_2 - k_F)$ brought by the diagonal part of the density matrix $n_E^{(-1)}$, which is from the slow varying density matrix part $\expval*{\rho(t)}$ in Eq.~\eqref{Eq: nE Boltzmann ansatz}.
The remaining two integrations over $\bm{k}$ and $\bm{k}_1$ is determined by the two factors $\Phi_2$ and $\Phi_7$.
When performing the integration over the magnitude of $k_3$ and $k_1$, the following identity is useful for the energy conservation part:
\begin{equation}
\begin{aligned}\label{Eq: Identity for energy denominators}
&\frac{1}{(\varepsilon_F^2-{\varepsilon^2_{\bm{k}_3}} + \mathrm{i}\,0^+)(\varepsilon_F^2-{\varepsilon^2_{\bm{k}_1}} - \mathrm{i}\,0^+)} - \\
&\frac{1}{(\varepsilon_F^2-{\varepsilon^2_{\bm{k}_3}} - \mathrm{i}\,0^+)(\varepsilon_F^2-{\varepsilon^2_{\bm{k}_1}} + \mathrm{i}\,0^+)} = \\
&2\pi \mathrm{i}\left( \frac{\delta(\varepsilon_F - \varepsilon_{\bm{k}_3})}{\varepsilon_F^2 - \varepsilon_{\bm{k}_1}^2} + \frac{\delta(\varepsilon_F - \varepsilon_{\bm{k}_1})}{\varepsilon_F^2 - \varepsilon_{\bm{k}_3}^2} \right) \,,
\end{aligned}  
\end{equation}
where we have ignored the band indices in all the superscripts for simplicity, and they all refer to the conduction band.
The denominators in Eq.~\eqref{Eq: Identity for energy denominators} can be simplified as $\hbar^2 v_F^2 k_F^2( 1 - (\hat{\bm{k}}-\hat{\bm{k}_1}+\hat{\bm{k}}_2)^2)$, which can be further simplified as:
\begin{equation}
-2 (1+\cos(\phi-\phi_2)-\cos(\phi-\phi_1)-\cos(\phi_2-\phi_1))
\end{equation}
Next, we need to evaluate the unit cell wavefunction overlaps in the numerator.
After the anti-symmetrization procedure, we find that only the imaginary part of the wavefunction overlaps product contributes to the Hall conductivity, and $\Im[\braket{u^+_{\bm{k}}}{u^+_{\bm{k}_1}} \braket{u^+_{\bm{k}_1}}{u^+_{\bm{k}_2}} \braket{u^+_{\bm{k}_2}}{u^+_{\bm{k}_3}} \braket{u^+_{\bm{k}_3}}{u^+_{\bm{k}}}]$ reads:
\begin{small}
\begin{equation}\label{Eq: Scattering element}
-\frac{m\hbar^2v_F^2k_F^2}{\varepsilon^3_F}(\sin(\phi-\phi_1)+\sin(\phi_1-\phi_2)+\sin(\phi_2-\phi)) \,.
\end{equation}
\end{small}
Combining all the factors together, we get the final expression for the Hall conductivity from the X diagram:
\begin{widetext}
\begin{equation}\label{Eq: X diagram conductivity before integration}
\sigma^{\mathrm{X}}_{xy} = - \frac{6e^2}{(2\pi)^3\hbar} \frac{m \varepsilon_F \hbar^2 v_F^2 k_F^2}{(\varepsilon_F^2 + 3 m^2)^2} \int \sin(\phi-\phi_2) \frac{\sin(\phi-\phi_1)+\sin(\phi_1-\phi_2)+\sin(\phi_2-\phi)}{1+\cos(\phi-\phi_2)-\cos(\phi-\phi_1)-\cos(\phi_2-\phi_1)} \dd{\phi}\dd{\phi_1}\dd{\phi_3} \,.
\end{equation}
\end{widetext}
The angular integration yields $-8\pi^3$, therefore, the final result for the Hall conductivity from the X diagram is:
\begin{equation}
\sigma_{xy}^{\mathrm{X}} = \frac{3 e^2}{h} \frac{m \varepsilon_F \hbar^2 v_F^2 k_F^2}{(\varepsilon_F^2 + 3 m^2)^2} \,.
\end{equation}

Next, we can consider the contributions from the $\Psi$-diagrams, where the self-energy corrections and the ladder dressing to the velocity operators will still hold, the collision mechanism will be a little different.
For the $\Psi$-diagrams, the collision terms usually includes three interaction potentials $V$ to the same side of the density matrix $\expval*{\rho(t)}$, and one $V$ on the other side, preserving the momentum conservation condition $\delta(\bm{k}-\bm{k}_1 + \bm{k}_2 - \bm{k}_3)$.
This structure determines that the slow varying density matrix $\expval*{\rho(t)}$ will pin the magnitude of $\bm{k}_3$ onto the Fermi surface $k_{F}$.
Particularly, for the massive Dirac model with white noise impurities, we notice that the angle integrand Eq.~\eqref{Eq: Scattering element} will change sign when we swap the $\phi_1$ and $\phi_2$ simultaneously, but will not change the sign of $\expval*{\rho(t)}$. Therefore, we notice that for this white noise model, the corresponding integrand in the conductivity integral for $\Psi$-diagram will vanish due to symmetry. This is just a reflection of the statistical isotropy of the Gaussian white noise model we considered. If an anisotropic impurity potential is used this term will generally be nonzero \cite{Ado2017}.

In parallel, here we also demonstrate how the same results can be obtained from the Keldysh formalism.
For simplifications of the calculations, we introduce the vector $\bm{d}_{\bm{k}} = (\hbar v_F k_x, \hbar v_F k_y, m)$, and a unit vector $\bm{n}_{\bm{k}} = \bm{d}_{\bm{k}}/\varepsilon_0$.
Furthermore, we notice that the projection operators to the $\pm$ bands can be expressed as:
\begin{equation}
P_{\pm}(\bm{k}) = \frac{1}{2} (1 \pm \bm{n}_{\bm{k}} \cdot \bm{\sigma}) \,.
\end{equation}
The free Green function is given by:
\begin{equation}
\me{G_0^{R/A}(\varepsilon)}{++}{\bm{k}\bm{k}} = \frac{1}{\varepsilon - \varepsilon^+_{\bm{k}} \pm \mathrm{i}\,0^+} \,.
\end{equation}
And the corresponding self-energy correction in the second-order Born approximation can be evaluated as:
\begin{equation}
\me{\Sigma^{R/A}(\varepsilon)}{++}{\bm{k}\bm{k}} = \frac{n_{\text{imp}} V_0^2}{(2\pi)^2} \int \me{G_0^{R/A}(\varepsilon)}{++}{\bm{k}_1\bm{k}_1} \norm{\braket{u^+_{\bm{k}}}{u^+_{\bm{k}_1}}}^2 \dd[2]{k_1} \,.
\end{equation}
After performing the integration, we get:
\begin{equation}
\tau^+_{\bm{k}} = \frac{\hbar}{2 \Im \me{\Sigma^{R}(\varepsilon)}{++}{\bm{k}\bm{k}}} \,,
\end{equation}
Here we do not repeat the evaluation of the transport relaxation time using the vertex correction, since the results must be the same as we obtained before.
which is the same as the result obtained from the Boltzmann equation approach.
Therefore, the conduction band impurity averaged Green's function reads:
\begin{equation}
\me{G^{R/A}(\varepsilon)}{++}{\bm{k}\bm{k}} = \frac{1}{\varepsilon_F - \varepsilon^+_{\bm{k}} \pm \mathrm{i}\,\hbar/(2\tau^+_{\bm{k}})} \,.
\end{equation}
Using Eq.~\eqref{Eq_Keldysh_longitudinal_conductivity}, the longitudinal conductivity can be evaluated as:
\begin{equation}
\sigma_{xx} = \frac{e^2\hbar}{(2\pi)^3} \int \tr[ v_x G^R \tilde{v}_x G^A ] \dd[2]{k} \,,
\end{equation}
where all the quantities are evaluated at the Fermi energy $\varepsilon_F$.
Using the identity Eq.~\eqref{Eq_Green_function_identity}, the integration can be simplified to:
\begin{equation}
\sigma_{xx} = \frac{e^2 \tau_{\text{tr}}}{(2\pi)^2} \int v_F^2 \cos^2\phi \delta(\varepsilon_F - \varepsilon^+_{\bm{k}}) \dd[2]{k} \,, 
\end{equation}
which recovers the same result as before.

To evaluate the Hall conductivity due to the $X$-diagrams using the Keldysh formalism, we use Eq.~\eqref{Eq_X_Keldysh_conductivity}:
\begin{equation}
\begin{aligned}
\sigma_{xy}^{X} = & 3e^2 v_F^2 \tau_{\text{TR}}^2 n_{\text{imp}}^2 V_0^4 \hbar \int \dd[2]{k} \dd[2]{k_1} \dd[2]{k_2} \dd[2]{k_3} \\
& \times \cos\phi_1\sin\phi \frac{\Im \tr[P_+(\bm{k}) P_+(\bm{k}_2) P_+(\bm{k}_1) P_+(\bm{k}_3)]}{\hbar^2 v_F^2 k_2^2 - \hbar^2 v_F^2 k_3^2} \\
& \times \delta(\varepsilon_F - \varepsilon^+_{\bm{k}}) \delta(\varepsilon_F - \varepsilon^+_{\bm{k}_1}) \delta(\bm{k}-\bm{k}_1+\bm{k}_2-\bm{k}_3) \,.
\end{aligned}
\end{equation}
This can be further simplified to:
\begin{equation}
\begin{aligned}
\sigma_{xy}^{X} = & \frac{3 e^2}{(2\pi)^4\hbar} v_F^2 \frac{\tau_{\text{tr}}^2}{\tau^2} n_{\text{imp}}^2 V_0^4 \int \dd{\phi} \dd{\phi_1} \dd{\phi_2} \cos\phi_1 \\
&\sin\phi \frac{\Im \tr[P_+(\bm{k}) P_+(\bm{k}_2) P_+(\bm{k}_1) P_+(\bm{k}+\bm{k}_1-\bm{k}_2)]}{\hbar^2 v_F^2 k_F^2 (1 - (\hat{\bm{k}}-\hat{\bm{k}_1}+\hat{\bm{k}}_2)^2)} \,,
\end{aligned}
\end{equation}
which can be anti-symmetrized to get the same structure as we calculated in Eq.~\eqref{Eq: X diagram conductivity before integration}, leading to the same final result for the Hall conductivity from the X diagram.

\section{Conclusion and Outlook}
\label{Section_4_Conclusion}

In this work, we have examined a systematic quantum kinetic framework to investigate impurity scattering effects up to the fourth order in the scattering potential.
By iteratively solving the density matrix equations in the length gauge, we explicitly derived the collision integrals corresponding to self-energy corrections, ladder vertex corrections, and, most crucially, the crossing terms.
Applying this formalism to the two-dimensional massive Dirac model with Gaussian white noise impurity, we found that the $\Psi$-type crossing diagrams vanish due to the statistical symmetry of the impurity potential.
In contrast, the X-type crossing diagrams lead to a finite anomalous Hall conductivity.
This contribution coexists with the intrinsic Berry curvature term, demonstrating that in the quantum transport regime, the extrinsic impurity effects can compare to the scaling behavior of intrinsic topology, a feature arising from the balance between impurity-enhanced quantum interference and impurity-induced lifetime reduction.

In the future the DMKE formalism can be extended along several directions. Firstly, the current restriction to Gaussian white noise can be relaxed. Our derivation of the multiple layer commutator structure is general; introducing non-Gaussian or magnetic impurities would break the symmetry that currently suppresses the $\Psi$-diagrams, potentially unlocking new skew-scattering contributions scaling with $\tau$ or $\tau^0$ that are relevant for magnetic textures. Secondly, this length gauge density matrix approach is naturally compatible with Wannier techniques, enabling quantitative multiband calculations for realistic materials following the prescription of Ref.~\cite{Culcer2017,Sekine2017}.
Unlike the Dirac model used here for demonstration, real topological materials possess complex multi-band structures where the interband matrix elements of the position operator play a decisive role.
By integrating our collision integrals into first-principles Wannier calculations, one can quantitatively evaluate the competition between Berry curvature and high-order scattering interference in realistic materials, providing a reliable theoretical tool for interpreting more transport experiments in complicated multiband materials.

\acknowledgments

RR acknowledges support from the UNSW Gordon Godfrey Bequest for Theoretical Physics visitors programme.

\newpage


\begin{thebibliography}{83}%
\makeatletter
\providecommand \@ifxundefined [1]{%
 \@ifx{#1\undefined}
}%
\providecommand \@ifnum [1]{%
 \ifnum #1\expandafter \@firstoftwo
 \else \expandafter \@secondoftwo
 \fi
}%
\providecommand \@ifx [1]{%
 \ifx #1\expandafter \@firstoftwo
 \else \expandafter \@secondoftwo
 \fi
}%
\providecommand \natexlab [1]{#1}%
\providecommand \enquote  [1]{``#1''}%
\providecommand \bibnamefont  [1]{#1}%
\providecommand \bibfnamefont [1]{#1}%
\providecommand \citenamefont [1]{#1}%
\providecommand \href@noop [0]{\@secondoftwo}%
\providecommand \href [0]{\begingroup \@sanitize@url \@href}%
\providecommand \@href[1]{\@@startlink{#1}\@@href}%
\providecommand \@@href[1]{\endgroup#1\@@endlink}%
\providecommand \@sanitize@url [0]{\catcode `\\12\catcode `\$12\catcode
  `\&12\catcode `\#12\catcode `\^12\catcode `\_12\catcode `\%12\relax}%
\providecommand \@@startlink[1]{}%
\providecommand \@@endlink[0]{}%
\providecommand \url  [0]{\begingroup\@sanitize@url \@url }%
\providecommand \@url [1]{\endgroup\@href {#1}{\urlprefix }}%
\providecommand \urlprefix  [0]{URL }%
\providecommand \Eprint [0]{\href }%
\providecommand \doibase [0]{https://doi.org/}%
\providecommand \selectlanguage [0]{\@gobble}%
\providecommand \bibinfo  [0]{\@secondoftwo}%
\providecommand \bibfield  [0]{\@secondoftwo}%
\providecommand \translation [1]{[#1]}%
\providecommand \BibitemOpen [0]{}%
\providecommand \bibitemStop [0]{}%
\providecommand \bibitemNoStop [0]{.\EOS\space}%
\providecommand \EOS [0]{\spacefactor3000\relax}%
\providecommand \BibitemShut  [1]{\csname bibitem#1\endcsname}%
\let\auto@bib@innerbib\@empty
\bibitem [{\citenamefont {Raimondi}\ \emph {et~al.}(2006)\citenamefont
  {Raimondi}, \citenamefont {Gorini}, \citenamefont {Schwab},\ and\
  \citenamefont {Dzierzawa}}]{Raimondi2006}%
  \BibitemOpen
  \bibfield  {author} {\bibinfo {author} {\bibfnamefont {R.}~\bibnamefont
  {Raimondi}}, \bibinfo {author} {\bibfnamefont {C.}~\bibnamefont {Gorini}},
  \bibinfo {author} {\bibfnamefont {P.}~\bibnamefont {Schwab}},\ and\ \bibinfo
  {author} {\bibfnamefont {M.}~\bibnamefont {Dzierzawa}},\ }\bibfield  {title}
  {\bibinfo {title} {Quasiclassical approach to the spin hall effect in the
  two-dimensional electron gas},\ }\href
  {https://doi.org/10.1103/PhysRevB.74.035340} {\bibfield  {journal} {\bibinfo
  {journal} {Phys. Rev. B}\ }\textbf {\bibinfo {volume} {74}},\ \bibinfo
  {pages} {035340} (\bibinfo {year} {2006})}\BibitemShut {NoStop}%
\bibitem [{\citenamefont {Culcer}\ \emph
  {et~al.}(2010{\natexlab{a}})\citenamefont {Culcer}, \citenamefont
  {Hankiewicz}, \citenamefont {Vignale},\ and\ \citenamefont
  {Winkler}}]{Culcer20102}%
  \BibitemOpen
  \bibfield  {author} {\bibinfo {author} {\bibfnamefont {D.}~\bibnamefont
  {Culcer}}, \bibinfo {author} {\bibfnamefont {E.~M.}\ \bibnamefont
  {Hankiewicz}}, \bibinfo {author} {\bibfnamefont {G.}~\bibnamefont
  {Vignale}},\ and\ \bibinfo {author} {\bibfnamefont {R.}~\bibnamefont
  {Winkler}},\ }\bibfield  {title} {\bibinfo {title} {Side jumps in the spin
  hall effect: Construction of the boltzmann collision integral},\ }\href
  {https://doi.org/10.1103/PhysRevB.81.125332} {\bibfield  {journal} {\bibinfo
  {journal} {Phys. Rev. B}\ }\textbf {\bibinfo {volume} {81}},\ \bibinfo
  {pages} {125332} (\bibinfo {year} {2010}{\natexlab{a}})}\BibitemShut
  {NoStop}%
\bibitem [{\citenamefont {Gao}\ \emph {et~al.}(2014)\citenamefont {Gao},
  \citenamefont {Yang},\ and\ \citenamefont {Niu}}]{Gao2014}%
  \BibitemOpen
  \bibfield  {author} {\bibinfo {author} {\bibfnamefont {Y.}~\bibnamefont
  {Gao}}, \bibinfo {author} {\bibfnamefont {S.~A.}\ \bibnamefont {Yang}},\ and\
  \bibinfo {author} {\bibfnamefont {Q.}~\bibnamefont {Niu}},\ }\bibfield
  {title} {\bibinfo {title} {Field induced positional shift of bloch electrons
  and its dynamical implications},\ }\href
  {https://doi.org/10.1103/PhysRevLett.112.166601} {\bibfield  {journal}
  {\bibinfo  {journal} {Phys. Rev. Lett.}\ }\textbf {\bibinfo {volume} {112}},\
  \bibinfo {pages} {166601} (\bibinfo {year} {2014})}\BibitemShut {NoStop}%
\bibitem [{\citenamefont {Du}\ \emph {et~al.}(2021)\citenamefont {Du},
  \citenamefont {Wang}, \citenamefont {Sun}, \citenamefont {Lu},\ and\
  \citenamefont {Xie}}]{Du2021}%
  \BibitemOpen
  \bibfield  {author} {\bibinfo {author} {\bibfnamefont {Z.~Z.}\ \bibnamefont
  {Du}}, \bibinfo {author} {\bibfnamefont {C.~M.}\ \bibnamefont {Wang}},
  \bibinfo {author} {\bibfnamefont {H.-P.}\ \bibnamefont {Sun}}, \bibinfo
  {author} {\bibfnamefont {H.-Z.}\ \bibnamefont {Lu}},\ and\ \bibinfo {author}
  {\bibfnamefont {X.~C.}\ \bibnamefont {Xie}},\ }\bibfield  {title} {\bibinfo
  {title} {Quantum theory of the nonlinear hall effect},\ }\href
  {https://doi.org/10.1038/s41467-021-25273-4} {\bibfield  {journal} {\bibinfo
  {journal} {Nature Communications}\ }\textbf {\bibinfo {volume} {12}},\
  \bibinfo {pages} {5038} (\bibinfo {year} {2021})}\BibitemShut {NoStop}%
\bibitem [{\citenamefont {Atencia}\ \emph
  {et~al.}(2022{\natexlab{a}})\citenamefont {Atencia}, \citenamefont {Niu},\
  and\ \citenamefont {Culcer}}]{Atencia2022}%
  \BibitemOpen
  \bibfield  {author} {\bibinfo {author} {\bibfnamefont {R.~B.}\ \bibnamefont
  {Atencia}}, \bibinfo {author} {\bibfnamefont {Q.}~\bibnamefont {Niu}},\ and\
  \bibinfo {author} {\bibfnamefont {D.}~\bibnamefont {Culcer}},\ }\bibfield
  {title} {\bibinfo {title} {Semiclassical response of disordered conductors:
  Extrinsic carrier velocity and spin and field-corrected collision integral},\
  }\href {https://doi.org/10.1103/PhysRevResearch.4.013001} {\bibfield
  {journal} {\bibinfo  {journal} {Phys. Rev. Res.}\ }\textbf {\bibinfo {volume}
  {4}},\ \bibinfo {pages} {013001} (\bibinfo {year}
  {2022}{\natexlab{a}})}\BibitemShut {NoStop}%
\bibitem [{\citenamefont {Valet}\ and\ \citenamefont
  {Raimondi}(2023)}]{Valet2023}%
  \BibitemOpen
  \bibfield  {author} {\bibinfo {author} {\bibfnamefont {T.}~\bibnamefont
  {Valet}}\ and\ \bibinfo {author} {\bibfnamefont {R.}~\bibnamefont
  {Raimondi}},\ }\bibfield  {title} {\bibinfo {title} {Semiclassical kinetic
  theory for systems with non-trivial quantum geometry and the expectation
  value of physical quantities},\ }\href
  {https://doi.org/10.1209/0295-5075/ace379} {\bibfield  {journal} {\bibinfo
  {journal} {Europhysics Letters}\ }\textbf {\bibinfo {volume} {143}},\
  \bibinfo {pages} {26004} (\bibinfo {year} {2023})}\BibitemShut {NoStop}%
\bibitem [{\citenamefont {Raimondi}\ and\ \citenamefont
  {Valet}(2025)}]{Raimondi2025}%
  \BibitemOpen
  \bibfield  {author} {\bibinfo {author} {\bibfnamefont {R.}~\bibnamefont
  {Raimondi}}\ and\ \bibinfo {author} {\bibfnamefont {T.}~\bibnamefont
  {Valet}},\ }\bibfield  {title} {\bibinfo {title} {Quantum kinetic theory of
  the spin hall effect for disordered graphene with rashba spin–orbit
  coupling},\ }\bibfield  {journal} {\bibinfo  {journal} {Condensed Matter}\
  }\textbf {\bibinfo {volume} {10}},\ \href
  {https://doi.org/10.3390/condmat10010004} {10.3390/condmat10010004} (\bibinfo
  {year} {2025})\BibitemShut {NoStop}%
\bibitem [{\citenamefont {Valet}\ and\ \citenamefont
  {Raimondi}(2025)}]{Valet2025}%
  \BibitemOpen
  \bibfield  {author} {\bibinfo {author} {\bibfnamefont {T.}~\bibnamefont
  {Valet}}\ and\ \bibinfo {author} {\bibfnamefont {R.}~\bibnamefont
  {Raimondi}},\ }\bibfield  {title} {\bibinfo {title} {Quantum kinetic theory
  of the linear response for weakly disordered multiband systems},\ }\href
  {https://doi.org/10.1103/PhysRevB.111.L041118} {\bibfield  {journal}
  {\bibinfo  {journal} {Phys. Rev. B}\ }\textbf {\bibinfo {volume} {111}},\
  \bibinfo {pages} {L041118} (\bibinfo {year} {2025})}\BibitemShut {NoStop}%
\bibitem [{\citenamefont {Culcer}\ \emph {et~al.}(2017)\citenamefont {Culcer},
  \citenamefont {Sekine},\ and\ \citenamefont {MacDonald}}]{Culcer2017}%
  \BibitemOpen
  \bibfield  {author} {\bibinfo {author} {\bibfnamefont {D.}~\bibnamefont
  {Culcer}}, \bibinfo {author} {\bibfnamefont {A.}~\bibnamefont {Sekine}},\
  and\ \bibinfo {author} {\bibfnamefont {A.~H.}\ \bibnamefont {MacDonald}},\
  }\bibfield  {title} {\bibinfo {title} {Interband coherence response to
  electric fields in crystals: Berry-phase contributions and disorder
  effects},\ }\href {https://doi.org/10.1103/PhysRevB.96.035106} {\bibfield
  {journal} {\bibinfo  {journal} {Phys. Rev. B}\ }\textbf {\bibinfo {volume}
  {96}},\ \bibinfo {pages} {035106} (\bibinfo {year} {2017})}\BibitemShut
  {NoStop}%
\bibitem [{\citenamefont {Atencia}\ \emph
  {et~al.}(2022{\natexlab{b}})\citenamefont {Atencia}, \citenamefont {Niu},\
  and\ \citenamefont {Culcer}}]{Rhonald2022}%
  \BibitemOpen
  \bibfield  {author} {\bibinfo {author} {\bibfnamefont {R.~B.}\ \bibnamefont
  {Atencia}}, \bibinfo {author} {\bibfnamefont {Q.}~\bibnamefont {Niu}},\ and\
  \bibinfo {author} {\bibfnamefont {D.}~\bibnamefont {Culcer}},\ }\bibfield
  {title} {\bibinfo {title} {Semiclassical response of disordered conductors:
  Extrinsic carrier velocity and spin and field-corrected collision integral},\
  }\href {https://doi.org/10.1103/PhysRevResearch.4.013001} {\bibfield
  {journal} {\bibinfo  {journal} {Phys. Rev. Res.}\ }\textbf {\bibinfo {volume}
  {4}},\ \bibinfo {pages} {013001} (\bibinfo {year}
  {2022}{\natexlab{b}})}\BibitemShut {NoStop}%
\bibitem [{\citenamefont {Atencia}\ \emph
  {et~al.}(2023{\natexlab{a}})\citenamefont {Atencia}, \citenamefont {Xiao},\
  and\ \citenamefont {Culcer}}]{Rhonald2023}%
  \BibitemOpen
  \bibfield  {author} {\bibinfo {author} {\bibfnamefont {R.~B.}\ \bibnamefont
  {Atencia}}, \bibinfo {author} {\bibfnamefont {D.}~\bibnamefont {Xiao}},\ and\
  \bibinfo {author} {\bibfnamefont {D.}~\bibnamefont {Culcer}},\ }\bibfield
  {title} {\bibinfo {title} {Disorder in the nonlinear anomalous hall effect of
  $\mathcal{PT}$-symmetric dirac fermions},\ }\href
  {https://doi.org/10.1103/PhysRevB.108.L201115} {\bibfield  {journal}
  {\bibinfo  {journal} {Phys. Rev. B}\ }\textbf {\bibinfo {volume} {108}},\
  \bibinfo {pages} {L201115} (\bibinfo {year}
  {2023}{\natexlab{a}})}\BibitemShut {NoStop}%
\bibitem [{\citenamefont {Ado}\ \emph {et~al.}(2015)\citenamefont {Ado},
  \citenamefont {Dmitriev}, \citenamefont {Ostrovsky},\ and\ \citenamefont
  {Titov}}]{Ado2015}%
  \BibitemOpen
  \bibfield  {author} {\bibinfo {author} {\bibfnamefont {I.~A.}\ \bibnamefont
  {Ado}}, \bibinfo {author} {\bibfnamefont {I.~A.}\ \bibnamefont {Dmitriev}},
  \bibinfo {author} {\bibfnamefont {P.~M.}\ \bibnamefont {Ostrovsky}},\ and\
  \bibinfo {author} {\bibfnamefont {M.}~\bibnamefont {Titov}},\ }\bibfield
  {title} {\bibinfo {title} {Anomalous hall effect with massive dirac
  fermions},\ }\href {https://doi.org/10.1209/0295-5075/111/37004} {\bibfield
  {journal} {\bibinfo  {journal} {Europhysics Letters}\ }\textbf {\bibinfo
  {volume} {111}},\ \bibinfo {pages} {37004} (\bibinfo {year}
  {2015})}\BibitemShut {NoStop}%
\bibitem [{\citenamefont {Ado}\ \emph {et~al.}(2016)\citenamefont {Ado},
  \citenamefont {Dmitriev}, \citenamefont {Ostrovsky},\ and\ \citenamefont
  {Titov}}]{Ado2016}%
  \BibitemOpen
  \bibfield  {author} {\bibinfo {author} {\bibfnamefont {I.~A.}\ \bibnamefont
  {Ado}}, \bibinfo {author} {\bibfnamefont {I.~A.}\ \bibnamefont {Dmitriev}},
  \bibinfo {author} {\bibfnamefont {P.~M.}\ \bibnamefont {Ostrovsky}},\ and\
  \bibinfo {author} {\bibfnamefont {M.}~\bibnamefont {Titov}},\ }\bibfield
  {title} {\bibinfo {title} {Anomalous hall effect in a 2d rashba
  ferromagnet},\ }\href {https://doi.org/10.1103/PhysRevLett.117.046601}
  {\bibfield  {journal} {\bibinfo  {journal} {Phys. Rev. Lett.}\ }\textbf
  {\bibinfo {volume} {117}},\ \bibinfo {pages} {046601} (\bibinfo {year}
  {2016})}\BibitemShut {NoStop}%
\bibitem [{\citenamefont {Ado}\ \emph {et~al.}(2017)\citenamefont {Ado},
  \citenamefont {Dmitriev}, \citenamefont {Ostrovsky},\ and\ \citenamefont
  {Titov}}]{Ado2017}%
  \BibitemOpen
  \bibfield  {author} {\bibinfo {author} {\bibfnamefont {I.~A.}\ \bibnamefont
  {Ado}}, \bibinfo {author} {\bibfnamefont {I.~A.}\ \bibnamefont {Dmitriev}},
  \bibinfo {author} {\bibfnamefont {P.~M.}\ \bibnamefont {Ostrovsky}},\ and\
  \bibinfo {author} {\bibfnamefont {M.}~\bibnamefont {Titov}},\ }\bibfield
  {title} {\bibinfo {title} {Sensitivity of the anomalous hall effect to
  disorder correlations},\ }\href {https://doi.org/10.1103/PhysRevB.96.235148}
  {\bibfield  {journal} {\bibinfo  {journal} {Phys. Rev. B}\ }\textbf {\bibinfo
  {volume} {96}},\ \bibinfo {pages} {235148} (\bibinfo {year}
  {2017})}\BibitemShut {NoStop}%
\bibitem [{\citenamefont {Zhang}\ and\ \citenamefont {Chen}(2023)}]{Zhang2023}%
  \BibitemOpen
  \bibfield  {author} {\bibinfo {author} {\bibfnamefont {J.-X.}\ \bibnamefont
  {Zhang}}\ and\ \bibinfo {author} {\bibfnamefont {W.}~\bibnamefont {Chen}},\
  }\bibfield  {title} {\bibinfo {title} {Anomalous hall effect in type-i weyl
  metals beyond the noncrossing approximation},\ }\href
  {https://doi.org/10.1103/PhysRevB.107.214204} {\bibfield  {journal} {\bibinfo
   {journal} {Phys. Rev. B}\ }\textbf {\bibinfo {volume} {107}},\ \bibinfo
  {pages} {214204} (\bibinfo {year} {2023})}\BibitemShut {NoStop}%
\bibitem [{\citenamefont {Gorini}\ \emph {et~al.}(2010)\citenamefont {Gorini},
  \citenamefont {Schwab}, \citenamefont {Raimondi},\ and\ \citenamefont
  {Shelankov}}]{Gorini2010}%
  \BibitemOpen
  \bibfield  {author} {\bibinfo {author} {\bibfnamefont {C.}~\bibnamefont
  {Gorini}}, \bibinfo {author} {\bibfnamefont {P.}~\bibnamefont {Schwab}},
  \bibinfo {author} {\bibfnamefont {R.}~\bibnamefont {Raimondi}},\ and\
  \bibinfo {author} {\bibfnamefont {A.~L.}\ \bibnamefont {Shelankov}},\
  }\bibfield  {title} {\bibinfo {title} {Non-abelian gauge fields in the
  gradient expansion: Generalized boltzmann and eilenberger equations},\ }\href
  {https://doi.org/10.1103/PhysRevB.82.195316} {\bibfield  {journal} {\bibinfo
  {journal} {Phys. Rev. B}\ }\textbf {\bibinfo {volume} {82}},\ \bibinfo
  {pages} {195316} (\bibinfo {year} {2010})}\BibitemShut {NoStop}%
\bibitem [{\citenamefont {Milletar\`{\i}}\ and\ \citenamefont
  {Ferreira}(2016)}]{Mirco2016}%
  \BibitemOpen
  \bibfield  {author} {\bibinfo {author} {\bibfnamefont {M.}~\bibnamefont
  {Milletar\`{\i}}}\ and\ \bibinfo {author} {\bibfnamefont {A.}~\bibnamefont
  {Ferreira}},\ }\bibfield  {title} {\bibinfo {title} {Quantum diagrammatic
  theory of the extrinsic spin hall effect in graphene},\ }\href
  {https://doi.org/10.1103/PhysRevB.94.134202} {\bibfield  {journal} {\bibinfo
  {journal} {Phys. Rev. B}\ }\textbf {\bibinfo {volume} {94}},\ \bibinfo
  {pages} {134202} (\bibinfo {year} {2016})}\BibitemShut {NoStop}%
\bibitem [{\citenamefont {Raimondi}\ \emph {et~al.}()\citenamefont {Raimondi},
  \citenamefont {Gorini},\ and\ \citenamefont {Tölle}}]{Raimondi2018}%
  \BibitemOpen
  \bibfield  {author} {\bibinfo {author} {\bibfnamefont {R.}~\bibnamefont
  {Raimondi}}, \bibinfo {author} {\bibfnamefont {C.}~\bibnamefont {Gorini}},\
  and\ \bibinfo {author} {\bibfnamefont {S.}~\bibnamefont {Tölle}},\ }\bibinfo
  {title} {Spin-charge coupling effects in a two-dimensional electron gas},\
  in\ \href {https://doi.org/10.1142/9789813234345_0005} {\emph {\bibinfo
  {booktitle} {Spin Orbitronics and Topological Properties of
  Nanostructures}}},\ pp.\ \bibinfo {pages} {80--109}\BibitemShut {NoStop}%
\bibitem [{\citenamefont {Lee}\ \emph {et~al.}(2004)\citenamefont {Lee},
  \citenamefont {Trionfi},\ and\ \citenamefont {Natelson}}]{Lee2004}%
  \BibitemOpen
  \bibfield  {author} {\bibinfo {author} {\bibfnamefont {S.}~\bibnamefont
  {Lee}}, \bibinfo {author} {\bibfnamefont {A.}~\bibnamefont {Trionfi}},\ and\
  \bibinfo {author} {\bibfnamefont {D.}~\bibnamefont {Natelson}},\ }\bibfield
  {title} {\bibinfo {title} {Quantum coherence in a ferromagnetic metal:
  Time-dependent conductance fluctuations},\ }\href
  {https://doi.org/10.1103/PhysRevB.70.212407} {\bibfield  {journal} {\bibinfo
  {journal} {Phys. Rev. B}\ }\textbf {\bibinfo {volume} {70}},\ \bibinfo
  {pages} {212407} (\bibinfo {year} {2004})}\BibitemShut {NoStop}%
\bibitem [{\citenamefont {Miyasato}\ \emph {et~al.}(2007)\citenamefont
  {Miyasato}, \citenamefont {Abe}, \citenamefont {Fujii}, \citenamefont
  {Asamitsu}, \citenamefont {Onoda}, \citenamefont {Onose}, \citenamefont
  {Nagaosa},\ and\ \citenamefont {Tokura}}]{Miyasato2007}%
  \BibitemOpen
  \bibfield  {author} {\bibinfo {author} {\bibfnamefont {T.}~\bibnamefont
  {Miyasato}}, \bibinfo {author} {\bibfnamefont {N.}~\bibnamefont {Abe}},
  \bibinfo {author} {\bibfnamefont {T.}~\bibnamefont {Fujii}}, \bibinfo
  {author} {\bibfnamefont {A.}~\bibnamefont {Asamitsu}}, \bibinfo {author}
  {\bibfnamefont {S.}~\bibnamefont {Onoda}}, \bibinfo {author} {\bibfnamefont
  {Y.}~\bibnamefont {Onose}}, \bibinfo {author} {\bibfnamefont
  {N.}~\bibnamefont {Nagaosa}},\ and\ \bibinfo {author} {\bibfnamefont
  {Y.}~\bibnamefont {Tokura}},\ }\bibfield  {title} {\bibinfo {title}
  {Crossover behavior of the anomalous hall effect and anomalous nernst effect
  in itinerant ferromagnets},\ }\href
  {https://doi.org/10.1103/PhysRevLett.99.086602} {\bibfield  {journal}
  {\bibinfo  {journal} {Phys. Rev. Lett.}\ }\textbf {\bibinfo {volume} {99}},\
  \bibinfo {pages} {086602} (\bibinfo {year} {2007})}\BibitemShut {NoStop}%
\bibitem [{\citenamefont {Tian}\ \emph {et~al.}(2009)\citenamefont {Tian},
  \citenamefont {Ye},\ and\ \citenamefont {Jin}}]{Tian2009}%
  \BibitemOpen
  \bibfield  {author} {\bibinfo {author} {\bibfnamefont {Y.}~\bibnamefont
  {Tian}}, \bibinfo {author} {\bibfnamefont {L.}~\bibnamefont {Ye}},\ and\
  \bibinfo {author} {\bibfnamefont {X.}~\bibnamefont {Jin}},\ }\bibfield
  {title} {\bibinfo {title} {Proper scaling of the anomalous hall effect},\
  }\href {https://doi.org/10.1103/PhysRevLett.103.087206} {\bibfield  {journal}
  {\bibinfo  {journal} {Phys. Rev. Lett.}\ }\textbf {\bibinfo {volume} {103}},\
  \bibinfo {pages} {087206} (\bibinfo {year} {2009})}\BibitemShut {NoStop}%
\bibitem [{\citenamefont {Nagaosa}\ \emph {et~al.}(2010)\citenamefont
  {Nagaosa}, \citenamefont {Sinova}, \citenamefont {Onoda}, \citenamefont
  {MacDonald},\ and\ \citenamefont {Ong}}]{Nagaosa2010}%
  \BibitemOpen
  \bibfield  {author} {\bibinfo {author} {\bibfnamefont {N.}~\bibnamefont
  {Nagaosa}}, \bibinfo {author} {\bibfnamefont {J.}~\bibnamefont {Sinova}},
  \bibinfo {author} {\bibfnamefont {S.}~\bibnamefont {Onoda}}, \bibinfo
  {author} {\bibfnamefont {A.~H.}\ \bibnamefont {MacDonald}},\ and\ \bibinfo
  {author} {\bibfnamefont {N.~P.}\ \bibnamefont {Ong}},\ }\bibfield  {title}
  {\bibinfo {title} {Anomalous hall effect},\ }\href
  {https://doi.org/10.1103/RevModPhys.82.1539} {\bibfield  {journal} {\bibinfo
  {journal} {Rev. Mod. Phys.}\ }\textbf {\bibinfo {volume} {82}},\ \bibinfo
  {pages} {1539} (\bibinfo {year} {2010})}\BibitemShut {NoStop}%
\bibitem [{\citenamefont {Morota}\ \emph {et~al.}(2011)\citenamefont {Morota},
  \citenamefont {Niimi}, \citenamefont {Ohnishi}, \citenamefont {Wei},
  \citenamefont {Tanaka}, \citenamefont {Kontani}, \citenamefont {Kimura},\
  and\ \citenamefont {Otani}}]{Morota2011}%
  \BibitemOpen
  \bibfield  {author} {\bibinfo {author} {\bibfnamefont {M.}~\bibnamefont
  {Morota}}, \bibinfo {author} {\bibfnamefont {Y.}~\bibnamefont {Niimi}},
  \bibinfo {author} {\bibfnamefont {K.}~\bibnamefont {Ohnishi}}, \bibinfo
  {author} {\bibfnamefont {D.~H.}\ \bibnamefont {Wei}}, \bibinfo {author}
  {\bibfnamefont {T.}~\bibnamefont {Tanaka}}, \bibinfo {author} {\bibfnamefont
  {H.}~\bibnamefont {Kontani}}, \bibinfo {author} {\bibfnamefont
  {T.}~\bibnamefont {Kimura}},\ and\ \bibinfo {author} {\bibfnamefont
  {Y.}~\bibnamefont {Otani}},\ }\bibfield  {title} {\bibinfo {title}
  {Indication of intrinsic spin hall effect in $4d$ and $5d$ transition
  metals},\ }\href {https://doi.org/10.1103/PhysRevB.83.174405} {\bibfield
  {journal} {\bibinfo  {journal} {Phys. Rev. B}\ }\textbf {\bibinfo {volume}
  {83}},\ \bibinfo {pages} {174405} (\bibinfo {year} {2011})}\BibitemShut
  {NoStop}%
\bibitem [{\citenamefont {Niimi}\ \emph {et~al.}(2011)\citenamefont {Niimi},
  \citenamefont {Morota}, \citenamefont {Wei}, \citenamefont {Deranlot},
  \citenamefont {Basletic}, \citenamefont {Hamzic}, \citenamefont {Fert},\ and\
  \citenamefont {Otani}}]{Niimi2011}%
  \BibitemOpen
  \bibfield  {author} {\bibinfo {author} {\bibfnamefont {Y.}~\bibnamefont
  {Niimi}}, \bibinfo {author} {\bibfnamefont {M.}~\bibnamefont {Morota}},
  \bibinfo {author} {\bibfnamefont {D.~H.}\ \bibnamefont {Wei}}, \bibinfo
  {author} {\bibfnamefont {C.}~\bibnamefont {Deranlot}}, \bibinfo {author}
  {\bibfnamefont {M.}~\bibnamefont {Basletic}}, \bibinfo {author}
  {\bibfnamefont {A.}~\bibnamefont {Hamzic}}, \bibinfo {author} {\bibfnamefont
  {A.}~\bibnamefont {Fert}},\ and\ \bibinfo {author} {\bibfnamefont
  {Y.}~\bibnamefont {Otani}},\ }\bibfield  {title} {\bibinfo {title} {Extrinsic
  spin hall effect induced by iridium impurities in copper},\ }\href
  {https://doi.org/10.1103/PhysRevLett.106.126601} {\bibfield  {journal}
  {\bibinfo  {journal} {Phys. Rev. Lett.}\ }\textbf {\bibinfo {volume} {106}},\
  \bibinfo {pages} {126601} (\bibinfo {year} {2011})}\BibitemShut {NoStop}%
\bibitem [{\citenamefont {Sinova}\ \emph {et~al.}(2015)\citenamefont {Sinova},
  \citenamefont {Valenzuela}, \citenamefont {Wunderlich}, \citenamefont
  {Back},\ and\ \citenamefont {Jungwirth}}]{Sinova2015}%
  \BibitemOpen
  \bibfield  {author} {\bibinfo {author} {\bibfnamefont {J.}~\bibnamefont
  {Sinova}}, \bibinfo {author} {\bibfnamefont {S.~O.}\ \bibnamefont
  {Valenzuela}}, \bibinfo {author} {\bibfnamefont {J.}~\bibnamefont
  {Wunderlich}}, \bibinfo {author} {\bibfnamefont {C.~H.}\ \bibnamefont
  {Back}},\ and\ \bibinfo {author} {\bibfnamefont {T.}~\bibnamefont
  {Jungwirth}},\ }\bibfield  {title} {\bibinfo {title} {Spin hall effects},\
  }\href {https://doi.org/10.1103/RevModPhys.87.1213} {\bibfield  {journal}
  {\bibinfo  {journal} {Rev. Mod. Phys.}\ }\textbf {\bibinfo {volume} {87}},\
  \bibinfo {pages} {1213} (\bibinfo {year} {2015})}\BibitemShut {NoStop}%
\bibitem [{\citenamefont {Musha}\ \emph {et~al.}(2019)\citenamefont {Musha},
  \citenamefont {Kanno},\ and\ \citenamefont {Ando}}]{Musha2019}%
  \BibitemOpen
  \bibfield  {author} {\bibinfo {author} {\bibfnamefont {A.}~\bibnamefont
  {Musha}}, \bibinfo {author} {\bibfnamefont {Y.}~\bibnamefont {Kanno}},\ and\
  \bibinfo {author} {\bibfnamefont {K.}~\bibnamefont {Ando}},\ }\bibfield
  {title} {\bibinfo {title} {Extrinsic-intrinsic crossover of the spin hall
  effect induced by alloying},\ }\href
  {https://doi.org/10.1103/PhysRevMaterials.3.054411} {\bibfield  {journal}
  {\bibinfo  {journal} {Phys. Rev. Mater.}\ }\textbf {\bibinfo {volume} {3}},\
  \bibinfo {pages} {054411} (\bibinfo {year} {2019})}\BibitemShut {NoStop}%
\bibitem [{\citenamefont {Chen}\ \emph {et~al.}(2010)\citenamefont {Chen},
  \citenamefont {Qin}, \citenamefont {Yang}, \citenamefont {Liu}, \citenamefont
  {Guan}, \citenamefont {Qu}, \citenamefont {Zhang}, \citenamefont {Shi},
  \citenamefont {Xie}, \citenamefont {Yang}, \citenamefont {Wu}, \citenamefont
  {Li},\ and\ \citenamefont {Lu}}]{Chen2010}%
  \BibitemOpen
  \bibfield  {author} {\bibinfo {author} {\bibfnamefont {J.}~\bibnamefont
  {Chen}}, \bibinfo {author} {\bibfnamefont {H.~J.}\ \bibnamefont {Qin}},
  \bibinfo {author} {\bibfnamefont {F.}~\bibnamefont {Yang}}, \bibinfo {author}
  {\bibfnamefont {J.}~\bibnamefont {Liu}}, \bibinfo {author} {\bibfnamefont
  {T.}~\bibnamefont {Guan}}, \bibinfo {author} {\bibfnamefont {F.~M.}\
  \bibnamefont {Qu}}, \bibinfo {author} {\bibfnamefont {G.~H.}\ \bibnamefont
  {Zhang}}, \bibinfo {author} {\bibfnamefont {J.~R.}\ \bibnamefont {Shi}},
  \bibinfo {author} {\bibfnamefont {X.~C.}\ \bibnamefont {Xie}}, \bibinfo
  {author} {\bibfnamefont {C.~L.}\ \bibnamefont {Yang}}, \bibinfo {author}
  {\bibfnamefont {K.~H.}\ \bibnamefont {Wu}}, \bibinfo {author} {\bibfnamefont
  {Y.~Q.}\ \bibnamefont {Li}},\ and\ \bibinfo {author} {\bibfnamefont
  {L.}~\bibnamefont {Lu}},\ }\bibfield  {title} {\bibinfo {title} {Gate-voltage
  control of chemical potential and weak antilocalization in
  ${\mathrm{bi}}_{2}{\mathrm{se}}_{3}$},\ }\href
  {https://doi.org/10.1103/PhysRevLett.105.176602} {\bibfield  {journal}
  {\bibinfo  {journal} {Phys. Rev. Lett.}\ }\textbf {\bibinfo {volume} {105}},\
  \bibinfo {pages} {176602} (\bibinfo {year} {2010})}\BibitemShut {NoStop}%
\bibitem [{\citenamefont {Culcer}\ \emph
  {et~al.}(2010{\natexlab{b}})\citenamefont {Culcer}, \citenamefont {Hwang},
  \citenamefont {Stanescu},\ and\ \citenamefont {Das~Sarma}}]{Culcer2010}%
  \BibitemOpen
  \bibfield  {author} {\bibinfo {author} {\bibfnamefont {D.}~\bibnamefont
  {Culcer}}, \bibinfo {author} {\bibfnamefont {E.~H.}\ \bibnamefont {Hwang}},
  \bibinfo {author} {\bibfnamefont {T.~D.}\ \bibnamefont {Stanescu}},\ and\
  \bibinfo {author} {\bibfnamefont {S.}~\bibnamefont {Das~Sarma}},\ }\bibfield
  {title} {\bibinfo {title} {Two-dimensional surface charge transport in
  topological insulators},\ }\href {https://doi.org/10.1103/PhysRevB.82.155457}
  {\bibfield  {journal} {\bibinfo  {journal} {Phys. Rev. B}\ }\textbf {\bibinfo
  {volume} {82}},\ \bibinfo {pages} {155457} (\bibinfo {year}
  {2010}{\natexlab{b}})}\BibitemShut {NoStop}%
\bibitem [{\citenamefont {Kim}\ \emph {et~al.}(2011)\citenamefont {Kim},
  \citenamefont {Brahlek}, \citenamefont {Bansal}, \citenamefont {Edrey},
  \citenamefont {Kapilevich}, \citenamefont {Iida}, \citenamefont {Tanimura},
  \citenamefont {Horibe}, \citenamefont {Cheong},\ and\ \citenamefont
  {Oh}}]{Kim2011}%
  \BibitemOpen
  \bibfield  {author} {\bibinfo {author} {\bibfnamefont {Y.~S.}\ \bibnamefont
  {Kim}}, \bibinfo {author} {\bibfnamefont {M.}~\bibnamefont {Brahlek}},
  \bibinfo {author} {\bibfnamefont {N.}~\bibnamefont {Bansal}}, \bibinfo
  {author} {\bibfnamefont {E.}~\bibnamefont {Edrey}}, \bibinfo {author}
  {\bibfnamefont {G.~A.}\ \bibnamefont {Kapilevich}}, \bibinfo {author}
  {\bibfnamefont {K.}~\bibnamefont {Iida}}, \bibinfo {author} {\bibfnamefont
  {M.}~\bibnamefont {Tanimura}}, \bibinfo {author} {\bibfnamefont
  {Y.}~\bibnamefont {Horibe}}, \bibinfo {author} {\bibfnamefont {S.-W.}\
  \bibnamefont {Cheong}},\ and\ \bibinfo {author} {\bibfnamefont
  {S.}~\bibnamefont {Oh}},\ }\bibfield  {title} {\bibinfo {title}
  {Thickness-dependent bulk properties and weak antilocalization effect in
  topological insulator bi${}_{2}$se${}_{3}$},\ }\href
  {https://doi.org/10.1103/PhysRevB.84.073109} {\bibfield  {journal} {\bibinfo
  {journal} {Phys. Rev. B}\ }\textbf {\bibinfo {volume} {84}},\ \bibinfo
  {pages} {073109} (\bibinfo {year} {2011})}\BibitemShut {NoStop}%
\bibitem [{\citenamefont {Steinberg}\ \emph {et~al.}(2011)\citenamefont
  {Steinberg}, \citenamefont {Lalo\"e}, \citenamefont {Fatemi}, \citenamefont
  {Moodera},\ and\ \citenamefont {Jarillo-Herrero}}]{Steinberg2011}%
  \BibitemOpen
  \bibfield  {author} {\bibinfo {author} {\bibfnamefont {H.}~\bibnamefont
  {Steinberg}}, \bibinfo {author} {\bibfnamefont {J.-B.}\ \bibnamefont
  {Lalo\"e}}, \bibinfo {author} {\bibfnamefont {V.}~\bibnamefont {Fatemi}},
  \bibinfo {author} {\bibfnamefont {J.~S.}\ \bibnamefont {Moodera}},\ and\
  \bibinfo {author} {\bibfnamefont {P.}~\bibnamefont {Jarillo-Herrero}},\
  }\bibfield  {title} {\bibinfo {title} {Electrically tunable surface-to-bulk
  coherent coupling in topological insulator thin films},\ }\href
  {https://doi.org/10.1103/PhysRevB.84.233101} {\bibfield  {journal} {\bibinfo
  {journal} {Phys. Rev. B}\ }\textbf {\bibinfo {volume} {84}},\ \bibinfo
  {pages} {233101} (\bibinfo {year} {2011})}\BibitemShut {NoStop}%
\bibitem [{\citenamefont {Ma}\ \emph {et~al.}(2019)\citenamefont {Ma},
  \citenamefont {Xu}, \citenamefont {Shen}, \citenamefont {MacNeill},
  \citenamefont {Fatemi}, \citenamefont {Chang}, \citenamefont {Mier~Valdivia},
  \citenamefont {Wu}, \citenamefont {Du}, \citenamefont {Hsu}, \citenamefont
  {Fang}, \citenamefont {Gibson}, \citenamefont {Watanabe}, \citenamefont
  {Taniguchi}, \citenamefont {Cava}, \citenamefont {Kaxiras}, \citenamefont
  {Lu}, \citenamefont {Lin}, \citenamefont {Fu}, \citenamefont {Gedik},\ and\
  \citenamefont {Jarillo-Herrero}}]{Qiong2019}%
  \BibitemOpen
  \bibfield  {author} {\bibinfo {author} {\bibfnamefont {Q.}~\bibnamefont
  {Ma}}, \bibinfo {author} {\bibfnamefont {S.-Y.}\ \bibnamefont {Xu}}, \bibinfo
  {author} {\bibfnamefont {H.}~\bibnamefont {Shen}}, \bibinfo {author}
  {\bibfnamefont {D.}~\bibnamefont {MacNeill}}, \bibinfo {author}
  {\bibfnamefont {V.}~\bibnamefont {Fatemi}}, \bibinfo {author} {\bibfnamefont
  {T.-R.}\ \bibnamefont {Chang}}, \bibinfo {author} {\bibfnamefont {A.~M.}\
  \bibnamefont {Mier~Valdivia}}, \bibinfo {author} {\bibfnamefont
  {S.}~\bibnamefont {Wu}}, \bibinfo {author} {\bibfnamefont {Z.}~\bibnamefont
  {Du}}, \bibinfo {author} {\bibfnamefont {C.-H.}\ \bibnamefont {Hsu}},
  \bibinfo {author} {\bibfnamefont {S.}~\bibnamefont {Fang}}, \bibinfo {author}
  {\bibfnamefont {Q.~D.}\ \bibnamefont {Gibson}}, \bibinfo {author}
  {\bibfnamefont {K.}~\bibnamefont {Watanabe}}, \bibinfo {author}
  {\bibfnamefont {T.}~\bibnamefont {Taniguchi}}, \bibinfo {author}
  {\bibfnamefont {R.~J.}\ \bibnamefont {Cava}}, \bibinfo {author}
  {\bibfnamefont {E.}~\bibnamefont {Kaxiras}}, \bibinfo {author} {\bibfnamefont
  {H.-Z.}\ \bibnamefont {Lu}}, \bibinfo {author} {\bibfnamefont
  {H.}~\bibnamefont {Lin}}, \bibinfo {author} {\bibfnamefont {L.}~\bibnamefont
  {Fu}}, \bibinfo {author} {\bibfnamefont {N.}~\bibnamefont {Gedik}},\ and\
  \bibinfo {author} {\bibfnamefont {P.}~\bibnamefont {Jarillo-Herrero}},\
  }\bibfield  {title} {\bibinfo {title} {Observation of the nonlinear hall
  effect under time-reversal-symmetric conditions},\ }\href
  {https://doi.org/10.1038/s41586-018-0807-6} {\bibfield  {journal} {\bibinfo
  {journal} {Nature}\ }\textbf {\bibinfo {volume} {565}},\ \bibinfo {pages}
  {337} (\bibinfo {year} {2019})}\BibitemShut {NoStop}%
\bibitem [{\citenamefont {Chichinadze}\ \emph {et~al.}(2025)\citenamefont
  {Chichinadze}, \citenamefont {Zhang}, \citenamefont {Lin}, \citenamefont
  {Morissette}, \citenamefont {Wang}, \citenamefont {Watanabe}, \citenamefont
  {Taniguchi}, \citenamefont {Vafek},\ and\ \citenamefont {Li}}]{Dmitry2025}%
  \BibitemOpen
  \bibfield  {author} {\bibinfo {author} {\bibfnamefont {D.~V.}\ \bibnamefont
  {Chichinadze}}, \bibinfo {author} {\bibfnamefont {N.~J.}\ \bibnamefont
  {Zhang}}, \bibinfo {author} {\bibfnamefont {J.-X.}\ \bibnamefont {Lin}},
  \bibinfo {author} {\bibfnamefont {E.}~\bibnamefont {Morissette}}, \bibinfo
  {author} {\bibfnamefont {X.}~\bibnamefont {Wang}}, \bibinfo {author}
  {\bibfnamefont {K.}~\bibnamefont {Watanabe}}, \bibinfo {author}
  {\bibfnamefont {T.}~\bibnamefont {Taniguchi}}, \bibinfo {author}
  {\bibfnamefont {O.}~\bibnamefont {Vafek}},\ and\ \bibinfo {author}
  {\bibfnamefont {J.~I.~A.}\ \bibnamefont {Li}},\ }\href
  {https://arxiv.org/abs/2411.11156} {\bibinfo {title} {Observation of giant
  nonlinear hall conductivity in bernal bilayer graphene}} (\bibinfo {year}
  {2025}),\ \Eprint {https://arxiv.org/abs/2411.11156} {arXiv:2411.11156
  [cond-mat.mes-hall]} \BibitemShut {NoStop}%
\bibitem [{\citenamefont {Ye}\ \emph {et~al.}(2018)\citenamefont {Ye},
  \citenamefont {Kang}, \citenamefont {Liu}, \citenamefont {von Cube},
  \citenamefont {Wicker}, \citenamefont {Suzuki}, \citenamefont {Jozwiak},
  \citenamefont {Bostwick}, \citenamefont {Rotenberg}, \citenamefont {Bell},
  \citenamefont {Fu}, \citenamefont {Comin},\ and\ \citenamefont
  {Checkelsky}}]{Linda2018}%
  \BibitemOpen
  \bibfield  {author} {\bibinfo {author} {\bibfnamefont {L.}~\bibnamefont
  {Ye}}, \bibinfo {author} {\bibfnamefont {M.}~\bibnamefont {Kang}}, \bibinfo
  {author} {\bibfnamefont {J.}~\bibnamefont {Liu}}, \bibinfo {author}
  {\bibfnamefont {F.}~\bibnamefont {von Cube}}, \bibinfo {author}
  {\bibfnamefont {C.~R.}\ \bibnamefont {Wicker}}, \bibinfo {author}
  {\bibfnamefont {T.}~\bibnamefont {Suzuki}}, \bibinfo {author} {\bibfnamefont
  {C.}~\bibnamefont {Jozwiak}}, \bibinfo {author} {\bibfnamefont
  {A.}~\bibnamefont {Bostwick}}, \bibinfo {author} {\bibfnamefont
  {E.}~\bibnamefont {Rotenberg}}, \bibinfo {author} {\bibfnamefont {D.~C.}\
  \bibnamefont {Bell}}, \bibinfo {author} {\bibfnamefont {L.}~\bibnamefont
  {Fu}}, \bibinfo {author} {\bibfnamefont {R.}~\bibnamefont {Comin}},\ and\
  \bibinfo {author} {\bibfnamefont {J.~G.}\ \bibnamefont {Checkelsky}},\
  }\bibfield  {title} {\bibinfo {title} {Massive dirac fermions in a
  ferromagnetic kagome metal},\ }\href {https://doi.org/10.1038/nature25987}
  {\bibfield  {journal} {\bibinfo  {journal} {Nature}\ }\textbf {\bibinfo
  {volume} {555}},\ \bibinfo {pages} {638} (\bibinfo {year}
  {2018})}\BibitemShut {NoStop}%
\bibitem [{\citenamefont {Chen}\ \emph {et~al.}(2021)\citenamefont {Chen},
  \citenamefont {Le}, \citenamefont {Fu}, \citenamefont {Lin}, \citenamefont
  {Schnelle}, \citenamefont {Sun},\ and\ \citenamefont {Felser}}]{Chen2021}%
  \BibitemOpen
  \bibfield  {author} {\bibinfo {author} {\bibfnamefont {D.}~\bibnamefont
  {Chen}}, \bibinfo {author} {\bibfnamefont {C.}~\bibnamefont {Le}}, \bibinfo
  {author} {\bibfnamefont {C.}~\bibnamefont {Fu}}, \bibinfo {author}
  {\bibfnamefont {H.}~\bibnamefont {Lin}}, \bibinfo {author} {\bibfnamefont
  {W.}~\bibnamefont {Schnelle}}, \bibinfo {author} {\bibfnamefont
  {Y.}~\bibnamefont {Sun}},\ and\ \bibinfo {author} {\bibfnamefont
  {C.}~\bibnamefont {Felser}},\ }\bibfield  {title} {\bibinfo {title} {Large
  anomalous hall effect in the kagome ferromagnet
  ${\mathrm{limn}}_{6}{\mathrm{sn}}_{6}$},\ }\href
  {https://doi.org/10.1103/PhysRevB.103.144410} {\bibfield  {journal} {\bibinfo
   {journal} {Phys. Rev. B}\ }\textbf {\bibinfo {volume} {103}},\ \bibinfo
  {pages} {144410} (\bibinfo {year} {2021})}\BibitemShut {NoStop}%
\bibitem [{\citenamefont {Liu}\ \emph {et~al.}(2018)\citenamefont {Liu},
  \citenamefont {Sun}, \citenamefont {Kumar}, \citenamefont {Muechler},
  \citenamefont {Sun}, \citenamefont {Jiao}, \citenamefont {Yang},
  \citenamefont {Liu}, \citenamefont {Liang}, \citenamefont {Xu}, \citenamefont
  {Kroder}, \citenamefont {S{\"u}{\ss}}, \citenamefont {Borrmann},
  \citenamefont {Shekhar}, \citenamefont {Wang}, \citenamefont {Xi},
  \citenamefont {Wang}, \citenamefont {Schnelle}, \citenamefont {Wirth},
  \citenamefont {Chen}, \citenamefont {Goennenwein},\ and\ \citenamefont
  {Felser}}]{Liu2025}%
  \BibitemOpen
  \bibfield  {author} {\bibinfo {author} {\bibfnamefont {E.}~\bibnamefont
  {Liu}}, \bibinfo {author} {\bibfnamefont {Y.}~\bibnamefont {Sun}}, \bibinfo
  {author} {\bibfnamefont {N.}~\bibnamefont {Kumar}}, \bibinfo {author}
  {\bibfnamefont {L.}~\bibnamefont {Muechler}}, \bibinfo {author}
  {\bibfnamefont {A.}~\bibnamefont {Sun}}, \bibinfo {author} {\bibfnamefont
  {L.}~\bibnamefont {Jiao}}, \bibinfo {author} {\bibfnamefont {S.-Y.}\
  \bibnamefont {Yang}}, \bibinfo {author} {\bibfnamefont {D.}~\bibnamefont
  {Liu}}, \bibinfo {author} {\bibfnamefont {A.}~\bibnamefont {Liang}}, \bibinfo
  {author} {\bibfnamefont {Q.}~\bibnamefont {Xu}}, \bibinfo {author}
  {\bibfnamefont {J.}~\bibnamefont {Kroder}}, \bibinfo {author} {\bibfnamefont
  {V.}~\bibnamefont {S{\"u}{\ss}}}, \bibinfo {author} {\bibfnamefont
  {H.}~\bibnamefont {Borrmann}}, \bibinfo {author} {\bibfnamefont
  {C.}~\bibnamefont {Shekhar}}, \bibinfo {author} {\bibfnamefont
  {Z.}~\bibnamefont {Wang}}, \bibinfo {author} {\bibfnamefont {C.}~\bibnamefont
  {Xi}}, \bibinfo {author} {\bibfnamefont {W.}~\bibnamefont {Wang}}, \bibinfo
  {author} {\bibfnamefont {W.}~\bibnamefont {Schnelle}}, \bibinfo {author}
  {\bibfnamefont {S.}~\bibnamefont {Wirth}}, \bibinfo {author} {\bibfnamefont
  {Y.}~\bibnamefont {Chen}}, \bibinfo {author} {\bibfnamefont {S.~T.~B.}\
  \bibnamefont {Goennenwein}},\ and\ \bibinfo {author} {\bibfnamefont
  {C.}~\bibnamefont {Felser}},\ }\bibfield  {title} {\bibinfo {title} {Giant
  anomalous hall effect in a ferromagnetic kagome-lattice semimetal},\ }\href
  {https://doi.org/10.1038/s41567-018-0234-5} {\bibfield  {journal} {\bibinfo
  {journal} {Nature Physics}\ }\textbf {\bibinfo {volume} {14}},\ \bibinfo
  {pages} {1125} (\bibinfo {year} {2018})}\BibitemShut {NoStop}%
\bibitem [{\citenamefont {Jungwirth}\ \emph {et~al.}(2002)\citenamefont
  {Jungwirth}, \citenamefont {Niu},\ and\ \citenamefont
  {MacDonald}}]{Jungwirth2002}%
  \BibitemOpen
  \bibfield  {author} {\bibinfo {author} {\bibfnamefont {T.}~\bibnamefont
  {Jungwirth}}, \bibinfo {author} {\bibfnamefont {Q.}~\bibnamefont {Niu}},\
  and\ \bibinfo {author} {\bibfnamefont {A.~H.}\ \bibnamefont {MacDonald}},\
  }\bibfield  {title} {\bibinfo {title} {Anomalous hall effect in ferromagnetic
  semiconductors},\ }\href {https://doi.org/10.1103/PhysRevLett.88.207208}
  {\bibfield  {journal} {\bibinfo  {journal} {Phys. Rev. Lett.}\ }\textbf
  {\bibinfo {volume} {88}},\ \bibinfo {pages} {207208} (\bibinfo {year}
  {2002})}\BibitemShut {NoStop}%
\bibitem [{\citenamefont {Culcer}\ \emph {et~al.}(2003)\citenamefont {Culcer},
  \citenamefont {MacDonald},\ and\ \citenamefont {Niu}}]{Culcer2003}%
  \BibitemOpen
  \bibfield  {author} {\bibinfo {author} {\bibfnamefont {D.}~\bibnamefont
  {Culcer}}, \bibinfo {author} {\bibfnamefont {A.}~\bibnamefont {MacDonald}},\
  and\ \bibinfo {author} {\bibfnamefont {Q.}~\bibnamefont {Niu}},\ }\bibfield
  {title} {\bibinfo {title} {Anomalous hall effect in paramagnetic
  two-dimensional systems},\ }\href
  {https://doi.org/10.1103/PhysRevB.68.045327} {\bibfield  {journal} {\bibinfo
  {journal} {Phys. Rev. B}\ }\textbf {\bibinfo {volume} {68}},\ \bibinfo
  {pages} {045327} (\bibinfo {year} {2003})}\BibitemShut {NoStop}%
\bibitem [{\citenamefont {Cullen}\ \emph {et~al.}(2021)\citenamefont {Cullen},
  \citenamefont {Bhalla}, \citenamefont {Marcellina}, \citenamefont
  {Hamilton},\ and\ \citenamefont {Culcer}}]{Cullen2021}%
  \BibitemOpen
  \bibfield  {author} {\bibinfo {author} {\bibfnamefont {J.~H.}\ \bibnamefont
  {Cullen}}, \bibinfo {author} {\bibfnamefont {P.}~\bibnamefont {Bhalla}},
  \bibinfo {author} {\bibfnamefont {E.}~\bibnamefont {Marcellina}}, \bibinfo
  {author} {\bibfnamefont {A.~R.}\ \bibnamefont {Hamilton}},\ and\ \bibinfo
  {author} {\bibfnamefont {D.}~\bibnamefont {Culcer}},\ }\bibfield  {title}
  {\bibinfo {title} {Generating a topological anomalous hall effect in a
  nonmagnetic conductor: An in-plane magnetic field as a direct probe of the
  berry curvature},\ }\href {https://doi.org/10.1103/PhysRevLett.126.256601}
  {\bibfield  {journal} {\bibinfo  {journal} {Phys. Rev. Lett.}\ }\textbf
  {\bibinfo {volume} {126}},\ \bibinfo {pages} {256601} (\bibinfo {year}
  {2021})}\BibitemShut {NoStop}%
\bibitem [{\citenamefont {Atencia}\ \emph
  {et~al.}(2023{\natexlab{b}})\citenamefont {Atencia}, \citenamefont {Xiao},\
  and\ \citenamefont {Culcer}}]{Atencia2023}%
  \BibitemOpen
  \bibfield  {author} {\bibinfo {author} {\bibfnamefont {R.~B.}\ \bibnamefont
  {Atencia}}, \bibinfo {author} {\bibfnamefont {D.}~\bibnamefont {Xiao}},\ and\
  \bibinfo {author} {\bibfnamefont {D.}~\bibnamefont {Culcer}},\ }\bibfield
  {title} {\bibinfo {title} {Disorder in the nonlinear anomalous hall effect of
  $\mathcal{PT}$-symmetric dirac fermions},\ }\href
  {https://doi.org/10.1103/PhysRevB.108.L201115} {\bibfield  {journal}
  {\bibinfo  {journal} {Phys. Rev. B}\ }\textbf {\bibinfo {volume} {108}},\
  \bibinfo {pages} {L201115} (\bibinfo {year}
  {2023}{\natexlab{b}})}\BibitemShut {NoStop}%
\bibitem [{\citenamefont {Sinova}\ \emph {et~al.}(2004)\citenamefont {Sinova},
  \citenamefont {Culcer}, \citenamefont {Niu}, \citenamefont {Sinitsyn},
  \citenamefont {Jungwirth},\ and\ \citenamefont {MacDonald}}]{Sinova2004}%
  \BibitemOpen
  \bibfield  {author} {\bibinfo {author} {\bibfnamefont {J.}~\bibnamefont
  {Sinova}}, \bibinfo {author} {\bibfnamefont {D.}~\bibnamefont {Culcer}},
  \bibinfo {author} {\bibfnamefont {Q.}~\bibnamefont {Niu}}, \bibinfo {author}
  {\bibfnamefont {N.~A.}\ \bibnamefont {Sinitsyn}}, \bibinfo {author}
  {\bibfnamefont {T.}~\bibnamefont {Jungwirth}},\ and\ \bibinfo {author}
  {\bibfnamefont {A.~H.}\ \bibnamefont {MacDonald}},\ }\bibfield  {title}
  {\bibinfo {title} {Universal intrinsic spin hall effect},\ }\href
  {https://doi.org/10.1103/PhysRevLett.92.126603} {\bibfield  {journal}
  {\bibinfo  {journal} {Phys. Rev. Lett.}\ }\textbf {\bibinfo {volume} {92}},\
  \bibinfo {pages} {126603} (\bibinfo {year} {2004})}\BibitemShut {NoStop}%
\bibitem [{\citenamefont {Culcer}\ \emph {et~al.}(2004)\citenamefont {Culcer},
  \citenamefont {Sinova}, \citenamefont {Sinitsyn}, \citenamefont {Jungwirth},
  \citenamefont {MacDonald},\ and\ \citenamefont {Niu}}]{Culcer2004}%
  \BibitemOpen
  \bibfield  {author} {\bibinfo {author} {\bibfnamefont {D.}~\bibnamefont
  {Culcer}}, \bibinfo {author} {\bibfnamefont {J.}~\bibnamefont {Sinova}},
  \bibinfo {author} {\bibfnamefont {N.~A.}\ \bibnamefont {Sinitsyn}}, \bibinfo
  {author} {\bibfnamefont {T.}~\bibnamefont {Jungwirth}}, \bibinfo {author}
  {\bibfnamefont {A.~H.}\ \bibnamefont {MacDonald}},\ and\ \bibinfo {author}
  {\bibfnamefont {Q.}~\bibnamefont {Niu}},\ }\bibfield  {title} {\bibinfo
  {title} {Semiclassical spin transport in spin-orbit-coupled bands},\ }\href
  {https://doi.org/10.1103/PhysRevLett.93.046602} {\bibfield  {journal}
  {\bibinfo  {journal} {Phys. Rev. Lett.}\ }\textbf {\bibinfo {volume} {93}},\
  \bibinfo {pages} {046602} (\bibinfo {year} {2004})}\BibitemShut {NoStop}%
\bibitem [{\citenamefont {Bhalla}\ \emph {et~al.}(2021)\citenamefont {Bhalla},
  \citenamefont {Deng}, \citenamefont {Wang}, \citenamefont {Wang},\ and\
  \citenamefont {Culcer}}]{Bhalla2021}%
  \BibitemOpen
  \bibfield  {author} {\bibinfo {author} {\bibfnamefont {P.}~\bibnamefont
  {Bhalla}}, \bibinfo {author} {\bibfnamefont {M.-X.}\ \bibnamefont {Deng}},
  \bibinfo {author} {\bibfnamefont {R.-Q.}\ \bibnamefont {Wang}}, \bibinfo
  {author} {\bibfnamefont {L.}~\bibnamefont {Wang}},\ and\ \bibinfo {author}
  {\bibfnamefont {D.}~\bibnamefont {Culcer}},\ }\bibfield  {title} {\bibinfo
  {title} {Nonlinear ballistic response of quantum spin hall edge states},\
  }\href {https://doi.org/10.1103/PhysRevLett.127.206801} {\bibfield  {journal}
  {\bibinfo  {journal} {Phys. Rev. Lett.}\ }\textbf {\bibinfo {volume} {127}},\
  \bibinfo {pages} {206801} (\bibinfo {year} {2021})}\BibitemShut {NoStop}%
\bibitem [{\citenamefont {Ma}\ \emph {et~al.}(2024)\citenamefont {Ma},
  \citenamefont {Cullen}, \citenamefont {Monir}, \citenamefont {Rahman},\ and\
  \citenamefont {Culcer}}]{Hongyang2024}%
  \BibitemOpen
  \bibfield  {author} {\bibinfo {author} {\bibfnamefont {H.}~\bibnamefont
  {Ma}}, \bibinfo {author} {\bibfnamefont {J.~H.}\ \bibnamefont {Cullen}},
  \bibinfo {author} {\bibfnamefont {S.}~\bibnamefont {Monir}}, \bibinfo
  {author} {\bibfnamefont {R.}~\bibnamefont {Rahman}},\ and\ \bibinfo {author}
  {\bibfnamefont {D.}~\bibnamefont {Culcer}},\ }\bibfield  {title} {\bibinfo
  {title} {Spin-hall effect in topological materials: evaluating the proper
  spin current in systems with arbitrary degeneracies},\ }\href
  {https://doi.org/10.1038/s44306-024-00057-w} {\bibfield  {journal} {\bibinfo
  {journal} {npj Spintronics}\ }\textbf {\bibinfo {volume} {2}},\ \bibinfo
  {pages} {55} (\bibinfo {year} {2024})}\BibitemShut {NoStop}%
\bibitem [{\citenamefont {Sato}\ \emph {et~al.}(2024)\citenamefont {Sato},
  \citenamefont {Haddad}, \citenamefont {Fulga}, \citenamefont {Assaad},\ and\
  \citenamefont {van~den Brink}}]{Toshihiro2024}%
  \BibitemOpen
  \bibfield  {author} {\bibinfo {author} {\bibfnamefont {T.}~\bibnamefont
  {Sato}}, \bibinfo {author} {\bibfnamefont {S.}~\bibnamefont {Haddad}},
  \bibinfo {author} {\bibfnamefont {I.~C.}\ \bibnamefont {Fulga}}, \bibinfo
  {author} {\bibfnamefont {F.~F.}\ \bibnamefont {Assaad}},\ and\ \bibinfo
  {author} {\bibfnamefont {J.}~\bibnamefont {van~den Brink}},\ }\bibfield
  {title} {\bibinfo {title} {Altermagnetic anomalous hall effect emerging from
  electronic correlations},\ }\href
  {https://doi.org/10.1103/PhysRevLett.133.086503} {\bibfield  {journal}
  {\bibinfo  {journal} {Phys. Rev. Lett.}\ }\textbf {\bibinfo {volume} {133}},\
  \bibinfo {pages} {086503} (\bibinfo {year} {2024})}\BibitemShut {NoStop}%
\bibitem [{\citenamefont {Go}\ \emph {et~al.}(2018)\citenamefont {Go},
  \citenamefont {Jo}, \citenamefont {Kim},\ and\ \citenamefont
  {Lee}}]{Dongwook2018}%
  \BibitemOpen
  \bibfield  {author} {\bibinfo {author} {\bibfnamefont {D.}~\bibnamefont
  {Go}}, \bibinfo {author} {\bibfnamefont {D.}~\bibnamefont {Jo}}, \bibinfo
  {author} {\bibfnamefont {C.}~\bibnamefont {Kim}},\ and\ \bibinfo {author}
  {\bibfnamefont {H.-W.}\ \bibnamefont {Lee}},\ }\bibfield  {title} {\bibinfo
  {title} {Intrinsic spin and orbital hall effects from orbital texture},\
  }\href {https://doi.org/10.1103/PhysRevLett.121.086602} {\bibfield  {journal}
  {\bibinfo  {journal} {Phys. Rev. Lett.}\ }\textbf {\bibinfo {volume} {121}},\
  \bibinfo {pages} {086602} (\bibinfo {year} {2018})}\BibitemShut {NoStop}%
\bibitem [{\citenamefont {Choi}\ \emph {et~al.}(2023)\citenamefont {Choi},
  \citenamefont {Jo}, \citenamefont {Ko}, \citenamefont {Go}, \citenamefont
  {Kim}, \citenamefont {Park}, \citenamefont {Kim}, \citenamefont {Min},
  \citenamefont {Choi},\ and\ \citenamefont {Lee}}]{Choi2023}%
  \BibitemOpen
  \bibfield  {author} {\bibinfo {author} {\bibfnamefont {Y.-G.}\ \bibnamefont
  {Choi}}, \bibinfo {author} {\bibfnamefont {D.}~\bibnamefont {Jo}}, \bibinfo
  {author} {\bibfnamefont {K.-H.}\ \bibnamefont {Ko}}, \bibinfo {author}
  {\bibfnamefont {D.}~\bibnamefont {Go}}, \bibinfo {author} {\bibfnamefont
  {K.-H.}\ \bibnamefont {Kim}}, \bibinfo {author} {\bibfnamefont {H.~G.}\
  \bibnamefont {Park}}, \bibinfo {author} {\bibfnamefont {C.}~\bibnamefont
  {Kim}}, \bibinfo {author} {\bibfnamefont {B.-C.}\ \bibnamefont {Min}},
  \bibinfo {author} {\bibfnamefont {G.-M.}\ \bibnamefont {Choi}},\ and\
  \bibinfo {author} {\bibfnamefont {H.-W.}\ \bibnamefont {Lee}},\ }\bibfield
  {title} {\bibinfo {title} {Observation of the orbital hall effect in a light
  metal ti},\ }\href {https://doi.org/10.1038/s41586-023-06101-9} {\bibfield
  {journal} {\bibinfo  {journal} {Nature}\ }\textbf {\bibinfo {volume} {619}},\
  \bibinfo {pages} {52} (\bibinfo {year} {2023})}\BibitemShut {NoStop}%
\bibitem [{\citenamefont {Lyalin}\ \emph {et~al.}(2023)\citenamefont {Lyalin},
  \citenamefont {Alikhah}, \citenamefont {Berritta}, \citenamefont {Oppeneer},\
  and\ \citenamefont {Kawakami}}]{Lyalin2023}%
  \BibitemOpen
  \bibfield  {author} {\bibinfo {author} {\bibfnamefont {I.}~\bibnamefont
  {Lyalin}}, \bibinfo {author} {\bibfnamefont {S.}~\bibnamefont {Alikhah}},
  \bibinfo {author} {\bibfnamefont {M.}~\bibnamefont {Berritta}}, \bibinfo
  {author} {\bibfnamefont {P.~M.}\ \bibnamefont {Oppeneer}},\ and\ \bibinfo
  {author} {\bibfnamefont {R.~K.}\ \bibnamefont {Kawakami}},\ }\bibfield
  {title} {\bibinfo {title} {Magneto-optical detection of the orbital hall
  effect in chromium},\ }\href {https://doi.org/10.1103/PhysRevLett.131.156702}
  {\bibfield  {journal} {\bibinfo  {journal} {Phys. Rev. Lett.}\ }\textbf
  {\bibinfo {volume} {131}},\ \bibinfo {pages} {156702} (\bibinfo {year}
  {2023})}\BibitemShut {NoStop}%
\bibitem [{\citenamefont {Tang}\ and\ \citenamefont {Bauer}(2024)}]{Tang2024}%
  \BibitemOpen
  \bibfield  {author} {\bibinfo {author} {\bibfnamefont {P.}~\bibnamefont
  {Tang}}\ and\ \bibinfo {author} {\bibfnamefont {G.~E.~W.}\ \bibnamefont
  {Bauer}},\ }\bibfield  {title} {\bibinfo {title} {Role of disorder in the
  intrinsic orbital hall effect},\ }\href
  {https://doi.org/10.1103/PhysRevLett.133.186302} {\bibfield  {journal}
  {\bibinfo  {journal} {Phys. Rev. Lett.}\ }\textbf {\bibinfo {volume} {133}},\
  \bibinfo {pages} {186302} (\bibinfo {year} {2024})}\BibitemShut {NoStop}%
\bibitem [{\citenamefont {Atencia}\ \emph
  {et~al.}(2024{\natexlab{a}})\citenamefont {Atencia}, \citenamefont {Arovas},\
  and\ \citenamefont {Culcer}}]{Atencia2024}%
  \BibitemOpen
  \bibfield  {author} {\bibinfo {author} {\bibfnamefont {R.~B.}\ \bibnamefont
  {Atencia}}, \bibinfo {author} {\bibfnamefont {D.~P.}\ \bibnamefont
  {Arovas}},\ and\ \bibinfo {author} {\bibfnamefont {D.}~\bibnamefont
  {Culcer}},\ }\bibfield  {title} {\bibinfo {title} {Intrinsic torque on the
  orbital angular momentum in an electric field},\ }\href
  {https://doi.org/10.1103/PhysRevB.110.035427} {\bibfield  {journal} {\bibinfo
   {journal} {Phys. Rev. B}\ }\textbf {\bibinfo {volume} {110}},\ \bibinfo
  {pages} {035427} (\bibinfo {year} {2024}{\natexlab{a}})}\BibitemShut
  {NoStop}%
\bibitem [{\citenamefont {Liu}\ and\ \citenamefont {Culcer}(2024)}]{Hong2024}%
  \BibitemOpen
  \bibfield  {author} {\bibinfo {author} {\bibfnamefont {H.}~\bibnamefont
  {Liu}}\ and\ \bibinfo {author} {\bibfnamefont {D.}~\bibnamefont {Culcer}},\
  }\bibfield  {title} {\bibinfo {title} {Dominance of extrinsic scattering
  mechanisms in the orbital hall effect: Graphene, transition metal
  dichalcogenides, and topological antiferromagnets},\ }\href
  {https://doi.org/10.1103/PhysRevLett.132.186302} {\bibfield  {journal}
  {\bibinfo  {journal} {Phys. Rev. Lett.}\ }\textbf {\bibinfo {volume} {132}},\
  \bibinfo {pages} {186302} (\bibinfo {year} {2024})}\BibitemShut {NoStop}%
\bibitem [{\citenamefont {Cullen}\ \emph {et~al.}(2025)\citenamefont {Cullen},
  \citenamefont {Arovas}, \citenamefont {Raimondi},\ and\ \citenamefont
  {Culcer}}]{James2025}%
  \BibitemOpen
  \bibfield  {author} {\bibinfo {author} {\bibfnamefont {J.~H.}\ \bibnamefont
  {Cullen}}, \bibinfo {author} {\bibfnamefont {D.~P.}\ \bibnamefont {Arovas}},
  \bibinfo {author} {\bibfnamefont {R.}~\bibnamefont {Raimondi}},\ and\
  \bibinfo {author} {\bibfnamefont {D.}~\bibnamefont {Culcer}},\ }\href
  {https://arxiv.org/abs/2505.02911} {\bibinfo {title} {Quantum geometry and
  dipolar dynamics in the orbital magneto-electric effect}} (\bibinfo {year}
  {2025}),\ \Eprint {https://arxiv.org/abs/2505.02911} {arXiv:2505.02911
  [cond-mat.mes-hall]} \BibitemShut {NoStop}%
\bibitem [{\citenamefont {Wunderlich}\ \emph {et~al.}(2005)\citenamefont
  {Wunderlich}, \citenamefont {Kaestner}, \citenamefont {Sinova},\ and\
  \citenamefont {Jungwirth}}]{Wunderlich2005}%
  \BibitemOpen
  \bibfield  {author} {\bibinfo {author} {\bibfnamefont {J.}~\bibnamefont
  {Wunderlich}}, \bibinfo {author} {\bibfnamefont {B.}~\bibnamefont
  {Kaestner}}, \bibinfo {author} {\bibfnamefont {J.}~\bibnamefont {Sinova}},\
  and\ \bibinfo {author} {\bibfnamefont {T.}~\bibnamefont {Jungwirth}},\
  }\bibfield  {title} {\bibinfo {title} {Experimental observation of the
  spin-hall effect in a two-dimensional spin-orbit coupled semiconductor
  system},\ }\href {https://doi.org/10.1103/PhysRevLett.94.047204} {\bibfield
  {journal} {\bibinfo  {journal} {Phys. Rev. Lett.}\ }\textbf {\bibinfo
  {volume} {94}},\ \bibinfo {pages} {047204} (\bibinfo {year}
  {2005})}\BibitemShut {NoStop}%
\bibitem [{\citenamefont {Onoda}\ \emph {et~al.}(2006)\citenamefont {Onoda},
  \citenamefont {Sugimoto},\ and\ \citenamefont {Nagaosa}}]{Onoda2006}%
  \BibitemOpen
  \bibfield  {author} {\bibinfo {author} {\bibfnamefont {S.}~\bibnamefont
  {Onoda}}, \bibinfo {author} {\bibfnamefont {N.}~\bibnamefont {Sugimoto}},\
  and\ \bibinfo {author} {\bibfnamefont {N.}~\bibnamefont {Nagaosa}},\
  }\bibfield  {title} {\bibinfo {title} {Intrinsic versus extrinsic anomalous
  hall effect in ferromagnets},\ }\href
  {https://doi.org/10.1103/PhysRevLett.97.126602} {\bibfield  {journal}
  {\bibinfo  {journal} {Phys. Rev. Lett.}\ }\textbf {\bibinfo {volume} {97}},\
  \bibinfo {pages} {126602} (\bibinfo {year} {2006})}\BibitemShut {NoStop}%
\bibitem [{\citenamefont {Onoda}\ \emph {et~al.}(2008)\citenamefont {Onoda},
  \citenamefont {Sugimoto},\ and\ \citenamefont {Nagaosa}}]{Onoda2008}%
  \BibitemOpen
  \bibfield  {author} {\bibinfo {author} {\bibfnamefont {S.}~\bibnamefont
  {Onoda}}, \bibinfo {author} {\bibfnamefont {N.}~\bibnamefont {Sugimoto}},\
  and\ \bibinfo {author} {\bibfnamefont {N.}~\bibnamefont {Nagaosa}},\
  }\bibfield  {title} {\bibinfo {title} {Quantum transport theory of anomalous
  electric, thermoelectric, and thermal hall effects in ferromagnets},\ }\href
  {https://doi.org/10.1103/PhysRevB.77.165103} {\bibfield  {journal} {\bibinfo
  {journal} {Phys. Rev. B}\ }\textbf {\bibinfo {volume} {77}},\ \bibinfo
  {pages} {165103} (\bibinfo {year} {2008})}\BibitemShut {NoStop}%
\bibitem [{\citenamefont {Xiao}\ \emph {et~al.}(2010)\citenamefont {Xiao},
  \citenamefont {Chang},\ and\ \citenamefont {Niu}}]{Xiao2010}%
  \BibitemOpen
  \bibfield  {author} {\bibinfo {author} {\bibfnamefont {D.}~\bibnamefont
  {Xiao}}, \bibinfo {author} {\bibfnamefont {M.-C.}\ \bibnamefont {Chang}},\
  and\ \bibinfo {author} {\bibfnamefont {Q.}~\bibnamefont {Niu}},\ }\bibfield
  {title} {\bibinfo {title} {Berry phase effects on electronic properties},\
  }\href {https://doi.org/10.1103/RevModPhys.82.1959} {\bibfield  {journal}
  {\bibinfo  {journal} {Rev. Mod. Phys.}\ }\textbf {\bibinfo {volume} {82}},\
  \bibinfo {pages} {1959} (\bibinfo {year} {2010})}\BibitemShut {NoStop}%
\bibitem [{\citenamefont {Niimi}\ and\ \citenamefont
  {Otani}(2015)}]{Niimi2015}%
  \BibitemOpen
  \bibfield  {author} {\bibinfo {author} {\bibfnamefont {Y.}~\bibnamefont
  {Niimi}}\ and\ \bibinfo {author} {\bibfnamefont {Y.}~\bibnamefont {Otani}},\
  }\bibfield  {title} {\bibinfo {title} {Reciprocal spin hall effects in
  conductors with strong spin–orbit coupling: a review},\ }\href
  {https://doi.org/10.1088/0034-4885/78/12/124501} {\bibfield  {journal}
  {\bibinfo  {journal} {Reports on Progress in Physics}\ }\textbf {\bibinfo
  {volume} {78}},\ \bibinfo {pages} {124501} (\bibinfo {year}
  {2015})}\BibitemShut {NoStop}%
\bibitem [{\citenamefont {Sekine}\ \emph {et~al.}(2017)\citenamefont {Sekine},
  \citenamefont {Culcer},\ and\ \citenamefont {MacDonald}}]{Sekine2017}%
  \BibitemOpen
  \bibfield  {author} {\bibinfo {author} {\bibfnamefont {A.}~\bibnamefont
  {Sekine}}, \bibinfo {author} {\bibfnamefont {D.}~\bibnamefont {Culcer}},\
  and\ \bibinfo {author} {\bibfnamefont {A.~H.}\ \bibnamefont {MacDonald}},\
  }\bibfield  {title} {\bibinfo {title} {Quantum kinetic theory of the chiral
  anomaly},\ }\href {https://doi.org/10.1103/PhysRevB.96.235134} {\bibfield
  {journal} {\bibinfo  {journal} {Phys. Rev. B}\ }\textbf {\bibinfo {volume}
  {96}},\ \bibinfo {pages} {235134} (\bibinfo {year} {2017})}\BibitemShut
  {NoStop}%
\bibitem [{\citenamefont {Zala}\ \emph {et~al.}(2001)\citenamefont {Zala},
  \citenamefont {Narozhny},\ and\ \citenamefont {Aleiner}}]{Zala2001}%
  \BibitemOpen
  \bibfield  {author} {\bibinfo {author} {\bibfnamefont {G.}~\bibnamefont
  {Zala}}, \bibinfo {author} {\bibfnamefont {B.~N.}\ \bibnamefont {Narozhny}},\
  and\ \bibinfo {author} {\bibfnamefont {I.~L.}\ \bibnamefont {Aleiner}},\
  }\bibfield  {title} {\bibinfo {title} {Interaction corrections at
  intermediate temperatures: Longitudinal conductivity and kinetic equation},\
  }\href {https://doi.org/10.1103/PhysRevB.64.214204} {\bibfield  {journal}
  {\bibinfo  {journal} {Phys. Rev. B}\ }\textbf {\bibinfo {volume} {64}},\
  \bibinfo {pages} {214204} (\bibinfo {year} {2001})}\BibitemShut {NoStop}%
\bibitem [{\citenamefont {Adroguer}\ \emph {et~al.}(2015)\citenamefont
  {Adroguer}, \citenamefont {Liu}, \citenamefont {Culcer},\ and\ \citenamefont
  {Hankiewicz}}]{Adroguer2015}%
  \BibitemOpen
  \bibfield  {author} {\bibinfo {author} {\bibfnamefont {P.}~\bibnamefont
  {Adroguer}}, \bibinfo {author} {\bibfnamefont {W.~E.}\ \bibnamefont {Liu}},
  \bibinfo {author} {\bibfnamefont {D.}~\bibnamefont {Culcer}},\ and\ \bibinfo
  {author} {\bibfnamefont {E.~M.}\ \bibnamefont {Hankiewicz}},\ }\bibfield
  {title} {\bibinfo {title} {Conductivity corrections for topological
  insulators with spin-orbit impurities: Hikami-larkin-nagaoka formula
  revisited},\ }\href {https://doi.org/10.1103/PhysRevB.92.241402} {\bibfield
  {journal} {\bibinfo  {journal} {Phys. Rev. B}\ }\textbf {\bibinfo {volume}
  {92}},\ \bibinfo {pages} {241402} (\bibinfo {year} {2015})}\BibitemShut
  {NoStop}%
\bibitem [{\citenamefont {Liu}\ \emph {et~al.}(2017)\citenamefont {Liu},
  \citenamefont {Hankiewicz},\ and\ \citenamefont {Culcer}}]{Liu2017}%
  \BibitemOpen
  \bibfield  {author} {\bibinfo {author} {\bibfnamefont {W.~E.}\ \bibnamefont
  {Liu}}, \bibinfo {author} {\bibfnamefont {E.~M.}\ \bibnamefont
  {Hankiewicz}},\ and\ \bibinfo {author} {\bibfnamefont {D.}~\bibnamefont
  {Culcer}},\ }\bibfield  {title} {\bibinfo {title} {Weak localization and
  antilocalization in topological materials with impurity spin-orbit
  interactions},\ }\bibfield  {journal} {\bibinfo  {journal} {Materials}\
  }\textbf {\bibinfo {volume} {10}},\ \href
  {https://doi.org/10.3390/ma10070807} {10.3390/ma10070807} (\bibinfo {year}
  {2017})\BibitemShut {NoStop}%
\bibitem [{\citenamefont {Liu}\ \emph {et~al.}(2025)\citenamefont {Liu},
  \citenamefont {Cullen}, \citenamefont {Arovas},\ and\ \citenamefont
  {Culcer}}]{Hong2025}%
  \BibitemOpen
  \bibfield  {author} {\bibinfo {author} {\bibfnamefont {H.}~\bibnamefont
  {Liu}}, \bibinfo {author} {\bibfnamefont {J.~H.}\ \bibnamefont {Cullen}},
  \bibinfo {author} {\bibfnamefont {D.~P.}\ \bibnamefont {Arovas}},\ and\
  \bibinfo {author} {\bibfnamefont {D.}~\bibnamefont {Culcer}},\ }\bibfield
  {title} {\bibinfo {title} {Quantum correction to the orbital hall effect},\
  }\href {https://doi.org/10.1103/PhysRevLett.134.036304} {\bibfield  {journal}
  {\bibinfo  {journal} {Phys. Rev. Lett.}\ }\textbf {\bibinfo {volume} {134}},\
  \bibinfo {pages} {036304} (\bibinfo {year} {2025})}\BibitemShut {NoStop}%
\bibitem [{\citenamefont {Bhalla}\ \emph {et~al.}(2023)\citenamefont {Bhalla},
  \citenamefont {Das}, \citenamefont {Agarwal},\ and\ \citenamefont
  {Culcer}}]{Bhalla2023}%
  \BibitemOpen
  \bibfield  {author} {\bibinfo {author} {\bibfnamefont {P.}~\bibnamefont
  {Bhalla}}, \bibinfo {author} {\bibfnamefont {K.}~\bibnamefont {Das}},
  \bibinfo {author} {\bibfnamefont {A.}~\bibnamefont {Agarwal}},\ and\ \bibinfo
  {author} {\bibfnamefont {D.}~\bibnamefont {Culcer}},\ }\bibfield  {title}
  {\bibinfo {title} {Quantum kinetic theory of nonlinear optical currents:
  Finite fermi surface and fermi sea contributions},\ }\href
  {https://doi.org/10.1103/PhysRevB.107.165131} {\bibfield  {journal} {\bibinfo
   {journal} {Phys. Rev. B}\ }\textbf {\bibinfo {volume} {107}},\ \bibinfo
  {pages} {165131} (\bibinfo {year} {2023})}\BibitemShut {NoStop}%
\bibitem [{\citenamefont {Pandey}\ \emph {et~al.}(2024)\citenamefont {Pandey},
  \citenamefont {Joy}, \citenamefont {Culcer},\ and\ \citenamefont
  {Bhalla}}]{Pandey2024}%
  \BibitemOpen
  \bibfield  {author} {\bibinfo {author} {\bibfnamefont {V.}~\bibnamefont
  {Pandey}}, \bibinfo {author} {\bibfnamefont {D.}~\bibnamefont {Joy}},
  \bibinfo {author} {\bibfnamefont {D.}~\bibnamefont {Culcer}},\ and\ \bibinfo
  {author} {\bibfnamefont {P.}~\bibnamefont {Bhalla}},\ }\bibfield  {title}
  {\bibinfo {title} {Longitudinal dc conductivity in dirac nodal line
  semimetals: Intrinsic and extrinsic contributions},\ }\href
  {https://doi.org/10.1103/PhysRevB.110.155108} {\bibfield  {journal} {\bibinfo
   {journal} {Phys. Rev. B}\ }\textbf {\bibinfo {volume} {110}},\ \bibinfo
  {pages} {155108} (\bibinfo {year} {2024})}\BibitemShut {NoStop}%
\bibitem [{\citenamefont {Atencia}\ \emph {et~al.}(2025)\citenamefont
  {Atencia}, \citenamefont {Liu}, \citenamefont {Loh},\ and\ \citenamefont
  {Culcer}}]{Rhonald2025}%
  \BibitemOpen
  \bibfield  {author} {\bibinfo {author} {\bibfnamefont {R.~B.}\ \bibnamefont
  {Atencia}}, \bibinfo {author} {\bibfnamefont {S.}~\bibnamefont {Liu}},
  \bibinfo {author} {\bibfnamefont {K.~P.}\ \bibnamefont {Loh}},\ and\ \bibinfo
  {author} {\bibfnamefont {D.}~\bibnamefont {Culcer}},\ }\href
  {https://arxiv.org/abs/2506.13869} {\bibinfo {title} {Room-temperature
  disorder-driven nonlinear transport in topological materials}} (\bibinfo
  {year} {2025}),\ \Eprint {https://arxiv.org/abs/2506.13869} {arXiv:2506.13869
  [cond-mat.mes-hall]} \BibitemShut {NoStop}%
\bibitem [{\citenamefont {Bhalla}\ \emph {et~al.}(2020)\citenamefont {Bhalla},
  \citenamefont {MacDonald},\ and\ \citenamefont {Culcer}}]{Bhalla2020}%
  \BibitemOpen
  \bibfield  {author} {\bibinfo {author} {\bibfnamefont {P.}~\bibnamefont
  {Bhalla}}, \bibinfo {author} {\bibfnamefont {A.~H.}\ \bibnamefont
  {MacDonald}},\ and\ \bibinfo {author} {\bibfnamefont {D.}~\bibnamefont
  {Culcer}},\ }\bibfield  {title} {\bibinfo {title} {Resonant photovoltaic
  effect in doped magnetic semiconductors},\ }\href
  {https://doi.org/10.1103/PhysRevLett.124.087402} {\bibfield  {journal}
  {\bibinfo  {journal} {Phys. Rev. Lett.}\ }\textbf {\bibinfo {volume} {124}},\
  \bibinfo {pages} {087402} (\bibinfo {year} {2020})}\BibitemShut {NoStop}%
\bibitem [{\citenamefont {Bhalla}\ \emph {et~al.}(2022)\citenamefont {Bhalla},
  \citenamefont {Das}, \citenamefont {Culcer},\ and\ \citenamefont
  {Agarwal}}]{Bhalla2022}%
  \BibitemOpen
  \bibfield  {author} {\bibinfo {author} {\bibfnamefont {P.}~\bibnamefont
  {Bhalla}}, \bibinfo {author} {\bibfnamefont {K.}~\bibnamefont {Das}},
  \bibinfo {author} {\bibfnamefont {D.}~\bibnamefont {Culcer}},\ and\ \bibinfo
  {author} {\bibfnamefont {A.}~\bibnamefont {Agarwal}},\ }\bibfield  {title}
  {\bibinfo {title} {Resonant second-harmonic generation as a probe of quantum
  geometry},\ }\href {https://doi.org/10.1103/PhysRevLett.129.227401}
  {\bibfield  {journal} {\bibinfo  {journal} {Phys. Rev. Lett.}\ }\textbf
  {\bibinfo {volume} {129}},\ \bibinfo {pages} {227401} (\bibinfo {year}
  {2022})}\BibitemShut {NoStop}%
\bibitem [{\citenamefont {Cullen}\ and\ \citenamefont
  {Culcer}(2023)}]{Cullen2023}%
  \BibitemOpen
  \bibfield  {author} {\bibinfo {author} {\bibfnamefont {J.~H.}\ \bibnamefont
  {Cullen}}\ and\ \bibinfo {author} {\bibfnamefont {D.}~\bibnamefont
  {Culcer}},\ }\bibfield  {title} {\bibinfo {title} {Spin-hall effect due to
  the bulk states of topological insulators: Extrinsic contribution to the
  proper spin current},\ }\href {https://doi.org/10.1103/PhysRevB.108.245418}
  {\bibfield  {journal} {\bibinfo  {journal} {Phys. Rev. B}\ }\textbf {\bibinfo
  {volume} {108}},\ \bibinfo {pages} {245418} (\bibinfo {year}
  {2023})}\BibitemShut {NoStop}%
\bibitem [{\citenamefont {Das}\ \emph {et~al.}(2023{\natexlab{a}})\citenamefont
  {Das}, \citenamefont {Lahiri}, \citenamefont {Atencia}, \citenamefont
  {Culcer},\ and\ \citenamefont {Agarwal}}]{Das2023}%
  \BibitemOpen
  \bibfield  {author} {\bibinfo {author} {\bibfnamefont {K.}~\bibnamefont
  {Das}}, \bibinfo {author} {\bibfnamefont {S.}~\bibnamefont {Lahiri}},
  \bibinfo {author} {\bibfnamefont {R.~B.}\ \bibnamefont {Atencia}}, \bibinfo
  {author} {\bibfnamefont {D.}~\bibnamefont {Culcer}},\ and\ \bibinfo {author}
  {\bibfnamefont {A.}~\bibnamefont {Agarwal}},\ }\bibfield  {title} {\bibinfo
  {title} {Intrinsic nonlinear conductivities induced by the quantum metric},\
  }\href {https://doi.org/10.1103/PhysRevB.108.L201405} {\bibfield  {journal}
  {\bibinfo  {journal} {Phys. Rev. B}\ }\textbf {\bibinfo {volume} {108}},\
  \bibinfo {pages} {L201405} (\bibinfo {year}
  {2023}{\natexlab{a}})}\BibitemShut {NoStop}%
\bibitem [{\citenamefont {Das}\ \emph {et~al.}(2024)\citenamefont {Das},
  \citenamefont {Ghorai}, \citenamefont {Culcer},\ and\ \citenamefont
  {Agarwal}}]{Das2024}%
  \BibitemOpen
  \bibfield  {author} {\bibinfo {author} {\bibfnamefont {K.}~\bibnamefont
  {Das}}, \bibinfo {author} {\bibfnamefont {K.}~\bibnamefont {Ghorai}},
  \bibinfo {author} {\bibfnamefont {D.}~\bibnamefont {Culcer}},\ and\ \bibinfo
  {author} {\bibfnamefont {A.}~\bibnamefont {Agarwal}},\ }\bibfield  {title}
  {\bibinfo {title} {Nonlinear valley hall effect},\ }\href
  {https://doi.org/10.1103/PhysRevLett.132.096302} {\bibfield  {journal}
  {\bibinfo  {journal} {Phys. Rev. Lett.}\ }\textbf {\bibinfo {volume} {132}},\
  \bibinfo {pages} {096302} (\bibinfo {year} {2024})}\BibitemShut {NoStop}%
\bibitem [{\citenamefont {Culcer}(2022)}]{Culcer2022AHE}%
  \BibitemOpen
  \bibfield  {author} {\bibinfo {author} {\bibfnamefont {D.}~\bibnamefont
  {Culcer}},\ }\href {https://doi.org/10.1016/B978-0-323-90800-9.00006-8}
  {\bibinfo {title} {The anomalous hall effect}} (\bibinfo {year} {2022}),\
  \bibinfo {note} {book chapter for Elsevier Encyclopedia of Condensed Matter
  Physics},\ \Eprint {https://arxiv.org/abs/2204.02434} {arXiv:2204.02434
  [cond-mat.mes-hall]} \BibitemShut {NoStop}%
\bibitem [{\citenamefont {Lee}\ and\ \citenamefont
  {Ramakrishnan}(1985)}]{Lee1985}%
  \BibitemOpen
  \bibfield  {author} {\bibinfo {author} {\bibfnamefont {P.~A.}\ \bibnamefont
  {Lee}}\ and\ \bibinfo {author} {\bibfnamefont {T.~V.}\ \bibnamefont
  {Ramakrishnan}},\ }\bibfield  {title} {\bibinfo {title} {Disordered
  electronic systems},\ }\href {https://doi.org/10.1103/RevModPhys.57.287}
  {\bibfield  {journal} {\bibinfo  {journal} {Rev. Mod. Phys.}\ }\textbf
  {\bibinfo {volume} {57}},\ \bibinfo {pages} {287} (\bibinfo {year}
  {1985})}\BibitemShut {NoStop}%
\bibitem [{\citenamefont {Rammer}\ and\ \citenamefont
  {Smith}(1986)}]{Rammer1986}%
  \BibitemOpen
  \bibfield  {author} {\bibinfo {author} {\bibfnamefont {J.}~\bibnamefont
  {Rammer}}\ and\ \bibinfo {author} {\bibfnamefont {H.}~\bibnamefont {Smith}},\
  }\bibfield  {title} {\bibinfo {title} {Quantum field-theoretical methods in
  transport theory of metals},\ }\href
  {https://doi.org/10.1103/RevModPhys.58.323} {\bibfield  {journal} {\bibinfo
  {journal} {Rev. Mod. Phys.}\ }\textbf {\bibinfo {volume} {58}},\ \bibinfo
  {pages} {323} (\bibinfo {year} {1986})}\BibitemShut {NoStop}%
\bibitem [{\citenamefont {Kovalev}\ \emph {et~al.}(2008)\citenamefont
  {Kovalev}, \citenamefont {V\'yborn\'y},\ and\ \citenamefont
  {Sinova}}]{Kovalev2008}%
  \BibitemOpen
  \bibfield  {author} {\bibinfo {author} {\bibfnamefont {A.~A.}\ \bibnamefont
  {Kovalev}}, \bibinfo {author} {\bibfnamefont {K.}~\bibnamefont
  {V\'yborn\'y}},\ and\ \bibinfo {author} {\bibfnamefont {J.}~\bibnamefont
  {Sinova}},\ }\bibfield  {title} {\bibinfo {title} {Hybrid skew scattering
  regime of the anomalous hall effect in rashba systems: Unifying keldysh,
  boltzmann, and kubo formalisms},\ }\href
  {https://doi.org/10.1103/PhysRevB.78.041305} {\bibfield  {journal} {\bibinfo
  {journal} {Phys. Rev. B}\ }\textbf {\bibinfo {volume} {78}},\ \bibinfo
  {pages} {041305} (\bibinfo {year} {2008})}\BibitemShut {NoStop}%
\bibitem [{\citenamefont {Kamenev}\ and\ \citenamefont
  {Levchenko}(2009)}]{Kamenev2009}%
  \BibitemOpen
  \bibfield  {author} {\bibinfo {author} {\bibfnamefont {A.}~\bibnamefont
  {Kamenev}}\ and\ \bibinfo {author} {\bibfnamefont {A.}~\bibnamefont
  {Levchenko}},\ }\bibfield  {title} {\bibinfo {title} {Keldysh technique and
  non-linear $\sigma$-model: basic principles and applications},\ }\href
  {https://doi.org/10.1080/00018730902850504} {\bibfield  {journal} {\bibinfo
  {journal} {Advances in Physics}\ }\textbf {\bibinfo {volume} {58}},\ \bibinfo
  {pages} {197} (\bibinfo {year} {2009})},\ \Eprint
  {https://arxiv.org/abs/https://doi.org/10.1080/00018730902850504}
  {https://doi.org/10.1080/00018730902850504} \BibitemShut {NoStop}%
\bibitem [{\citenamefont {Kovalev}\ \emph {et~al.}(2009)\citenamefont
  {Kovalev}, \citenamefont {Tserkovnyak}, \citenamefont {V\'yborn\'y},\ and\
  \citenamefont {Sinova}}]{Kovalev2009}%
  \BibitemOpen
  \bibfield  {author} {\bibinfo {author} {\bibfnamefont {A.~A.}\ \bibnamefont
  {Kovalev}}, \bibinfo {author} {\bibfnamefont {Y.}~\bibnamefont
  {Tserkovnyak}}, \bibinfo {author} {\bibfnamefont {K.}~\bibnamefont
  {V\'yborn\'y}},\ and\ \bibinfo {author} {\bibfnamefont {J.}~\bibnamefont
  {Sinova}},\ }\bibfield  {title} {\bibinfo {title} {Transport theory for
  disordered multiple-band systems: Anomalous hall effect and anisotropic
  magnetoresistance},\ }\href {https://doi.org/10.1103/PhysRevB.79.195129}
  {\bibfield  {journal} {\bibinfo  {journal} {Phys. Rev. B}\ }\textbf {\bibinfo
  {volume} {79}},\ \bibinfo {pages} {195129} (\bibinfo {year}
  {2009})}\BibitemShut {NoStop}%
\bibitem [{\citenamefont {Gorini}\ \emph {et~al.}(2012)\citenamefont {Gorini},
  \citenamefont {Raimondi},\ and\ \citenamefont {Schwab}}]{Gorini2012}%
  \BibitemOpen
  \bibfield  {author} {\bibinfo {author} {\bibfnamefont {C.}~\bibnamefont
  {Gorini}}, \bibinfo {author} {\bibfnamefont {R.}~\bibnamefont {Raimondi}},\
  and\ \bibinfo {author} {\bibfnamefont {P.}~\bibnamefont {Schwab}},\
  }\bibfield  {title} {\bibinfo {title} {Onsager relations in a two-dimensional
  electron gas with spin-orbit coupling},\ }\href
  {https://doi.org/10.1103/PhysRevLett.109.246604} {\bibfield  {journal}
  {\bibinfo  {journal} {Phys. Rev. Lett.}\ }\textbf {\bibinfo {volume} {109}},\
  \bibinfo {pages} {246604} (\bibinfo {year} {2012})}\BibitemShut {NoStop}%
\bibitem [{\citenamefont {Xiao}\ \emph {et~al.}(2019)\citenamefont {Xiao},
  \citenamefont {Du},\ and\ \citenamefont {Niu}}]{Xiao2019}%
  \BibitemOpen
  \bibfield  {author} {\bibinfo {author} {\bibfnamefont {C.}~\bibnamefont
  {Xiao}}, \bibinfo {author} {\bibfnamefont {Z.~Z.}\ \bibnamefont {Du}},\ and\
  \bibinfo {author} {\bibfnamefont {Q.}~\bibnamefont {Niu}},\ }\bibfield
  {title} {\bibinfo {title} {Theory of nonlinear hall effects: Modified
  semiclassics from quantum kinetics},\ }\href
  {https://doi.org/10.1103/PhysRevB.100.165422} {\bibfield  {journal} {\bibinfo
   {journal} {Phys. Rev. B}\ }\textbf {\bibinfo {volume} {100}},\ \bibinfo
  {pages} {165422} (\bibinfo {year} {2019})}\BibitemShut {NoStop}%
\bibitem [{\citenamefont {Nandy}\ and\ \citenamefont
  {Sodemann}(2019)}]{Nandy2019}%
  \BibitemOpen
  \bibfield  {author} {\bibinfo {author} {\bibfnamefont {S.}~\bibnamefont
  {Nandy}}\ and\ \bibinfo {author} {\bibfnamefont {I.}~\bibnamefont
  {Sodemann}},\ }\bibfield  {title} {\bibinfo {title} {Symmetry and quantum
  kinetics of the nonlinear hall effect},\ }\href
  {https://doi.org/10.1103/PhysRevB.100.195117} {\bibfield  {journal} {\bibinfo
   {journal} {Phys. Rev. B}\ }\textbf {\bibinfo {volume} {100}},\ \bibinfo
  {pages} {195117} (\bibinfo {year} {2019})}\BibitemShut {NoStop}%
\bibitem [{\citenamefont {Atencia}\ \emph
  {et~al.}(2024{\natexlab{b}})\citenamefont {Atencia}, \citenamefont
  {Agarwal},\ and\ \citenamefont {Culcer}}]{Rhonald2024}%
  \BibitemOpen
  \bibfield  {author} {\bibinfo {author} {\bibfnamefont {R.~B.}\ \bibnamefont
  {Atencia}}, \bibinfo {author} {\bibfnamefont {A.}~\bibnamefont {Agarwal}},\
  and\ \bibinfo {author} {\bibfnamefont {D.}~\bibnamefont {Culcer}},\
  }\bibfield  {title} {\bibinfo {title} {Orbital angular momentum of bloch
  electrons: equilibrium formulation, magneto-electric phenomena, and the
  orbital hall effect},\ }\href {https://doi.org/10.1080/23746149.2024.2371972}
  {\bibfield  {journal} {\bibinfo  {journal} {Advances in Physics: X}\ }\textbf
  {\bibinfo {volume} {9}},\ \bibinfo {pages} {2371972} (\bibinfo {year}
  {2024}{\natexlab{b}})},\ \Eprint
  {https://arxiv.org/abs/https://doi.org/10.1080/23746149.2024.2371972}
  {https://doi.org/10.1080/23746149.2024.2371972} \BibitemShut {NoStop}%
\bibitem [{\citenamefont {Das}\ \emph {et~al.}(2023{\natexlab{b}})\citenamefont
  {Das}, \citenamefont {Lahiri}, \citenamefont {Atencia}, \citenamefont
  {Culcer},\ and\ \citenamefont {Agarwal}}]{Kamal2023}%
  \BibitemOpen
  \bibfield  {author} {\bibinfo {author} {\bibfnamefont {K.}~\bibnamefont
  {Das}}, \bibinfo {author} {\bibfnamefont {S.}~\bibnamefont {Lahiri}},
  \bibinfo {author} {\bibfnamefont {R.~B.}\ \bibnamefont {Atencia}}, \bibinfo
  {author} {\bibfnamefont {D.}~\bibnamefont {Culcer}},\ and\ \bibinfo {author}
  {\bibfnamefont {A.}~\bibnamefont {Agarwal}},\ }\bibfield  {title} {\bibinfo
  {title} {Intrinsic nonlinear conductivities induced by the quantum metric},\
  }\href {https://doi.org/10.1103/PhysRevB.108.L201405} {\bibfield  {journal}
  {\bibinfo  {journal} {Phys. Rev. B}\ }\textbf {\bibinfo {volume} {108}},\
  \bibinfo {pages} {L201405} (\bibinfo {year}
  {2023}{\natexlab{b}})}\BibitemShut {NoStop}%
\bibitem [{\citenamefont {Aversa}\ and\ \citenamefont
  {Sipe}(1995)}]{Aversa1995}%
  \BibitemOpen
  \bibfield  {author} {\bibinfo {author} {\bibfnamefont {C.}~\bibnamefont
  {Aversa}}\ and\ \bibinfo {author} {\bibfnamefont {J.~E.}\ \bibnamefont
  {Sipe}},\ }\bibfield  {title} {\bibinfo {title} {Nonlinear optical
  susceptibilities of semiconductors: Results with a length-gauge analysis},\
  }\href {https://doi.org/10.1103/PhysRevB.52.14636} {\bibfield  {journal}
  {\bibinfo  {journal} {Phys. Rev. B}\ }\textbf {\bibinfo {volume} {52}},\
  \bibinfo {pages} {14636} (\bibinfo {year} {1995})}\BibitemShut {NoStop}%
\bibitem [{\citenamefont {Morimoto}\ and\ \citenamefont
  {Nagaosa}(2016)}]{Takahiro2016}%
  \BibitemOpen
  \bibfield  {author} {\bibinfo {author} {\bibfnamefont {T.}~\bibnamefont
  {Morimoto}}\ and\ \bibinfo {author} {\bibfnamefont {N.}~\bibnamefont
  {Nagaosa}},\ }\bibfield  {title} {\bibinfo {title} {Topological nature of
  nonlinear optical effects in solids},\ }\href
  {https://doi.org/10.1126/sciadv.1501524} {\bibfield  {journal} {\bibinfo
  {journal} {Science Advances}\ }\textbf {\bibinfo {volume} {2}},\ \bibinfo
  {pages} {e1501524} (\bibinfo {year} {2016})},\ \Eprint
  {https://arxiv.org/abs/https://www.science.org/doi/pdf/10.1126/sciadv.1501524}
  {https://www.science.org/doi/pdf/10.1126/sciadv.1501524} \BibitemShut
  {NoStop}%
\bibitem [{\citenamefont {Parks}\ \emph {et~al.}(2023)\citenamefont {Parks},
  \citenamefont {Moloney},\ and\ \citenamefont {Brabec}}]{Parks2023}%
  \BibitemOpen
  \bibfield  {author} {\bibinfo {author} {\bibfnamefont {A.~M.}\ \bibnamefont
  {Parks}}, \bibinfo {author} {\bibfnamefont {J.~V.}\ \bibnamefont {Moloney}},\
  and\ \bibinfo {author} {\bibfnamefont {T.}~\bibnamefont {Brabec}},\
  }\bibfield  {title} {\bibinfo {title} {Gauge invariant formulation of the
  semiconductor bloch equations},\ }\href
  {https://doi.org/10.1103/PhysRevLett.131.236902} {\bibfield  {journal}
  {\bibinfo  {journal} {Phys. Rev. Lett.}\ }\textbf {\bibinfo {volume} {131}},\
  \bibinfo {pages} {236902} (\bibinfo {year} {2023})}\BibitemShut {NoStop}%
\end{thebibliography}
\end{document}